\documentclass[1p,sort&compress]{elsarticle}
\pdfoutput=1   
\usepackage{float}
\usepackage{amsmath,amssymb}
\usepackage{mathtools,slashed}
\usepackage{bm}
\usepackage{comment}
\usepackage{xcolor}
\usepackage{subcaption}
\usepackage{array}
\usepackage{multirow}
\usepackage{url}
\usepackage{color}
\usepackage[T1]{fontenc}
\usepackage[latin9]{inputenc}
\usepackage{graphicx}
\usepackage{esint}
\usepackage{hyperref}
\usepackage{atbegshi}
\usepackage{lipsum}
\newcommand{\gbw}{g_{\Delta,{\rm BW}}}
\newcommand{\gloeft}{g_{\Delta N\pi }}
\def\epi{\epsilon_\pi}
\def\Lchi{\Lambda_\chi}

\journal{Nuclear Physics B}

\begin{document}

\begin{flushright}
MIT-CTP/5455\\DESY-22-140
\end{flushright}

\begin{frontmatter}
\title{Elastic nucleon-pion scattering at $m_{\pi} = 200~{\rm MeV}$\\ from lattice QCD} 

\author[1]{John Bulava}
\ead{john.bulava@desy.de}
\affiliation[1]{organization={Deutsches Elektronen-Synchrotron DESY},
                addressline={Platanenallee 6},
                postcode={15738},
                city={Zeuthen},
                country={Germany}}

\author[2]{Andrew D. Hanlon}
\affiliation[2]{organization={Physics Department, Brookhaven National Laboratory},
                postcode={11973},
                city={Upton, New York},
                country={USA}}
\author[3,4]{Ben H\"{o}rz}
\affiliation[3]{organization={Nuclear Science Division, Lawrence Berkeley National Laboratory},
                postcode={94720},
                city={Berkeley, CA},
                country={USA}}
\affiliation[4]{organization={Intel Deutschland GmbH},
                addressline={Dornacher Str. 1},
                postcode={85622},
                city={Feldkirchen},
                country={Germany}}

\author[5]{Colin Morningstar}
\ead{cmorning@andrew.cmu.edu}
\affiliation[5]{organization={Dept.~of Physics, Carnegie Mellon University},
                postcode={15213},
                city={Pittsburgh, PA},
                country={USA}}

\author[6]{Amy Nicholson}
\affiliation[6]{organization={Dept.~of Physics and Astronomy, University of North Carolina},
                postcode={27516},
                city={Chapel Hill, NC},
                country={USA}}

\author[7]{Fernando Romero-L\'{o}pez}
\ead{fernando@mit.edu}
\affiliation[7]{organization={Center for Theoretical Physics, Massachusetts Inst.~of Technology},
                postcode={02139},
                city={Cambridge, MA},
                country={USA}}

\author[5]{Sarah Skinner}
\ead{sarahski@andrew.cmu.edu}

\author[8,3]{Pavlos Vranas}
\affiliation[8]{organization={Physics Division, Lawrence Livermore National Laboratory},
                postcode={94550},
                city={Livermore, CA},
                country={USA}}

\author[3]{Andr\'{e} Walker-Loud}

\date{Sept.~6, 2022}

\begin{abstract}
Elastic nucleon-pion scattering amplitudes are computed using lattice QCD on a single ensemble of gauge field 
configurations with $N_{\rm f} = 2+1$ dynamical quark flavors and $m_{\pi} = 200~{\rm  MeV}$. The $s$-wave 
scattering lengths with both total isospins $I=1/2$ and $I=3/2$ are inferred from the finite-volume spectrum 
below the inelastic threshold together with the $I=3/2$ $p$-wave containing the  $\Delta(1232)$ resonance. 
The amplitudes are well-described by the effective range expansion with parameters constrained by fits to 
the finite-volume energy levels, enabling a determination of the $I=3/2$ scattering length with statistical 
errors below 5\%, while the $I = 1/2$ scattering length is somewhat less precisely evaluated. Systematic 
errors due to excited states and the influence of higher partial waves are controlled, providing a step 
toward future computations down to physical light quark masses with multiple lattice spacings and volumes. 
\end{abstract}


\end{frontmatter}

\newpage

\section{Introduction}

Nucleon-pion ($N\pi)$ scattering is a fundamental nuclear physics process.  Because the pion is the 
lightest hadron, pion exchange between nucleons governs the long-range nuclear force and contributes 
to the binding of protons and neutrons into atomic nuclei.  Nucleon-pion scattering also gives rise 
to the narrow $\Delta(1232)$ resonance which influences many nuclear processes, including 
lepton-nucleon and lepton-nucleus scattering relevant to a range of electron-nucleus and 
neutrino-nucleus scattering experiments.

While $N\pi$ scattering is well understood experimentally and phenomenologically, such as through 
the Roy-Steiner equations~\cite{RuizdeElvira:2017stg}, the ability to determine the amplitudes directly 
from quantum chromodynamics (QCD) is hampered by its non-perturbative nature at low energies. After 
QCD was established as the underlying theory of the strong nuclear force, chiral perturbation theory 
($\chi$PT)~\cite{Gasser:1983yg} and chiral-EFT~\cite{Weinberg:1990rz,Weinberg:1991um} were developed 
to systematically describe the low-energy dynamics of pions and nucleons in an effective field theory 
(EFT) framework. For a recent review, see Ref.~\cite{Hammer:2019poc}. While EFT methods are generally 
effective in treating low-energy hadron scattering processes, a number of challenges can only be 
addressed with first-principles QCD calculations, for which lattice QCD is an essential 
non-perturbative tool. 

For example, many of the low-energy constants (LECs) of nuclear EFTs are difficult to determine from 
experimental information alone.  Lattice QCD can assist in the determination of LECs by carrying out
computations at a variety of quark masses and by computing processes which are difficult to measure 
experimentally, such as hyperon-nucleon and three-nucleon interactions, as well as short-distance 
matrix elements of electroweak and beyond-the-Standard Model operators.  See recent reviews for further 
discussion~\cite{Drischler:2019xuo,Tews:2020hgp,Davoudi:2020ngi,Cirigliano:2022oqy,Cirigliano:2022rmf}.  
The beneficial interplay between EFTs and lattice computations is already developing for meson-meson 
scattering~\cite{Niehus:2020gmf, Molina:2020qpw,Mai:2019pqr, Guo:2018zvl, Guo:2018kno, Bolton:2015psa}, 
but few lattice studies of meson-baryon scattering amplitudes currently exist.

Another important issue concerns the convergence of EFTs, which are asymptotic expansions in small 
momenta and light quark masses with convergence not guaranteed at the physical quark masses.  Lattice 
QCD has already provided numerical evidence that $SU(2)$ baryon $\chi$PT is not converging at or 
slightly above the physical pion mass for the nucleon mass and axial 
coupling~\cite{Walker-Loud:2008rui,Walker-Loud:2013yua,Chang:2018uxx}.  
Including explicit $\Delta$ degrees of freedom may improve convergence of $SU(2)$ baryon $\chi$PT, but 
introduces a plethora of additional unknown LECs. Lattice QCD calculations of $N\pi$ scattering at 
various pion masses can help verify the convergence pattern and whether it is improved with explicit 
$\Delta$s~\cite{Siemens:2016jwj}, as well as constrain the additional LECs.  $N\pi$ scattering is 
additionally important because of the current tension between lattice QCD determinations of the 
nucleon-pion sigma term~\cite{QCDSF-UKQCD:2011qop,Yang:2015uis,Alexandrou:2019brg,Borsanyi:2020bpd} 
$\sigma_{\pi N}$ and phenomenological determinations~\cite{Hoferichter:2015dsa,RuizdeElvira:2017stg} 
(see Ref.~\cite{Gupta:2021ahb} for a possible resolution). $\sigma_{\pi N}$ plays an important role 
in the analysis of direct dark matter detection experiments~\cite{Hoferichter:2016ocj}.
Controlled lattice QCD calculations of $N \pi$ scattering may help understand this tension. 

As a final example, a future prospect for lattice QCD is the determination of inputs to models of 
neutrino-nucleus scattering cross sections to aid  next-generation experiments, like 
DUNE~\cite{Abi:2020wmh} and Hyper-K~\cite{Hyper-Kamiokande:2018ofw}, designed to measure unknown 
properties associated with neutrino oscillations.  The importance of lattice QCD input was recently 
highlighted by current lattice QCD results for elastic nucleon form factors~\cite{Meyer:2022mix}.
The frontier for these lattice QCD applications is the $\Delta$-resonance and pion-production 
contributions to inelastic $\nu N$ structure.  To carry out this program, it is essential to first 
demonstrate control of $N\pi$ scattering, a necessary component of nucleon inelastic resonant structure.

Lattice QCD calculations of two-pion systems are well established (for a recent review, see 
Ref.~\cite{Briceno:2017max}), and there are now numerous three-meson 
results~\cite{Horz:2019rrn,Blanton:2019vdk,Hansen:2020otl,Fischer:2020jzp,Alexandru:2020xqf,Blanton:2021llb,Mai:2021nul}.
In contrast, there are few nucleon-pion scattering studies in lattice QCD. Ref.~\cite{Lang:2012db} 
included nucleon-pion operators to extract the spectrum in the $I=1/2$ channel at a single lattice 
spacing using a $16^3\times 32$ lattice volume with $m_\pi\approx266$~MeV and obtained an estimate 
of the scattering length with a significance of roughly four standard deviations.  
Refs.~\cite{Andersen:2017una} and \cite{Alexandrou:2017mpi} each employ a single ensemble with
$m_{\pi}\gtrsim 250~{\rm MeV}$ to evaluate scattering phase shifts relevant to the $\Delta$ resonance, 
but neither presented statistically significant results for the scattering lengths. There is also 
older work which employs the quenched approximation~\cite{Fukugita:1994ve} and preliminary 
unpublished results for the $I=3/2$ amplitudes~\cite{Verduci:2014btc,Mohler:2012nh,Pittler:2021bqw}. 
The determination of finite-volume nucleon-pion energies in Ref.~\cite{Lang:2016hnn} is performed 
close to the physical quark masses, but scattering amplitudes are not computed. Lattice computations 
of meson-baryon scattering lengths in other systems have also been 
performed~\cite{Detmold:2015qwf,Torok:2009dg}.

Recent advances in lattice QCD computations of multi-hadron scattering amplitudes are due in part to 
stochastic algorithms employing Laplacian-Heaviside (LapH) smearing to efficiently compute 
timeslice-to-timeslice quark propagators~\cite{Peardon:2009gh, Morningstar:2011ka} which enable 
definite momentum projections of the constituent hadrons in multi-hadron interpolators and the 
evaluation of all needed Wick contraction topologies.  Recently, these algorithms have been 
successfully applied to meson-baryon scattering amplitudes~\cite{Andersen:2017una, Lang:2016hnn}. 
Alternatively, Ref.~\cite{Alexandrou:2017mpi} employs sequential sources, while the scattering 
channels in Refs.~\cite{Detmold:2015qwf,Torok:2009dg} are chosen to avoid same-time valence quark 
propagation and can be straightforwardly implemented with point-to-all. The LapH approach has also 
been employed to three-meson~\cite{Blanton:2019igq,Horz:2019rrn,Culver:2019vvu,Alexandru:2020xqf,Brett:2021wyd,Mai:2021nul,Hansen:2020otl,Blanton:2021llb,Fischer:2020jzp} 
and two-baryon~\cite{Francis:2018qch,Green:2021qol,Horz:2020zvv} amplitudes.

This work is part of an ongoing project to obtain $N\pi$ scattering amplitudes from lattice QCD, 
which requires computations using several Monte Carlo ensembles to reach the physical pion mass and 
extrapolate to the continuum limit. Nucleon-pion correlation functions in lattice QCD suffer from an 
exponential degradation in the signal-to-noise ratio with increasing time separation, which hampers 
the determination of nucleon-pion energies from the large-time asymptotics.  This difficulty worsens 
as the quark mass is decreased to its physical value. One important objective of this work is to 
determine if the stochastic-LapH approach of Ref.~\cite{Morningstar:2011ka} is viable for computing 
nucleon-pion scattering amplitudes close to the physical values of the quark masses.  Another 
objective is to compare two different methods~\cite{Morningstar:2017spu} of extracting the $K$-matrix 
from finite-volume energies.  The results presented here extend those of Ref.~\cite{Andersen:2017una}. 
An update with increased statistics on the same $m_{\pi} = 280~{\rm MeV}$ ensemble used in 
Ref.~\cite{Andersen:2017una} is not included here due to instabilities discovered in the gauge 
generation of that ensemble, as detailed in Ref.~\cite{Mohler:2020txx}.

Both the total isospin $I=1/2$ and $I=3/2$ scattering lengths at light quark masses corresponding 
to $m_{\pi} = 200~{\rm MeV}$ are computed in this work. The results are
\begin{gather}
\label{e:slen}
    m_{\pi}a_0^{3/2} = -0.2735(81) \, , \qquad m_{\pi}a_0^{1/2} = 0.142(22),
\end{gather}
where the errors are statistical only. The Breit-Wigner parameters for the $\Delta(1232)$-resonance 
are also determined from the $I=3/2$, $J^P = 3/2^+$ partial wave
\begin{gather}
    \frac{m_{\Delta}}{m_{\pi}} = 6.257(35) , \qquad \gbw = 14.41(53) ,
\end{gather}
where the corresponding scattering phase shift is shown in Fig.~\ref{fig:deltaphase}. Since only a single
ensemble of gauge field configurations is employed, the estimation of systematic errors due to the 
finite lattice size, lattice spacing, and unphysically large light quark mass is left for future work.
However, systematic errors due to the determination of finite-volume energies, the reduced symmetries 
of the periodic simulation volume, and the parametrization of the amplitudes are addressed. The methods 
presented here therefore provide a step toward the lattice determination of the nucleon-pion scattering
lengths at the physical point with controlled statistical and systematic errors.

The remainder of this work is organized as follows. Sec.~\ref{s:fv} discusses the effects of the finite 
spatial volume, including the corresponding reduction in symmetry and the relation between finite-volume 
energies and infinite-volume scattering amplitudes. Sec.~\ref{s:comp} presents the computational 
framework, including the lattice regularization and simulation, the measurement of correlation functions, 
and the determination of the spectrum from them. Results for the amplitudes are presented and discussed 
in Sec.~\ref{s:res}, while Sec.~\ref{s:conc} concludes.

\section{Finite-volume formalism}
\label{s:fv}

The Euclidean metric with which lattice QCD simulations are necessarily performed complicates the 
determination of scattering amplitudes. It was shown long ago by Maiani and Testa~\cite{Maiani:1990ca} 
that the direct application of an asymptotic formalism to Euclidean correlation functions does not 
yield on-shell scattering amplitudes away from threshold. Instead, lattice QCD computations exploit 
the finite spatial volume to relate scattering amplitudes to the shift of multi-hadron energies from 
their non-interacting values~\cite{Luscher:1990ux}.  See Ref.~\cite{Bruno:2020kyl} for a more complete 
investigation of the Maiani-Testa theorem, and Refs.~\cite{Hansen:2017mnd,Bulava:2019kbi} for an 
alternative approach to computing scattering amplitudes from Euclidean correlation functions based 
on Ref.~\cite{Barata:1990rn}.

This section summarizes the relationship between finite-volume spectra and elastic nucleon-pion 
scattering amplitudes. Due to the reduced symmetry of the periodic spatial volume, this 
relationship is not one-to-one and  generally involves a parametrization of the lowest partial 
wave amplitudes with parameters constrained by a fit to the entire finite-volume spectrum. Symmetry 
breaking due to the finite lattice spacing is also present, but ignored. At fixed physical volume 
and quark masses, the continuum limit of the finite volume spectrum exists and is assumed for this 
discussion.

For a particular total momentum $\boldsymbol{P}$, the relationship between the finite-volume 
center-of-mass energies $E_{\rm cm}$ determined in lattice QCD and elastic nucleon-pion scattering 
amplitudes specified in the well-known $K$-matrix is given by the determinantal equation  
\begin{align}\label{e:det}
{\rm det}[ \tilde{K}^{-1}(E_{\rm cm}) - B^{\boldsymbol{P}}(E_{\rm cm})] 
  + {\rm O}({\rm e}^{-M L}) = 0 \ ,
\end{align}
where $\tilde{K}$ is proportional to the $K$-matrix and $B^{\boldsymbol{P}}(E_{\rm cm})$ is the 
so-called box matrix, using the notation of Ref.~\cite{Morningstar:2017spu}.  
This relationship holds below the nucleon-pion-pion threshold, up to 
corrections which vanish exponentially for asymptotically large $ML$, where $L$ is the side length 
of the cubic box of volume $L^3$ and $M$ the smallest relevant energy scale. The determinant is 
taken over all scattering channels specified by total angular momentum $J$, the projection of $J$ 
along the $z$-axis $m_J$, and the orbital angular momentum $\ell$. For elastic nucleon-pion 
scattering the total spin $S=1/2$ is fixed, and therefore not indicated explicitly.  The $K$-matrix 
is diagonal in $J$ and $m_J$, and, for elastic nucleon-pion scattering, additionally diagonal in 
$\ell$.  The $\tilde{K}$-matrix in Eq.~(\ref{e:det}) explicitly includes threshold-barrier 
factors which are integral powers of $q_{\rm cm} = \sqrt{\boldsymbol{q}_{\rm cm}^2}$, with 
\begin{gather}
\boldsymbol{q}_{\rm cm}^2 = \frac{E_{\rm cm}^2}{4} - \frac{m_{\pi}^2 + m_{\rm N}^2}{2} 
   + \frac{(m_\pi^2 - m_{\rm N}^2)^2}{4E_{\rm cm}^2}  \ ,
\end{gather}
so that $\tilde{K}^{-1}$ is smooth near the nucleon-pion threshold. Each diagonal element of 
$\tilde{K}$ is associated with a particular partial wave specified by $J^P$, where $P$ is the 
parity, or equivalently $(2J,\ell)$, so that
\begin{gather}
\tilde{K}^{-1}_{J\ell,J'\ell'} = \delta_{JJ'}\delta_{\ell\ell'} 
   q_{\rm cm}^{2\ell + 1} {\rm cot} \, \delta_{J\ell}(E_{\rm cm}) \ ,
\end{gather}
where $\delta_{J\ell}(E_{\rm cm})$ is the scattering phase shift.

The box matrix   
$B^{\boldsymbol{P}}(E_{\rm cm})$ encodes the reduced symmetries of the periodic spatial volume, 
and is in general dense in all indices. The finite-volume energies used to constrain $K$ from 
Eq.~(\ref{e:det}) possess the quantum numbers associated with symmetries of the box, namely a 
particular irreducible representation of the finite-volume little group for the total spatial 
momentum $\boldsymbol{P} = \frac{2\pi}{L} \boldsymbol{d}$, with $\boldsymbol{d}\in\mathbb{Z}^3$.     
The matrices in Eq.~(\ref{e:det}) are therefore block-diagonalized in the basis of finite-volume 
irreducible representations (irreps), with each energy analyzed using a single (infinite-dimensional) 
block. Since the subduction from infinite-volume partial waves to finite-volume irreps is not in 
general one-to-one, an additional occurrence index $n$ is required to specify the matrix elements 
in each block. A particular block is denoted by the finite-volume irrep $\Lambda(\boldsymbol{d}^2)$ 
and a row of this irrep $\lambda$. Since the spectrum is independent of the row $\lambda$, this 
index is henceforth omitted. For a particular block, the block-diagonalized box-matrix is denoted 
$B^{\Lambda(\boldsymbol{d}^2)}_{J\ell n, J'\ell'n'}$. After transforming to the block diagonal 
matrix, the $\tilde{K}$ matrix has the form given by Eq.~(35) in Ref.~\cite{Morningstar:2017spu}.  

In practical applications, the matrices in Eq.~(\ref{e:det}) are truncated to some maximum orbital 
angular momentum $\ell_{\rm max}$. Threshold-barrier arguments ensure that, at fixed $E_{\rm cm}$, 
higher partial waves are suppressed by powers of $q_{\rm cm}$, but systematic errors due to finite 
$\ell_{\rm max}$ must be assessed.  The expressions for all elements of $B^{\Lambda}(d^2)$ relevant 
for this work are given in Ref.~\cite{Morningstar:2017spu}, although some are present already in 
Ref.~\cite{Gockeler:2012yj}. The occurrence pattern of lowest-lying partial waves in the 
finite-volume irreps is given in Table~\ref{t:irreps}.

\begin{table}[t]
\begin{center}
\begin{tabular}{c|c|c|c}
$\boldsymbol{d}$ & $\Lambda$ & dim. & contributing $(2J,\ell)^{n_{\rm occ}}$ 
for $\ell_{\rm max}=2$ \\ \hline
$(0,0,0)$ & $G_{1{\rm u}}$ & 2 & $(1,0)$\\
& $G_{1{\rm g}}$ & 2 & $(1,1)$\\
& $H_{{\rm g}}$ &  4 & $(3,1)$, $(5,2)$ \\
& $H_{{\rm u}}$ &  4 & $(3,2)$, $5,2)$ \\
& $G_{2{\rm g}}$ & 2 & $(5,2)$ \\ 
\hline
 $(0,0,n)$ & $G_1$ & 2 & $(1,0)$, $(1,1)$, $(3,1)$, $(3,2)$, $(5,2)$ \\
 & $G_2$ & 2 & $(3,1)$, $(3,2)$, $(5,2)^2$\\
\hline
 $(0,n,n)$ & $G$ & 2 & $(1,0)$, $(1,1)$, $(3,1)^2$, $(3,2)^2$, $(5,2)^3$ \\
 \hline 
$(n,n,n)$ & $G$ & 2 & $(1,0)$, $(1,1)$, $(3,1)$, $(3,2)$, $(5,2)^2$ \\ 
 & $F_{1}$ & 1 &  $(3,1)$, $(3,2)$, $(5,2)$ \\ 
 & $F_{2}$ & 1 &  $(3,1)$, $(3,2)$, $(5,2)$ \\ 
\end{tabular}
\end{center}
\caption{\label{t:irreps} A list of the lowest contributing partial waves for each 
 irrep of the finite-volume little group $\Lambda$ in momentum class $\boldsymbol{d}$ employed 
 in this work. All partial waves with $\ell\le\ell_{\rm max}$ for $\ell_{\rm max}=2$ are shown 
 and each partial wave is denoted by $(2J, \ell)$. The superscript $n_{\rm occ}$ denotes the 
 number of multiple occurrences (subductions) of the partial wave in the irrep. The pattern of 
 partial wave mixing is evidently more complicated for irreps with non-zero total momentum.}
\end{table}

Employing this formalism for nucleon-pion scattering presents additional difficulties compared 
to simpler scattering processes. First, due to the non-zero nucleon spin, two partial waves 
contribute for each non-zero $\ell$, one with $J = \ell + 1/2$ and the other with $J = \ell - 1/2$. 
Secondly, off-diagonal elements of the box matrix induce mixings of different partial waves in the 
quantization condition. For $\ell_{\rm max} = 2$, energies $E_{\rm cm}$ in irreps with 
$\boldsymbol{d}^2 = 0$ determine the quantity 
$q_{\rm cm}^{2\ell+1}{\rm cot} \, \delta_{J\ell}(E_{\rm cm})$ for $s$- and $p$-waves, while 
these partial waves cannot be unambiguously isolated for levels in irreps with non-zero total 
momentum. This complication necessitates global fits of all energies to determine the desired 
partial waves, which are discussed in Sec.~\ref{s:res}. 

\section{Spectrum computation details}
\label{s:comp}

This section details the numerical determination of finite-volume nucleon-pion energies used to 
constrain the $\ell \le 2$ partial waves of the $I=1/2$ and $I=3/2$ elastic nucleon-pion scattering 
amplitudes. Properties of the single ensemble of gauge field configurations are given in Sec.~\ref{s:ens}, 
and computation of the nucleon-pion correlation functions from them is discussed in Sec.~\ref{s:laph}. 
The subsequent determination of the finite-volume spectra from the correlation functions is detailed 
in Sec.~\ref{s:fit}. 

\subsection{Ensemble details}\label{s:ens}

This computation uses the D200 ensemble of QCD gauge configurations generated by the Coordinated 
Lattice Simulations (CLS) consortium~\cite{Bruno:2014jqa}, whose properties are summarized in 
Table~\ref{tab:comp_deets}. It was generated using the tree-level improved L$\ddot{\textup{u}}$scher-Weisz 
gauge action~\cite{Luscher:1984xn} and a non-perturbatively O(\textit{a})-improved Wilson fermion 
action~\cite{Bulava:2013cta}. Open temporal boundary conditions~\cite{Luscher:2011kk} are employed to 
reduce the autocorrelation of the global topological charge. However, all interpolating fields must 
be sufficiently far from the boundaries to reduce spurious contributions to the fall-off of temporal 
correlation functions. An analysis of the zero-momentum single-pion and $\rho$-meson correlators 
in Ref.~\cite{Andersen:2018mau} suggests that a minimum distance of $m_{\pi}t_{\rm bnd} \gtrsim 2$ is 
sufficient to keep temporal boundary effects below the statistical errors in the determination of 
energies. The time ranges for the correlators employed here are such that 
$m_{\pi}t_{\rm bnd} \gtrsim 2.3$. 

\begin{table}[t]
\begin{center}
\begin{tabular}{c@{\hskip 12pt}c@{\hskip 12pt}c@{\hskip 12pt}c@{\hskip 12pt}c@{\hskip 12pt}}
 $a [\textup{fm}]$ & $(L/a)^3 \times T/a$ & $N_{\rm meas}$ & $am_\pi$ &  $am_{\rm K}$ \\ \hline
     0.0633(4)(6) & $64^3 \times 128$ & 2000 & 0.06617(33) & 0.15644(16) 
\end{tabular}
\\[5mm]
\begin{tabular}{c@{\hskip 12pt}c@{\hskip 12pt}c@{\hskip 12pt}}
 $af_{\pi}$ & $af_{\rm K}$ & $am_{\rm N}$ \\ \hline
 0.04233(16) & 0.04928(21) &  0.3148(23)
\end{tabular}
\end{center}
\caption{Parameters of the D200 ensemble produced by the CLS consortium~\cite{Bruno:2014jqa}. The 
lattice spacing $a$ is from Ref.~\cite{Strassberger:2021tsu} (with both statistical and systematic errors) 
and follows the strategy of Ref.~\cite{Bruno:2016plf}. The number of gauge configurations employed here 
is specified by $N_{\rm meas}$. The pion mass $m_\pi$ and nucleon mass $m_{\rm N}$ determinations are 
discussed in Sec.~\ref{s:fit}.  The kaon mass, denoted $m_{\rm K}$, and the pion and kaon decay constants, 
denoted $f_{\pi}$ and $f_{\rm K}$, are taken from Ref.~\cite{Bruno:2016plf}. 
\label{tab:comp_deets}}
\end{table}

A complete description of the algorithm used to generate the D200 ensemble is presented in 
Ref.~\cite{Bruno:2014jqa}, but some details relevant for the present work are given below. All CLS 
ensembles use twisted-mass reweighting~\cite{Luscher:2008tw} for the degenerate light quark doublet 
and the  Rational Hybrid Monte Carlo (RHMC) approximation for the strange quark~\cite{Clark:2006fx}. 
Both representations of the fermion determinants require reweighting factors to change the simulated 
action to the desired distribution. All primary observables are therefore re-weighted according to
\begin{equation}\label{e:rw}
 \langle A \rangle = \frac{\langle AW\rangle_{\rm W}}{\langle W \rangle_{\rm W}}
\end{equation}
where $\langle ...\rangle_{\rm W}$ denotes the ensemble average with respect to the simulated action. 
$W$ is the product of two factors $W=W_0 W_1$, where $W_0$ and $W_1$ are the reweighting factors for 
the light and strange quark actions. They are estimated stochastically on each gauge configuration 
as in Ref.~\cite{Bruno:2014jqa}. 

The lattice scale is determined at a fixed value of the gauge coupling according to the massless 
scheme described in Ref.~\cite{Bruno:2016plf} and updated in Ref.~\cite{Strassberger:2021tsu}. 
Specifically, the kaon decay constant $f_{\rm K}$ is enforced to take its physical value at the 
physical point where the pion and kaon masses take their physical values. This point is identified 
along a trajectory in which the bare light- and strange-quark masses are varied, keeping the sum of 
the (renormalized) quark masses fixed. The heavier-than-physical pion mass $m_{\pi} = 200~{\rm MeV}$ 
for the D200 therefore results in $m_{\rm K} = 480~{\rm MeV}$, which is less than the physical value.
In practice, the bare quark mass tuning satisfies the trajectory condition only approximately. In 
order to correct any mistuning \emph{a posteriori}, Ref.~\cite{Bruno:2016plf} applies slight shifts 
to the quark masses to ensure the trajectory condition is respected in the scale determination. No 
such shift is applied here.    

\begin{figure}[t]
\centering
\includegraphics[width=\linewidth]{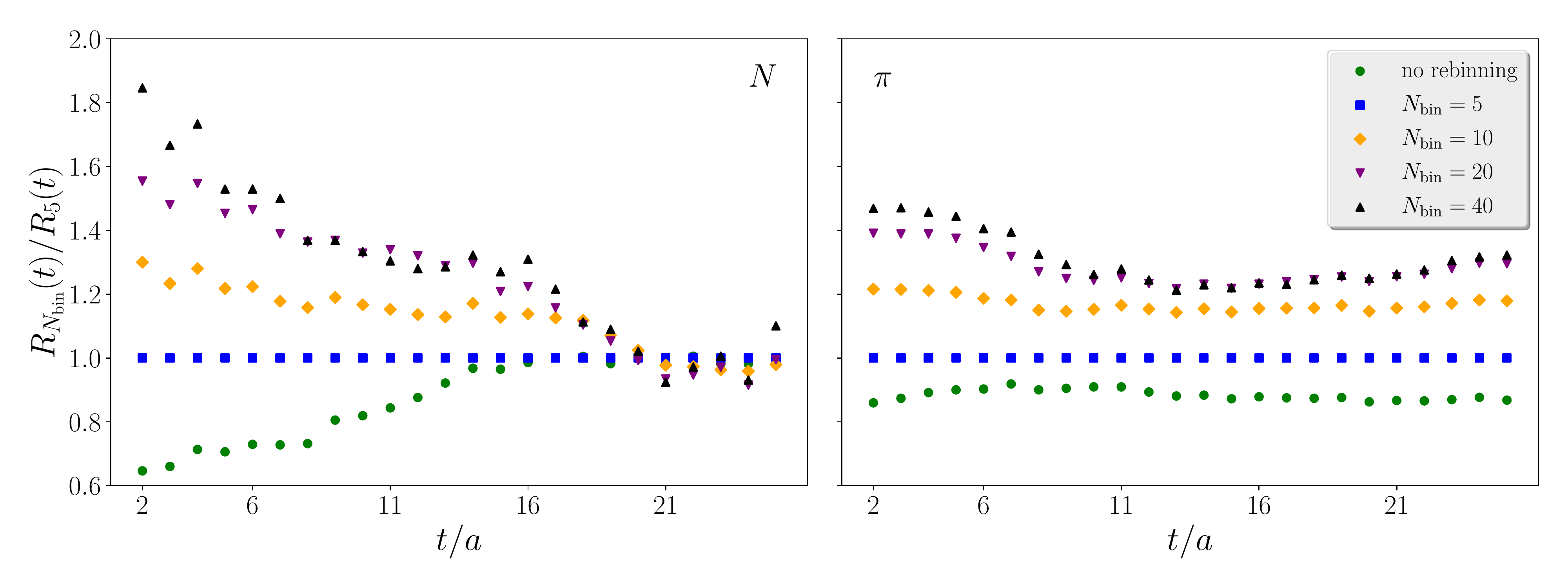}
\caption{Relative error of the zero-momentum nucleon (left) and pion (right) correlators, 
denoted $R_{N_{\textup{bin}}}(t)$, for several bin sizes $N_{\textup{bin}}$. All points are 
normalized by the $N_{\rm bin} = 5$ value with errors estimated using the bootstrap procedure 
with $N_\textup{B}=1000$ resamples. All subsequent analysis, which does not employ any 
correlation functions with $t/a \lesssim 8-10$, ignores autocorrelation and uses $N_{\rm bin}=20$.
\label{fig:rebin_check}}
\end{figure}

In this study, correlation function measurements  are separated by four molecular dynamics units (MDU's).  
To check for autocorrelations, the original measurements are binned by averaging  $N_\textup{bin}$ 
consecutive gauge configurations. The dependence of the relative errors on $N_{\rm bin}$ for the 
single-nucleon and single-pion correlators is shown in Figure \ref{fig:rebin_check}. Although 
evidence of autocorrelation  remains for $t/a\lesssim 8-10$ between $N_\textup{bin} = 20$ and $40$, 
these early timeslices are not used in the analysis, suggesting that $N_\textup{bin} = 20$ is sufficient 
to account for any autocorrelations in our energy determinations.

\subsection{Correlation function construction}
\label{s:laph}

The determination of finite-volume nucleon-pion energies requires a diverse set of temporal correlation 
functions measured on the D200 gauge field ensemble. In addition to diagonal correlation functions 
between single-pion and single-nucleon interpolating operators, correlation matrices between all 
operators in each irrep are required. For the $I=3/2$ irreps in Table~\ref{t:irreps} where the 
resonant $(2J,\ell) = (3, 1)$ partial wave contributes, single-baryon operators are included in addition 
to nucleon-pion operators resulting in additional valence quark-line topologies. These topologies include 
those with lines that start and end on the same timeslice. 

\begin{table}[t]
\begin{center}
\begin{tabular}{c  c  c  c  c  c  c}
$N_\textup{D}$ & $(\rho,n_\rho)$ & $N_\textup{ev}$ & $N_\textup{R}^\textup{fix}$ 
   & $N_\textup{R}^\textup{rel}$ & Noise dilution & $N_{t_0}$\\ \hline
 2176 & (0.1,36) & 448 & 6 & 2 & (TF,SF,LI16$)_\textup{fix}$(TI8,SF,LI16$)_\textup{rel}$ & 4
\end{tabular}
\end{center}
\caption{Parameters of the stochastic LapH implementation used to compute temporal correlators in this 
work. $N_D$ is the number of Dirac matrix inversions required per configuration and $(\rho,n_\rho)$ the 
stout smearing parameters for the spatial links in the gauge-covariant Laplace operator. $N_\textup{ev}$ 
denotes the dimension of the LapH subspace. $N_\textup{R}^{\rm fix}$ and $N_{\rm R}^{\rm rel}$ are the 
number of stochastic sources for fixed and relative quark lines, respectively. Notation used to specify 
the dilution scheme for each line type is explained in the text, and the number of source times on each 
configuration is $N_{t_0}$. }
\label{fig:laph_deets}
\end{table}

Our operator construction is described in Ref.~\cite{Morningstar:2013bda} and our method of evaluating 
the temporal correlators is detailed in Ref.~\cite{Morningstar:2011ka}. Well-designed multi-hadron 
interpolators are comprised of individual constituent hadrons each having definite momenta.  Evaluating 
the temporal correlations of such operators requires all-to-all quark propagators, where all elements of 
the Dirac matrix inverse are computed. The stochastic-LapH approach~\cite{Morningstar:2011ka} enables 
the efficient treatment of this inverse, provided at least one of the quark fields is LapH 
smeared~\cite{Peardon:2009gh}. This smearing procedure is effected by a projection onto the space 
spanned by the $N_\textup{ev}$ lowest eigenmodes of the gauge-covariant three-dimensional Laplace 
operator in terms of link variables which are stout smeared~\cite{Morningstar:2003gk} with parameters 
$(\rho, n_{\rho})$. Although the $N_\textup{ev}$ required to maintain a constant smearing radius grows 
with the spatial volume, the growth of the number of Dirac matrix inversions $N_\textup{D}$ can be 
significantly reduced with the introduction of stochastic estimators in the LapH subspace. Such 
estimators are specified by the number of dilution projectors in the time (`T'), spin (`S'), and 
Laplacian eigenvector (`L') indices, for which `F' denotes full dilution and `I$n$' some number of 
uniformly interlaced projectors. Different dilution schemes are used for fixed-time quark lines, 
denoted `fix', which propagate between different timeslices, and relative-time lines (`rel') which 
start and end at the same time.  In this work, the relative-time quark lines were only used at the 
sink time, while the fixed-time lines were used for quark propagation starting and ending at the 
source time. Both the dilution schemes and the number of stochastic sources used for each type of 
line are specified in Table~\ref{fig:laph_deets}. Source times $t_0=35, 64$ were used for correlations 
going forwards in time, and $t_0=64, 92$ were used for correlations going backwards in time.

A beneficial property of the stochastic estimators is the factorization of the inverse of the Dirac 
matrix, which enables correlation construction to proceed in three steps: (1) Dirac matrix inversion, 
(2) hadron sink/source construction, and (3) correlation function formation. After determining the 
stochastically-estimated propagators in step (1), the hadron source/sink tensors are computed in 
step (2). These tensors are subsequently reused to construct many different correlation functions 
in step (3), which consists of optimized~\cite{Horz:2019rrn} tensor contractions. Averages over the
$N_{t_0}= 4$ different source times (two for forward propagation 
and two for backward propagation), all possible permutations of the available noise sources in a 
given Wick contraction, all total momenta of equal magnitude, and all equivalent irrep rows are 
performed to increase statistics. 

\subsection{Determination of finite-volume energies}\label{s:fit}

\begin{figure}[t]
\centering
\includegraphics[width=\textwidth]{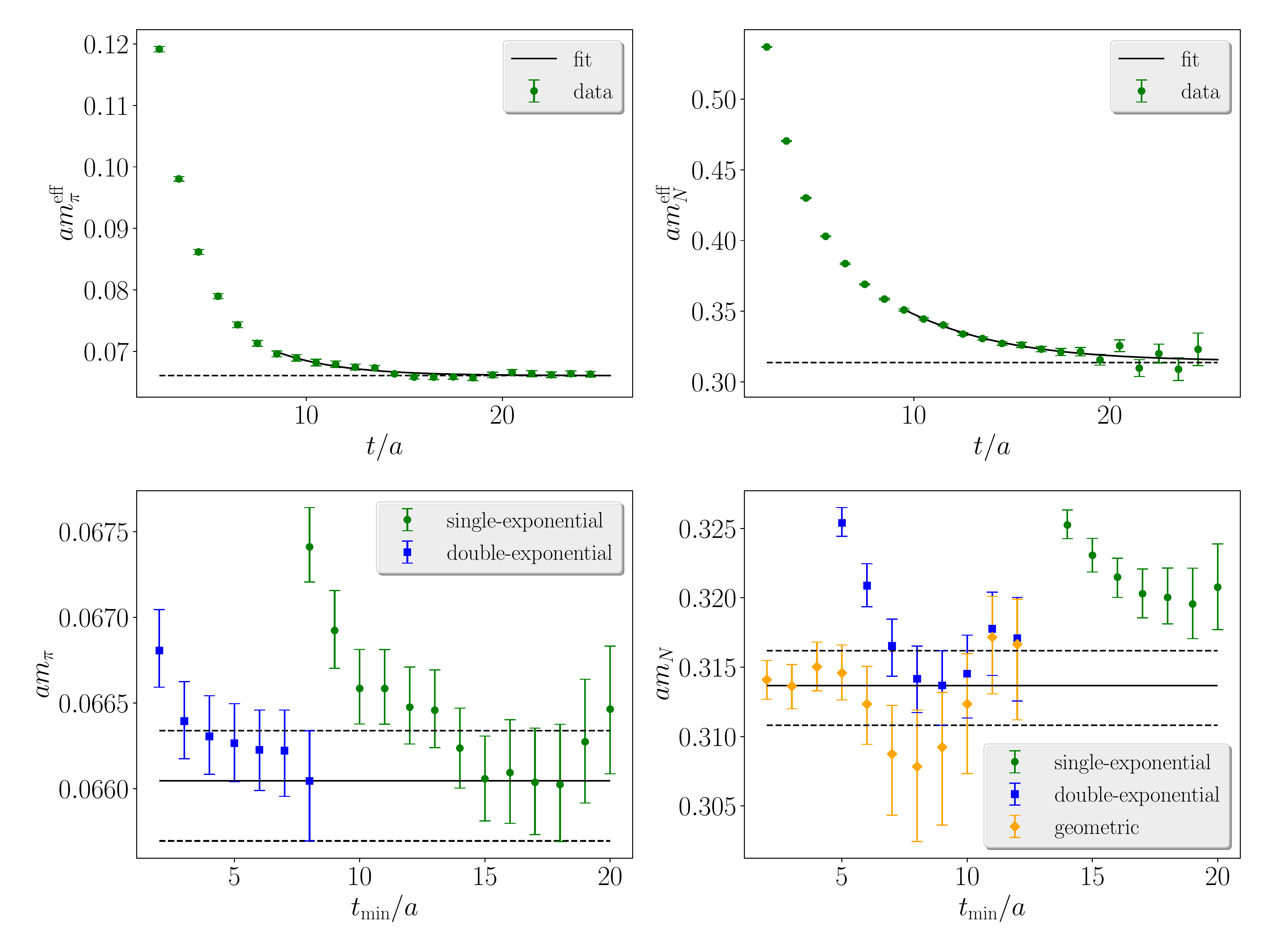}
\caption{\label{f:mnpi} Fits to determine $am_{\pi}$ and $am_{\rm N}$. 
Bottom row: Variation of the fit range $[t_{\rm min},t_{\rm max}]$ with $t_{\rm max} = 25a$ for 
correlated-$\chi^2$ fits. Both single- and double-exponential fits are shown and the horizontal 
band indicates the two-exponential fit range chosen so that statistical errors are dominant. 
Top row: the chosen two-exponential fits overlayed on standard effective masses for $am_{\pi}$ 
(left) and $am_{\rm N}$ (right).} 
\end{figure}

Once the correlation functions computed as described in Sec.~\ref{s:laph} are available, the 
determination of finite-volume energies can commence. From the (binned) correlator and reweighting 
factor measurements, the reweighted correlation functions are computed as secondary observables according 
to Eq.~(\ref{e:rw}). Their statistical errors and covariances are used in fits to determine energies and 
estimated by the bootstrap procedure with $N_{\rm B} = 800$ samples. 

In order to ensure that $t_{\rm bnd}$ is sufficiently large, a maximum time separation 
$t_{\rm max} = 25a$ is enforced globally in the analysis. Energies are determined from 
correlated-$\chi^2$ fits to both single- and two-exponential fit forms, which are additionally compared 
to a ``geometric series'' form
\begin{gather}\label{e:geom}
    C(t) = \frac{A{\rm e}^{-Et}}{1 - B{\rm e}^{-Mt}},
\end{gather}
which consists of four free parameters. We also explored a ``multi-exponential'' variant of the 
geometric series, with the replacement $Be^{-Mt}\rightarrow \sum_n B_n e^{-M_nt}$.

The application of our approach to determining the nucleon and pion masses is shown in Fig.~\ref{f:mnpi}.
As usual, a fit range is desired so that statistical errors on the energies are larger than systematic 
ones. This optimal range is selected according to several criteria. First, a good fit quality 
$q\gtrsim 0.2-0.3$ is enforced to ensure that the fit describes the data within the usual 68\% confidence 
interval quoted for statistical errors. Second, the absence of any  statistically significant change in 
the energy upon variation of $t_{\rm min}$ around the chosen fit range further suggests that the 
asymptotic large-time behavior is applicable. Finally, consistency across different fit forms supports 
the hypothesis that the energy determination is statistics limited. For the pion, consistency between 
single- and two-exponential fits, as well as the mild variation with $t_{\rm min}$, suggests that 
statistical errors are dominant. As is evident in $am^{\rm eff}_{\rm N}$, single-exponential fits 
for $am_{\rm N}$ are unsuitable, but consistency between the double-exponential and geometric fit 
forms is reassuring.  

For the nucleon-pion channels, excited state energies are determined in addition to ground states. 
This requires a large basis of interpolating operators in each irrep and two-point correlations 
between them. The resulting $N_{\rm op}\times N_{\rm op}$ hermitian matrix, denoted $C_{ij}(t)$, is 
rotated~\cite{Michael:1982gb} using eigenvectors $v_n(t_0, t_{\rm d})$ of the generalized eigenvalue 
problem (GEVP)
\begin{gather}\label{e:gevp}
    C(t_{\rm d})v_n(t_0, t_{\rm d}) = \lambda_n(t_0, t_{\rm d}) C(t_0) v_n(t_0, t_{\rm d}).
\end{gather}
In our single-pivot approach,
the correlation matrix is rotated by these vectors for all $t$, and the diagonal elements of the 
rotated matrix, denoted $D_n(t)$, are correlators with optimal overlap onto the lowest $N_{\rm op}$ 
states. Although diagonalizing separately on each time-slice ensures that the optimized correlators 
are increasingly dominated by the desired state for asymptotically large 
times~\cite{Luscher:1990ck, Blossier:2009kd}, in practice, the single-pivot method produces nearly 
identical results if $(t_0,t_{\rm d})$ are chosen appropriately. Systematic errors related to this 
are controlled by ensuring that extracted energies are insensitive to $(t_0,t_{\rm d})$ and 
$N_{\rm op}$ and by ensuring that the rotated correlation matrix remains diagonal within statistical 
errors for all time separations $t>t_{\rm d}$. The advantages of diagonalizing on a single set of times 
include a better signal-to-noise ratio for large times and no need for eigenvector pinning in which 
eigenvectors are re-ordered for diagonalizations at different times and bootstrap samples.

After forming the optimized correlators, the following ratio is taken 
\begin{gather}\label{e:ratio}
    R_n(t) = \frac{D_n(t)}{C_{\pi}(\boldsymbol{d}_{\pi}^2, t)\,C_{\rm N}(\boldsymbol{d}_{\rm N}^2, t) } ,
\end{gather}
with $\boldsymbol{d}_{\pi}^2$ and $\boldsymbol{d}_{\rm N}^2$  chosen so that 
\begin{gather}
    E_n^{\rm non.\, int.} = \sqrt{m_{\pi}^2 + \left(\frac{2\pi \boldsymbol{d}_{\pi}}{L}
    \right)^2} + \sqrt{m_{\rm N}^2 + \left(\frac{2\pi \boldsymbol{d}_{\rm N}}{L} \right)^2}
\end{gather}
corresponds to the closest non-interacting energy. The ratio $R_n(t)$ is then fit to the 
single-exponential ansatz $R_n(t) = A_n \textup{e}^{-\Delta E_n t}$ to determine the energy shift 
$a\Delta E_n$, from which the lab-frame energy is reconstructed 
$aE_n^\textup{lab} = a\Delta E_n + aE_n^\textup{non.int.}$. Although the ratio fits enable somewhat 
smaller $t_\textup{min}$ when $\Delta E_n$ is small, they offer little advantage for states which are 
significantly shifted from non-interacting levels. Nonetheless, ratio 
fits are employed for all levels in the nucleon-pion irreps, and are typically consistent with single- 
and double-exponential fits directly to $D_n(t)$.  

\begin{figure}[t]
\centering
\includegraphics[width=0.49\linewidth]{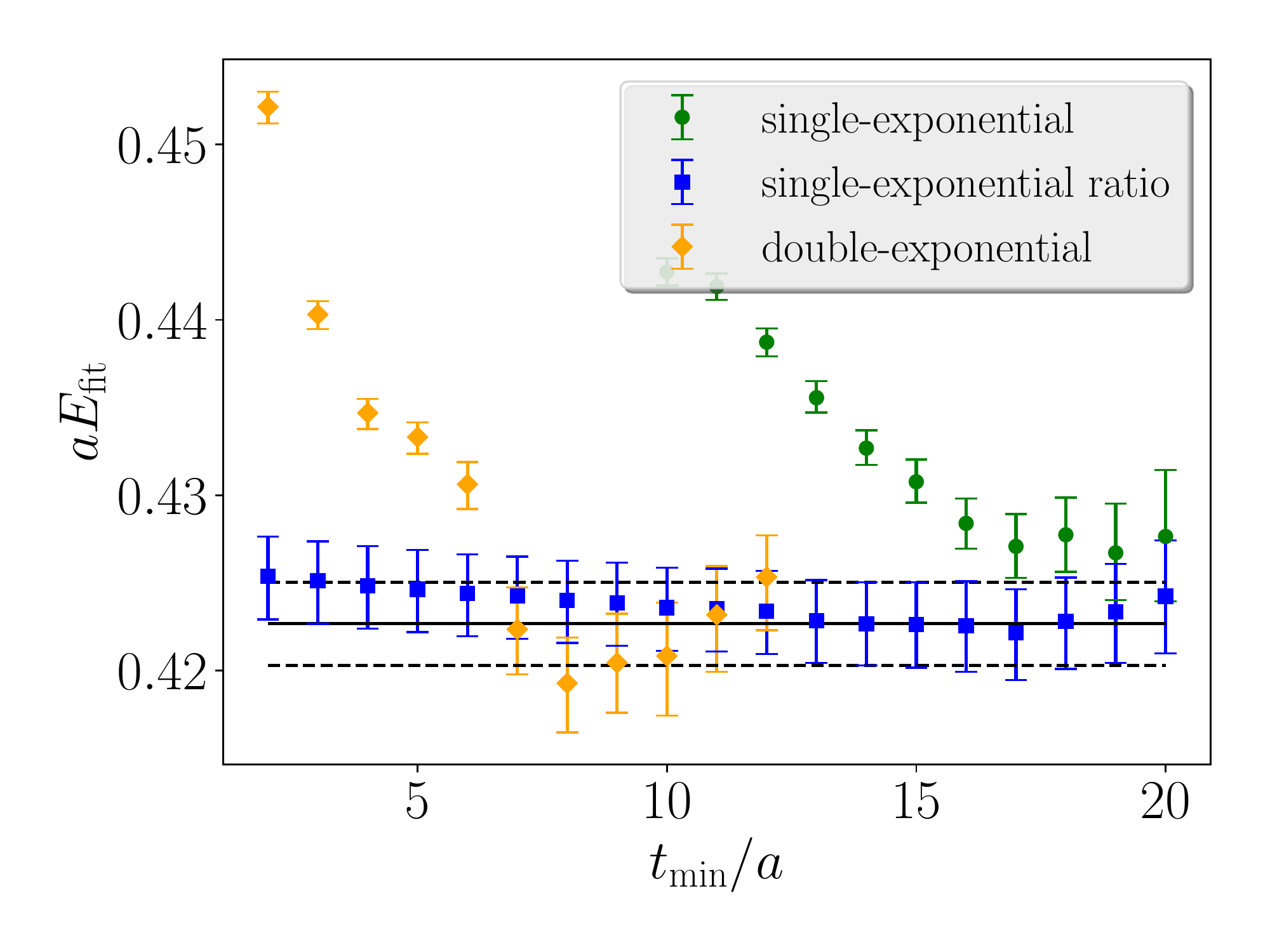}
\includegraphics[width=0.43\linewidth]{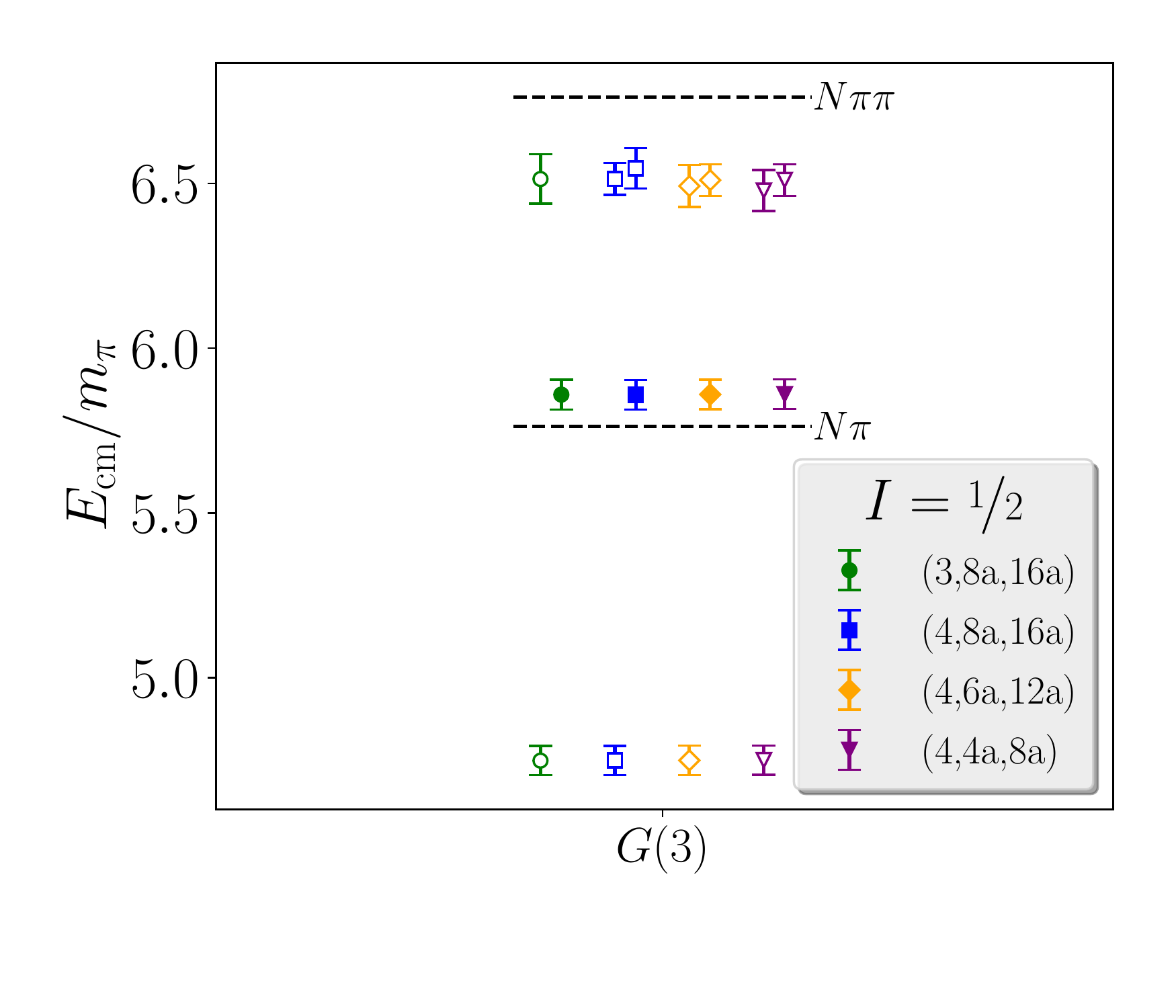}
\caption{\label{fig:spectrum_determination} Example determination of the spectrum in the 
$I=1/2$ $G(3)$ irrep. Left: $t_{\rm min}$-plot for the second level including single- and 
two-exponential fits to $D_n(t)$ as well as single-exponential fits to the ratio $R_n(t)$ in 
Eq.~(\ref{e:ratio}). All fits employ the GEVP of Eq.~(\ref{e:gevp}) with $N_{\rm op} = 4$ and 
$(t_0, t_{\rm d}) = (8a,16a)$. The horizontal band represents the chosen $t_{\rm min}$ from the 
ratio fit. Right: stability under variation of the GEVP parameters for the entire spectrum in 
this irrep. For each level $t_{\rm min}$ is fixed while $(N_{\rm op}, t_0, t_{\rm d})$ are 
varied as shown in the legend. The elastic and inelastic thresholds are represented by dotted 
lines.} 
\end{figure}

A sample illustration of the procedure for nucleon-pion energies is shown in 
Fig.~\ref{fig:spectrum_determination} for the second level in the $I=1/2$ $G(3)$ irrep. Due to partial 
wave mixing, the single-nucleon state is also present in this irrep. The GEVP is therefore required to 
properly isolate the desired higher-lying nucleon-pion energies. Analogous plots for all levels are 
given in \ref{s:appendix_gevp} and~\ref{s:appendix_tmin} for the GEVP- and 
$t_{\rm min}$-stability plots, respectively.

The spectra resulting from this analysis are shown in Figs.~\ref{fig:isodoublet_spectrum} and 
\ref{fig:isoquartet_spectrum} for the $I=1/2$ and $I=3/2$ channels, respectively.

\section{Scattering parameter results and discussion}
\label{s:res}

This section details the determination of the scattering parameters from the finite-volume energies. 
The parameterizations of the $K$ matrix elements are presented, and best-fit values for the parameters 
are summarized.  Lastly, a comparison with chiral perturbation theory is made.

\subsection{Scattering parameter determinations}

\begin{figure}[t]
\centering
\begin{subfigure}{1.0\linewidth}
    \centering
    \includegraphics[width=0.8\linewidth]{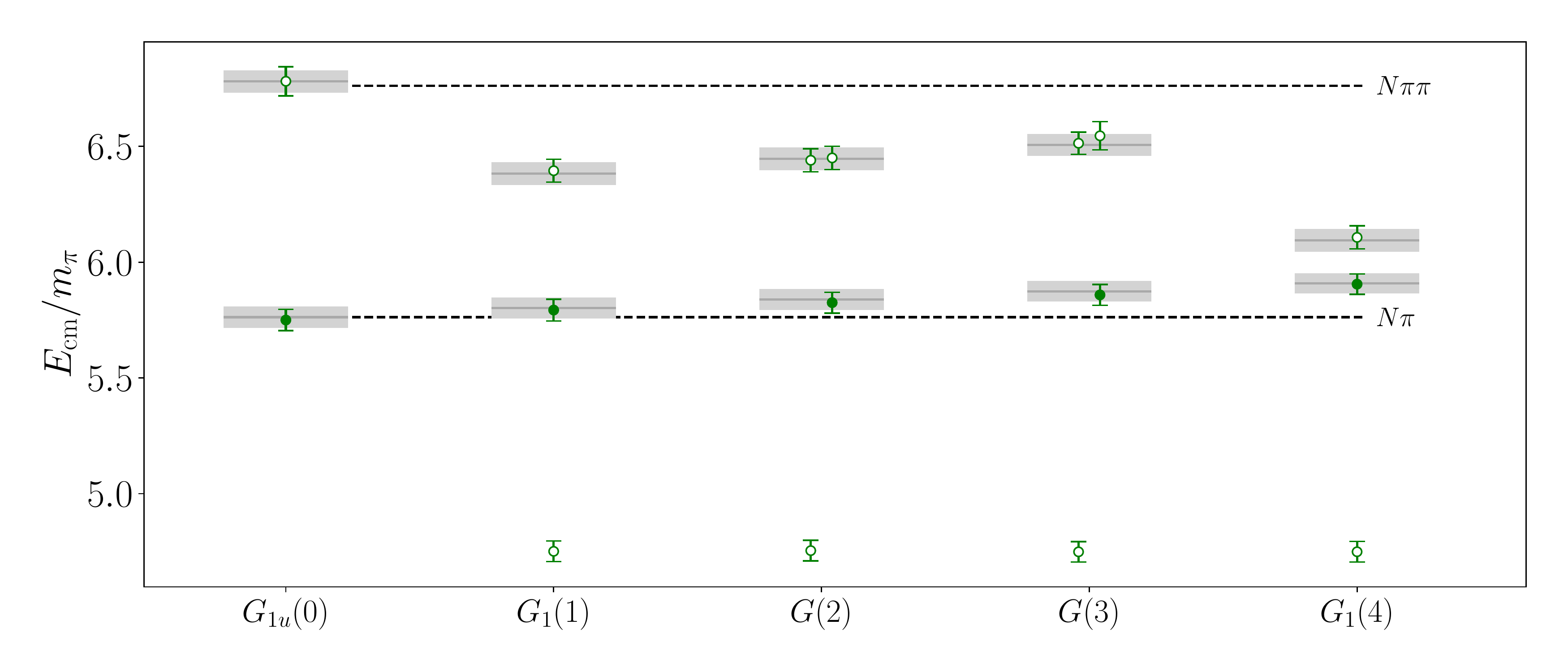}
    \caption{\label{fig:isodoublet_spectrum}The $I=1/2$ spectrum.}
\end{subfigure}
\begin{subfigure}{1.0\linewidth}
    \centering
    \includegraphics[width=0.8\linewidth]{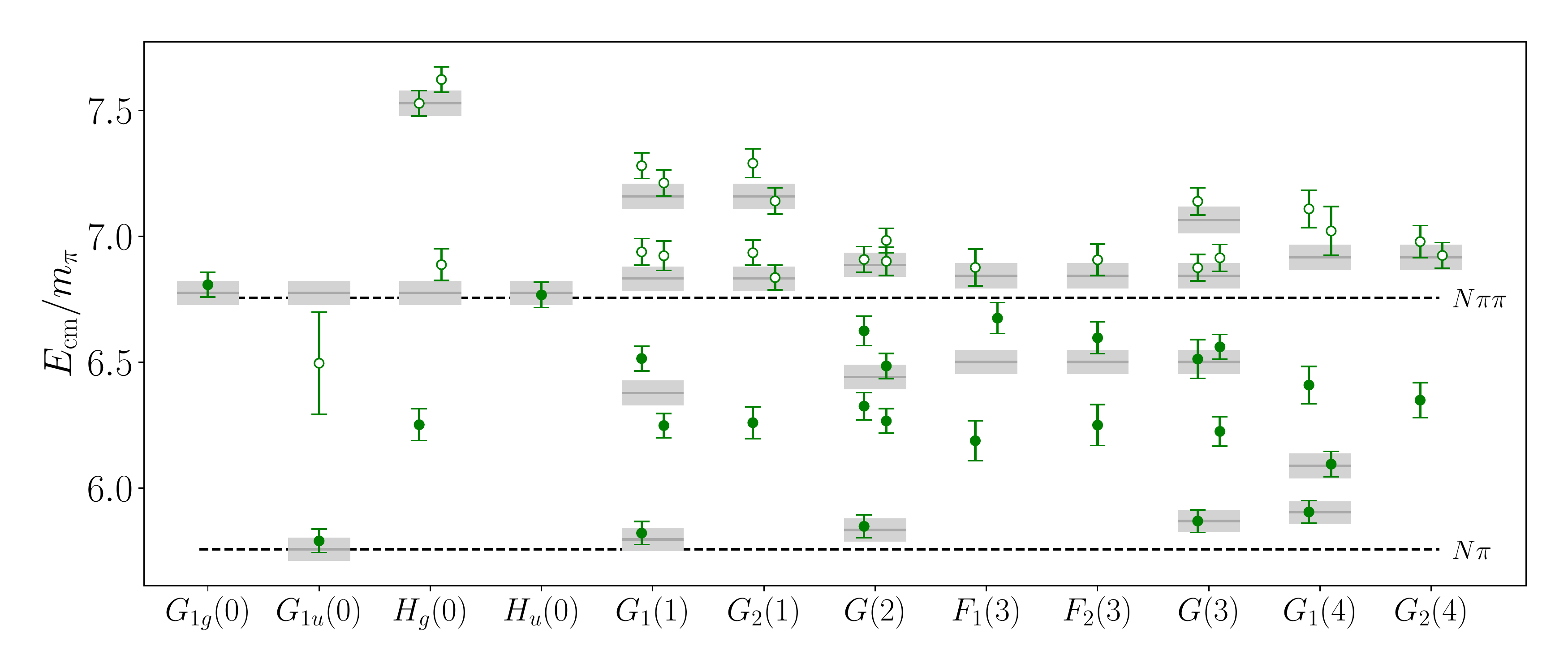}
    \caption{\label{fig:isoquartet_spectrum}The $I=3/2$ spectrum.}
\end{subfigure}

\caption{\label{fig:spectrums} The low-lying $I=1/2$ and $I=3/2$ nucleon-pion spectra in the
center-of-momentum frame on the D200 ensemble described in Table~\ref{tab:comp_deets}. Each column 
corresponds to a particular irrep $\Lambda$ of the little group of total momentum 
$\boldsymbol{P}^2=(2\pi/L)^2\boldsymbol{d}^2$, denoted $\Lambda(\textbf{\textup{d}}^2)$. Dashed 
lines indicate the boundaries of the elastic region. Solid lines and shaded regions indicate 
non-interacting $N\pi$ levels and their associated statistical errors. Levels employed in 
subsequent fits to constrain the scattering amplitudes are shown with solid symbols. 
For $I=3/2$, all well-constrained levels with overlap below the $N\pi\pi$ threshold 
are included. For the $I=1/2$ channel, we are only interested in the scattering length, so only the 
ground state in each irrep is used to determine the $I=1/2$ amplitude near the $N\pi$ threshold.}
\end{figure}

The energies shown in Figs.~\ref{fig:isodoublet_spectrum} and~\ref{fig:isoquartet_spectrum} are next 
used to determine scattering amplitudes via the relations in Sec.~\ref{s:fv}.  Although these relations 
are only applicable to energies below the nucleon-pion-pion threshold, the slow growth of three-body 
phase space near threshold suppresses corrections to Eq.~(\ref{e:det}) and the coupling of 
nucleon-pion-pion states to our operator basis is naively suppressed by the spatial volume, so energies 
somewhat above the inelastic threshold are expected to be appropriate for inclusion in our global fits.
Nevertheless, we restrict our attention to center-of-mass energies below or within one standard 
deviation of the threshold $E_{\rm cm} = 2m_\pi + m_{\rm N}$.

The goal of this analysis is a parametrization of the $J^P = 1/2^-$ partial wave for both isospins, 
and the $3/2^+$ wave with $I=3/2$. As discussed in Sec.~\ref{s:fv}, energies from irreps with zero total 
momentum directly provide points for these partial waves up to corrections from $\ell \ge 3$ contributions. 
However, mixing among the desired waves, as well as with others, generally occurs for energies in irreps 
with non-zero total momentum. The zero-momentum points are therefore a useful guide when plotted together 
with the partial wave fits.

Each partial wave is parametrized using the effective range expansion.  For the $I=3/2$, $J^P=3/2^+$ wave, 
the next-to-leading order is included
\begin{equation}
    \frac{q_{\rm cm}^{3}}{m_\pi^{3}} \cot \delta_{3/2^+}  = \frac{6 \pi \sqrt{s} }{m^3_\pi \gbw^2} 
    ( m_\Delta^2 - s ),
\end{equation}
where $\sqrt{s} = E_{\rm cm}=\sqrt{m_\pi^2 + q_{\rm cm}^2} + \sqrt{m_N^2 + q_{\rm cm}^2}$, and the 
effective range fit parameters are reorganized to form the conventional  Breit-Wigner properties of 
the $\Delta(1232)$ resonance, denoted $\gbw^2$ and $m_\Delta$. For the other waves, the leading term 
in the effective range expansion is sufficient
\begin{equation}
    \frac{q_{\rm cm}^{2\ell+1}}{m_\pi^{2\ell+1}} \cot \delta^I_{J^P}  = \frac{\sqrt{s}}{m_\pi A^I_{J^P}},
\end{equation}
where the overall $\sqrt{s}$ factors are adopted from standard continuum analysis~\cite{Yndurain:2007qm}, 
and the single fit parameter $A_{J^P}^I$ is trivially related to the scattering length
\begin{equation}
   m_\pi^{2\ell+1} a_{J^P}^{I} =  \frac{m_\pi }{m_\pi + m_N} A^I_{J^P} .
\end{equation}

Two different fit strategies are employed to determine the parameters from the finite-volume energies.
The first, called the ``spectrum method''~\cite{Guo:2012hv}, obtains best-fit values of the model 
parameters $\{p_n\}$ by minimizing
\begin{equation}\label{e:chsq}
    \chi^{2}\left(\left\{p_{n}\right\}\right)=\sum_{i j}\left(\frac{q_{{\rm cm} ,i}^2}{m_{\pi}^2}
    -\frac{q_{{\rm cm} ,i}^{2,\textrm{QC}}}{m_{\pi}^2}\left(\left\{p_{n}\right\}\right)\right) 
    C_{i j}^{-1} \left(\frac{q_{{\rm cm} ,j}^2}{m_{\pi}^2}-\frac{q_{{\rm cm} ,j}^{2,
    \textrm{QC}}}{m_{\pi}^2}\left(\left\{p_{n}\right\}\right)\right),
\end{equation}
where the $q_{{\rm cm},i}^2$ are the center-of-mass momenta squared computed from lattice QCD, with 
covariance matrix $C$, and $q_{{\rm cm},i}^{2,\textrm{QC}}\left(\left\{p_{n}\right\}\right)$ are the 
center-of-mass momenta squared evaluated from the model fit form for a given choice of parameters 
$\left\{p_{n}\right\}$. The fact that the model depends on $m_{\rm N}/m_{\pi}$, and is therefore not 
independent of the data to be fit, complicates the evaluation of the covariance matrix $C$.  As 
discussed in Ref.~\cite{Morningstar:2017spu}, a simple way to avoid this complication so that $C$ is 
just the covariance matrix of the data is to introduce model parameters for $m_{\pi}L$ and the ratio 
$m_{\rm N}/m_{\pi}$, and include appropriate additional terms in the $\chi^2$ of Eq.~(\ref{e:chsq}). 
Given the relatively small errors on $m_{\rm N}/m_{\pi}$ and $m_\pi L$, these additional terms have 
little effect on the fit parameters and the resultant $\chi^2$, and are subsequently ignored. Note 
that the evaluation of the $(q_{{\rm cm},i}^{2, {\rm QC}}/m_{\pi}^2)(\{p_n\})$ for a particular 
choice of the parameters requires the determination of roots of Eq.~(\ref{e:det}), a procedure which 
can be delicate for closely-spaced energies.

The second method, called the ``determinant residual'' method~\cite{Morningstar:2017spu}, employs the 
determinants of Eq.~(\ref{e:det}) themselves as the residuals to be minimized. These determinants
depend on the fit parameters through the $K$-matrix, which are adjusted to minimize the residuals and 
best satisfy Eq.~(\ref{e:det}). This approach avoids the subtleties associated with root-finding, but 
has other difficulties. For the spectrum method, the covariance between the residuals is, to a good 
approximation, simply the covariance between the $q_{{\rm cm}, i}^2/m_{\pi}^2$, which can be estimated 
once and does not depend on the fit parameters. Conversely, for the determinant residual method, the 
covariance must be re-estimated whenever the parameters are changed. Since the statistical errors on 
the determinant are typically larger than those on $q_{{\rm cm}, i}^2/m_{\pi}^2$, this approach is 
less sensitive to higher partial waves, and results in a smaller $\chi^2$ compared to the spectrum 
method.

\begin{table}[t]
\begin{center}
\begin{tabular}{c@{\hskip 5pt}c@{\hskip 4pt}c@{\hskip 8pt}c@{\hskip 8pt}c@{\hskip 8pt}c@{\hskip 8pt}c@{\hskip 8pt}c@{\hskip 8pt}c@{\hskip 8pt}c}
Fit & $N_{\rm pw}$ & $A_{1/2^-}$ & $\gbw$ & $M_\Delta/ M_\pi$ 
             & $A_{1/2^+}$ & $A_{3/2^-}$ & $A_{5/2^-}$ & $\chi^2$ & dofs \\ \hline
SP & 2 & $-1.56(4)$ & 13.8(6) & 6.281(16) & --- & --- & --- & 44.38 & $23-3$  \\ 
DR & 2 & $-1.57(5)$ & 14.4(5) & 6.257(36) & --- & --- & --- & 14.91 & $23-3$  \\ 
SP & 5 &$-1.53(4)$ & 14.7(7) & 6.290(18) & $-0.19(6)$ & $-0.46(12)$ & 0.37(10)  & 30.17 & $25-6$   \\ 
\end{tabular}
\end{center}
\caption{\label{t:i32} Results for the fits in the $I=3/2$ channel. $N_{\rm pw}$ is the number of 
partial waves included in the fit. Two different fit forms are included, the one denoted 
$N_{\rm pw} = 2$ includes only the desired partial waves, namely $J^P = 1/2^-$ and $3/2^+$, while 
the one with $N_{\rm pw}=5$ includes all $s$-, $p$-, and $d$-waves, employing the two 
additional energy levels in the  $G_{\rm 1g}(0)$ and $H_{\rm u}(0)$ irreps.
For the $N_{\rm pw}=2$ fit, 
results from the determinant-residual method, denoted `DR', are shown in addition to the spectrum 
method, denoted `SP'.  }
\label{tab:fitsquartet}
\end{table}

\begin{table}[t]
\begin{center}
\begin{tabular}{c@{\hskip 12pt}c@{\hskip 12pt}c@{\hskip 12pt}c@{\hskip 12pt}c}
Fit & $N_{\rm pw}$ & $A_{1/2^-}$ & $\chi^2$ & dofs \\ \hline
SP & 1 & 0.82(12) & 1.68 & $5-1$  \\ 
DR & 1 & 0.92(22) & 1.72 & $5-1$  \\ 
SP & 1 & 0.82(13) & 0.79 & $4-1$  \\ 
\end{tabular}
\end{center}
\caption{\label{t:i12} Results for fits to the $I=1/2$ spectrum in Fig.~\ref{fig:isodoublet_spectrum}. 
$N_{\rm pw}$ is the number of partial waves included in the fit. Due to the small number of levels, 
all fits include only the desired $J^P=1/2^-$ partial wave. Nonetheless, the effect of the omitted 
$p$-waves is estimated by removing the $G_1(4)$ level, which evidently has little influence on the 
result. `SP' refers to the spectrum method, and `DR' refers to the determinant-residual method.}
\label{tab:fitsdoublet}
\end{table}

For the $I=3/2$ fits, the $J^P=1/2^+$, $3/2^-$, and $5/2^-$ partial waves are added to the spectrum 
method fits along with the ground states in the $G_{\rm 1g}(0)$ and $H_{\rm u}(0)$ irreps. The 
$I=3/2$ spectrum in the $G_{\rm 2g}(0)$ irrep was not computed, and irreps in the $I=1/2$ channel 
which do not contain the $s$-wave were also omitted. This choice was made for computational simplicity, 
although these irreps may be beneficial to further constrain higher partial waves in future work.  
The determinant residual method was found to be less able to constrain higher partial 
waves and was only used in fits that included just the $J^P=1/2^-$, $3/2^+$ waves.
Nonetheless, the consistency between these 
different fitting methods, as well as those including higher partial waves, suggest that uncertainties 
on amplitude parameters are statistics dominated.
 
For the $I=1/2$ channel, $\ell_{\rm max}=0$ is employed. Although the small number of levels precludes 
a sophisticated estimate of the effect of higher partial waves, the influence of the omitted $p$-waves 
can be explored by examining the influence of the highest-lying level on the fit.  Table~\ref{t:i12} 
indicates that the effective range is insensitive to the omission of the lowest-lying nucleon-pion 
level in the $G_{1}(4)$ irrep. These $I=1/2$ fits are also insensitive to an additional term in the 
effective range expansion, and exhibit no statistically significant difference between the spectrum 
and determinant-residual methods.

\begin{figure}[t]
\centering
\begin{subfigure}{0.5\textwidth}
\centering
    \includegraphics[width=\linewidth]{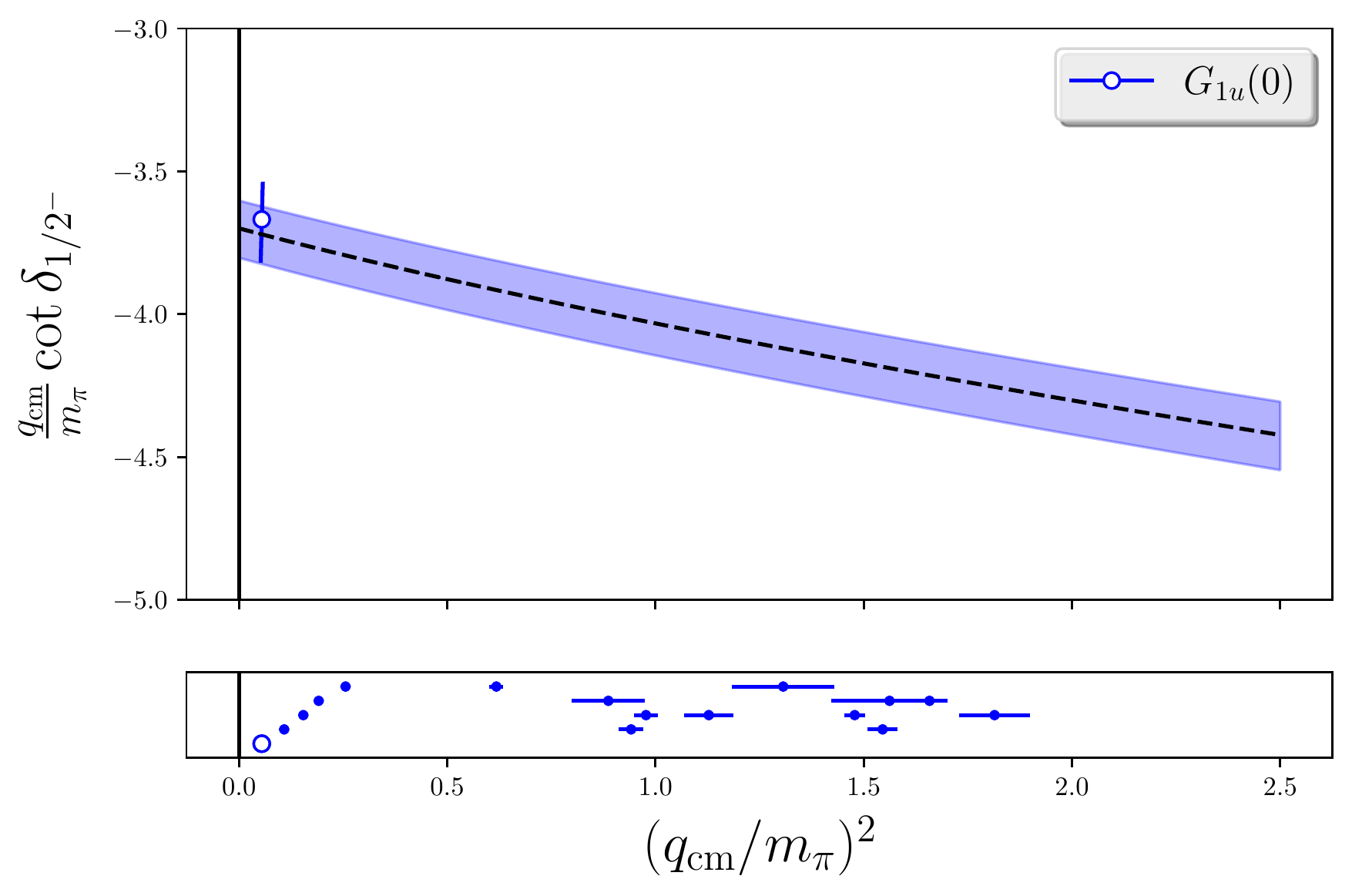}
    \caption{$J^P=1/2^-$}
\end{subfigure}%
\begin{subfigure}{0.5\textwidth}
\centering
    \includegraphics[width=\linewidth]{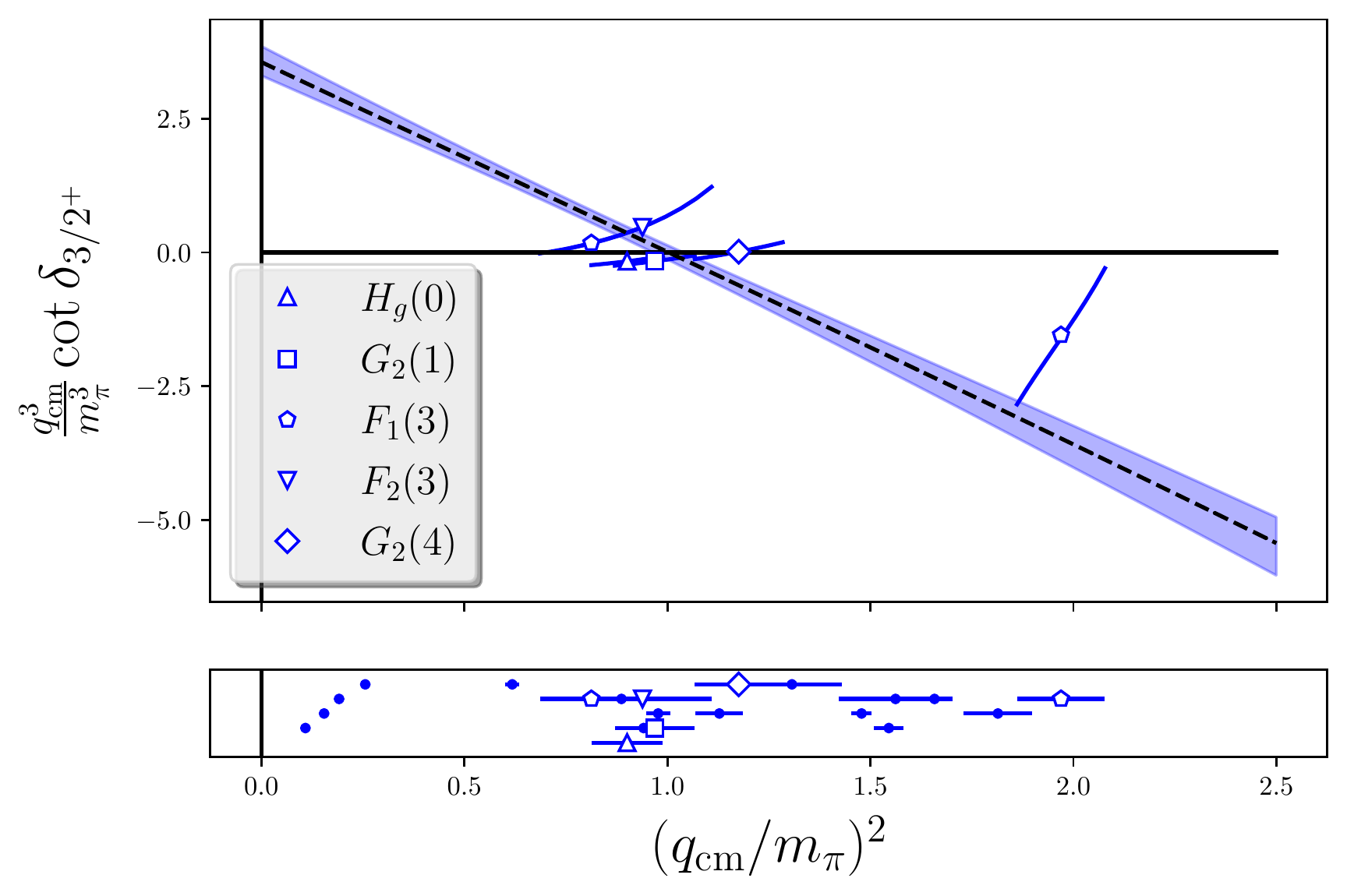}
    \caption{$J^P=3/2^+$}
\end{subfigure}
\caption{
The results from fits to the $I=3/2$ spectrum in Fig.~\ref{fig:isoquartet_spectrum} using the 
spectrum method including the $J^P=1/2^-, 3/2^+$ partial waves only, omitting the 
$H_{\rm u}(0)$ and $G_{1\rm g}(0)$ irreps. 
The lower panel of each partial wave shows the squares of the center-of-mass momenta of the
finite-volume levels which contribute to fitting that partial wave.  Most levels, shown with solid symbols,
contribute to both partial waves, so solving for the partial wave phase shift
shown in the upper panel cannot be done.  When a particular level couples
only to the partial wave shown, a phase shift point can be obtained from the energy level and is shown 
in the upper panel. Hollow symbols indicate such levels. For clarity, the levels in the lower panel are 
vertically spaced according to the (integer-valued) total momentum $\boldsymbol{d}^2$.
}
\label{fig:fitsI32}
\end{figure}

\begin{figure}[t]
\centering
\begin{subfigure}{0.5\textwidth}
\centering
    \includegraphics[width=\linewidth]{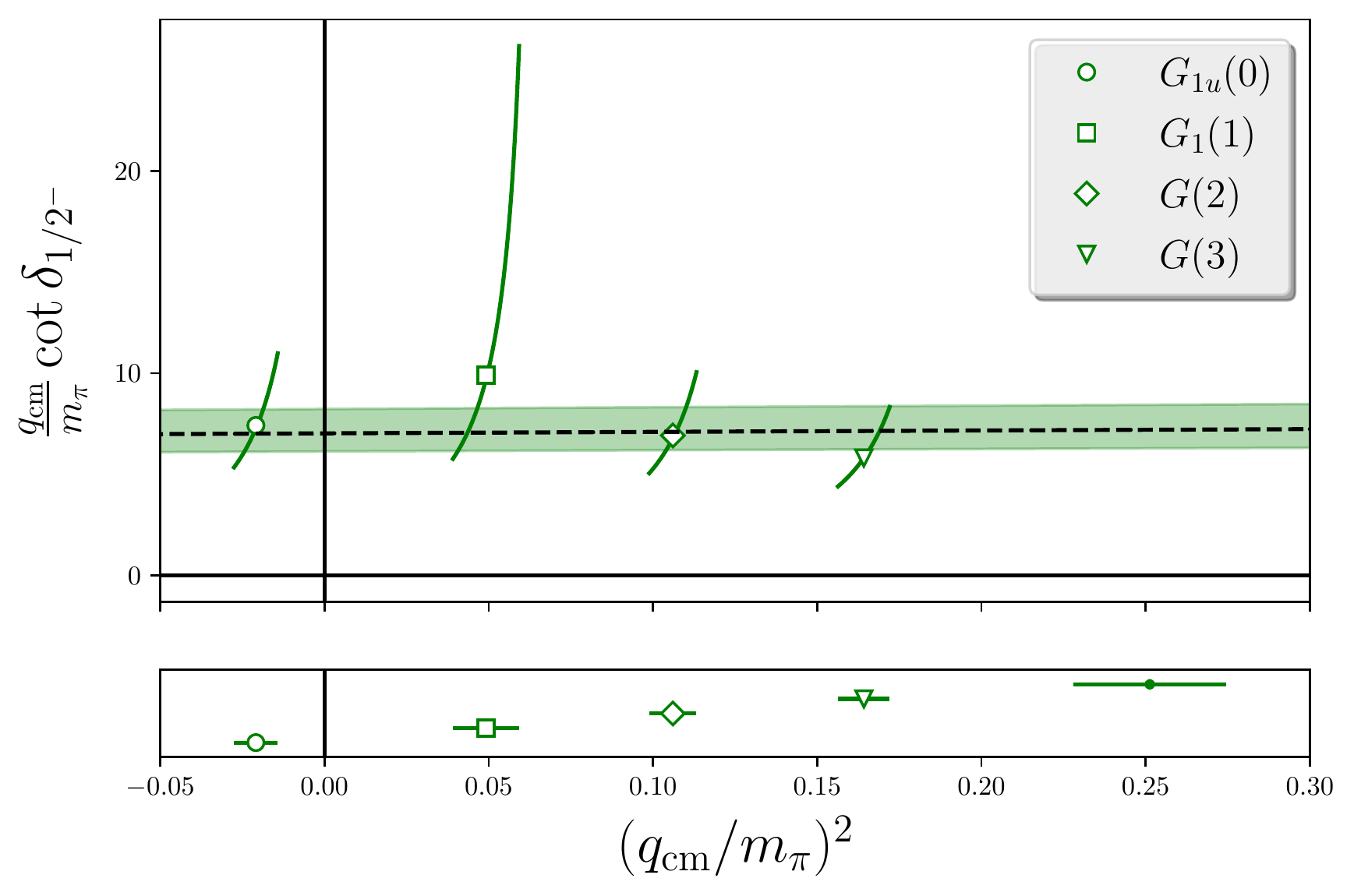}
\end{subfigure}%
\caption{The result of fits to the $I=1/2$ spectrum in Fig.~\ref{fig:isodoublet_spectrum} to determine 
the scattering length of the $J^P=1/2^-$ wave. As in Fig.~\ref{fig:fitsI32}, the lower panel shows the 
input spectra. For $\ell_{\rm max}=0$, even levels with total non-zero momentum result in phase shift 
point in the upper panel.
The level with largest $q_{\rm cm}^2$ is not shown in the upper panel due to its large error.
}
\label{fig:fitsI12}
\end{figure}

\begin{figure}[tp]
\centering
\begin{subfigure}{0.7\textwidth}
\centering
    \includegraphics[width=\linewidth]{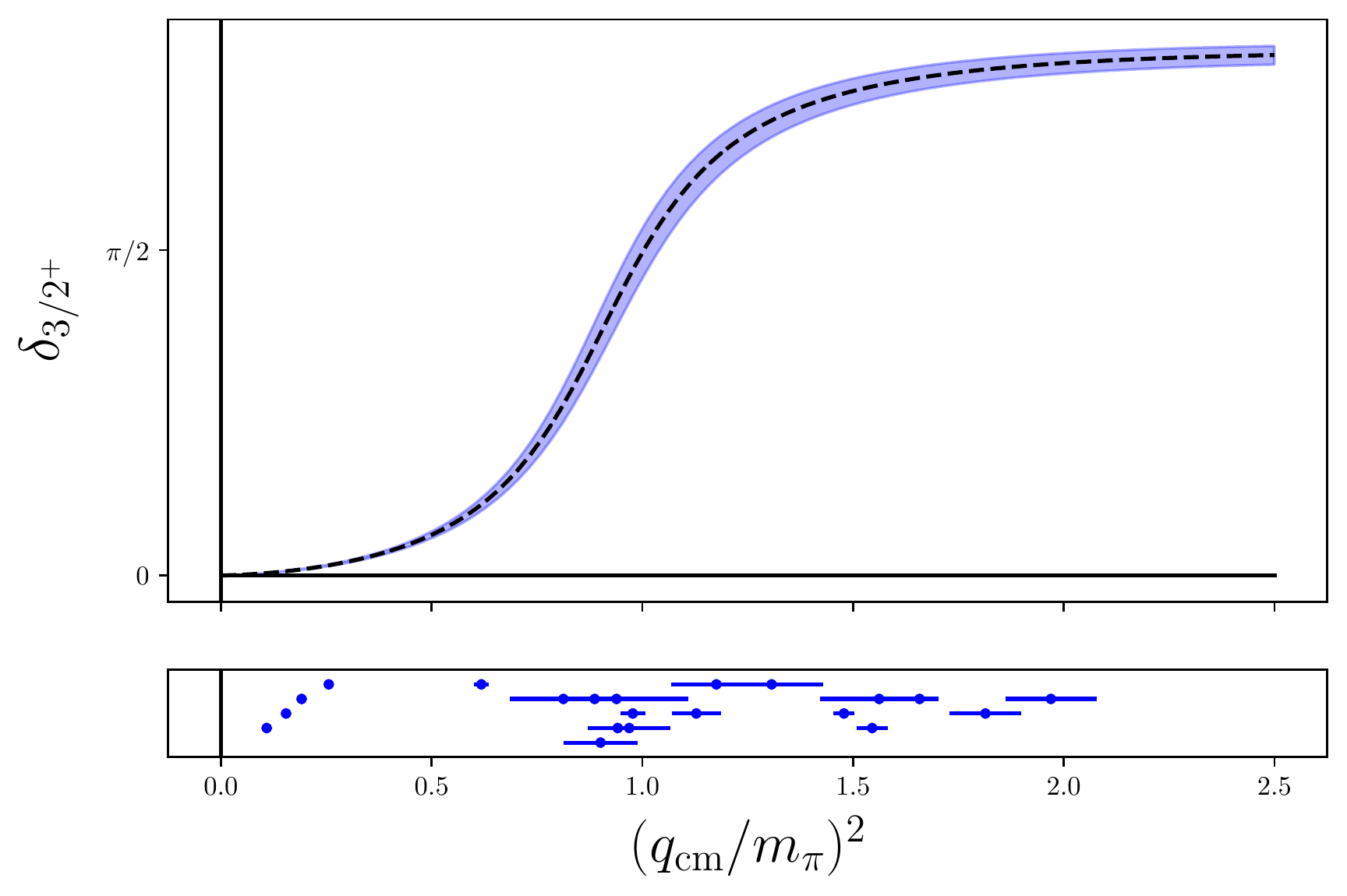}
\end{subfigure}%
\caption{Scattering phase shift of the $I=3/2$, $J^P = 3/2^+$ partial wave containing the $\Delta(1232)$ 
resonance. The curve is obtained from a fit of the finite-volume energies shown in the lower panel using 
Eq.~(\ref{e:det}) and a  Breit-Wigner form. The energies are computed on the single $N_{\rm f} = 2+1$ 
lattice QCD gauge field ensemble with $a=0.065$ fm and $m_\pi=200$~MeV described in 
Table~\ref{tab:comp_deets}.
Levels used in the fit are shown in the lower panel, similar to Figs.~\ref{fig:fitsI32} 
and \ref{fig:fitsI12}, but no data points are shown in the upper panel to more clearly show the final fit form.
}
\label{fig:deltaphase}
\end{figure}

Results from fits using both the spectrum and determinant-residual methods including various partial 
waves are given in Tables~\ref{t:i32} and~\ref{t:i12} for  $I=3/2$ and $I=1/2$, respectively. In 
addition to the desired partial waves, fits using the spectrum method are mildly sensitive to the 
$J^P = 1/2^+$, $3/2^-$, and $5/2^-$ waves with $I=3/2$. Although not included in the table, the 
determination of the effective range for both isospins is robust to the addition of the next term 
in the effective range expansion. Results for the partial waves from the fit including only the 
desired partial waves are shown with the points from the total-zero momentum irreps in 
Figs.~\ref{fig:fitsI32} and~\ref{fig:fitsI12} for the $I=3/2$ and $I=1/2$ partial waves, respectively. 
The phase shift $\delta_{3/2^+}$ has the characteristic profile of the $\Delta(1232)$ resonance and is 
shown in Fig.~\ref{fig:deltaphase}. Since the scattering length is the only desired parameter from 
the $I=1/2$ spectrum, only the lowest nucleon-pion levels from each irrep are included in the fit, 
as denoted by the solid symbols in Fig.~\ref{fig:isodoublet_spectrum}. Full exploration of the 
elastic $I=1/2$ spectrum likely requires additional operators beyond the scope of this work, due 
to the strongly-interacting $J^P=1/2^+$ wave containing the $N(1440)$ Roper resonance.

\begin{figure}[tp]
\centering
\begin{subfigure}{\linewidth}
    \centering
     \includegraphics[width=0.9\linewidth]{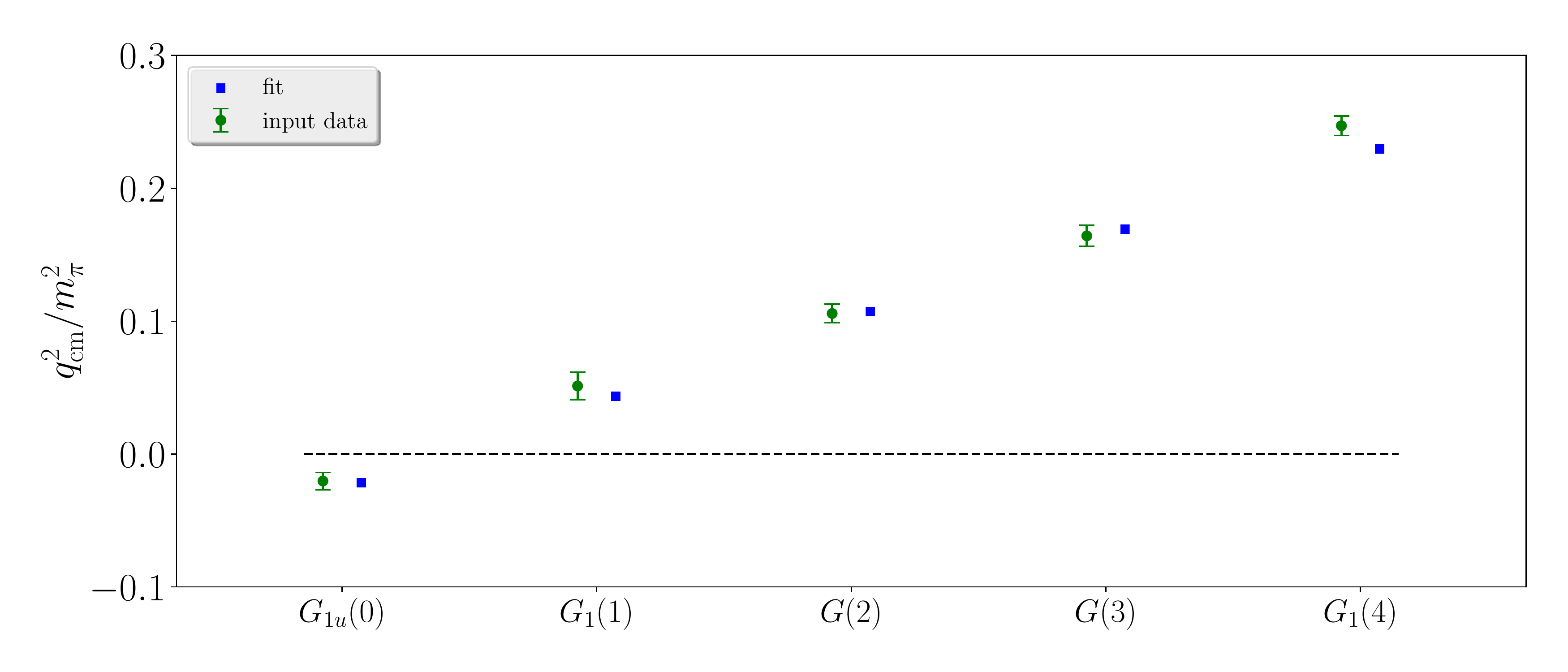}
     \label{fig:predicted_levels_isodoublet}
     \caption{The $I=1/2$ spectrum compared with model values.}
\end{subfigure}
\begin{subfigure}{\linewidth}
    \centering
     \includegraphics[width=0.9\linewidth]{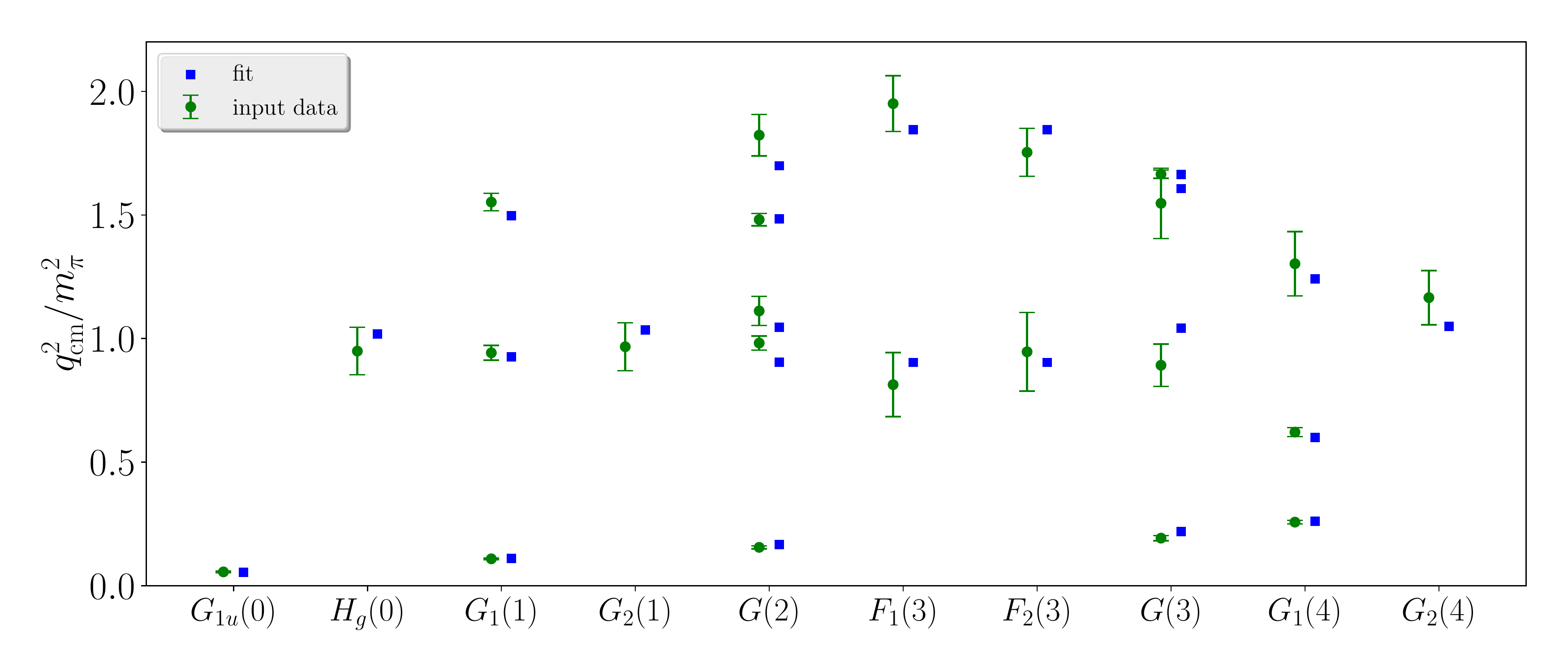}
     \label{fig:predicted_levels_isoquartet}
     \caption{The $I=3/2$ spectrum compared with model values.}
\end{subfigure}
\caption{The center-of-mass momentum $q_{\rm cm}^2/m_{\pi}^2$ for the $I=1/2$ and $I=3/2$ spectra 
together with model values from amplitude fits employing the spectrum method with $N_{\rm pw} = 2$ 
partial waves for $I=3/2$. For $I=1/2$, only the $s$-wave is included and  the fit to all five points 
is shown.
\label{fig:predicted_levels}}
\end{figure}

The spectrum method enables an additional visualization of the quality of fits to the finite-volume spectra.
The residual is constructed using model values of $q_{\rm cm}^{2, {\rm QC}}/m_{\pi}^2$ which depend on the 
parameters and can be compared with the input data from the spectrum. Such comparisons are shown in 
Fig.~\ref{fig:predicted_levels} for both the $I=1/2$ and $I=3/2$ spectra. Although not shown explicitly 
on the plot, the ground states in $G_1(1)$, $G(2)$, $G(3)$, and $G_1(4)$ with $I=3/2$ are sensitive to 
the $J^P=3/2^+$ partial wave. The $\ell_{\rm max}=0$ approximation significantly increases the $\chi^2$ 
for these levels. Conversely, these levels therefore place significant constraints on the near-threshold 
behavior of the $3/2^+$ wave, in contrast to the higher-lying levels in the $H_g(0)$, $G_2(1)$, $F_1(3)$, 
$F_2(3)$, and $G_2(4)$ irreps. The ground states in the $G_{1\rm g}(0)$ and $H_{\rm u}(0)$ irreps are not 
shown on the plot, and only included in the $N_{\rm pw} = 5$ fit in Table~\ref{t:i32}.

\begin{figure}[t]
    \centering
    \includegraphics[width=0.49\linewidth]{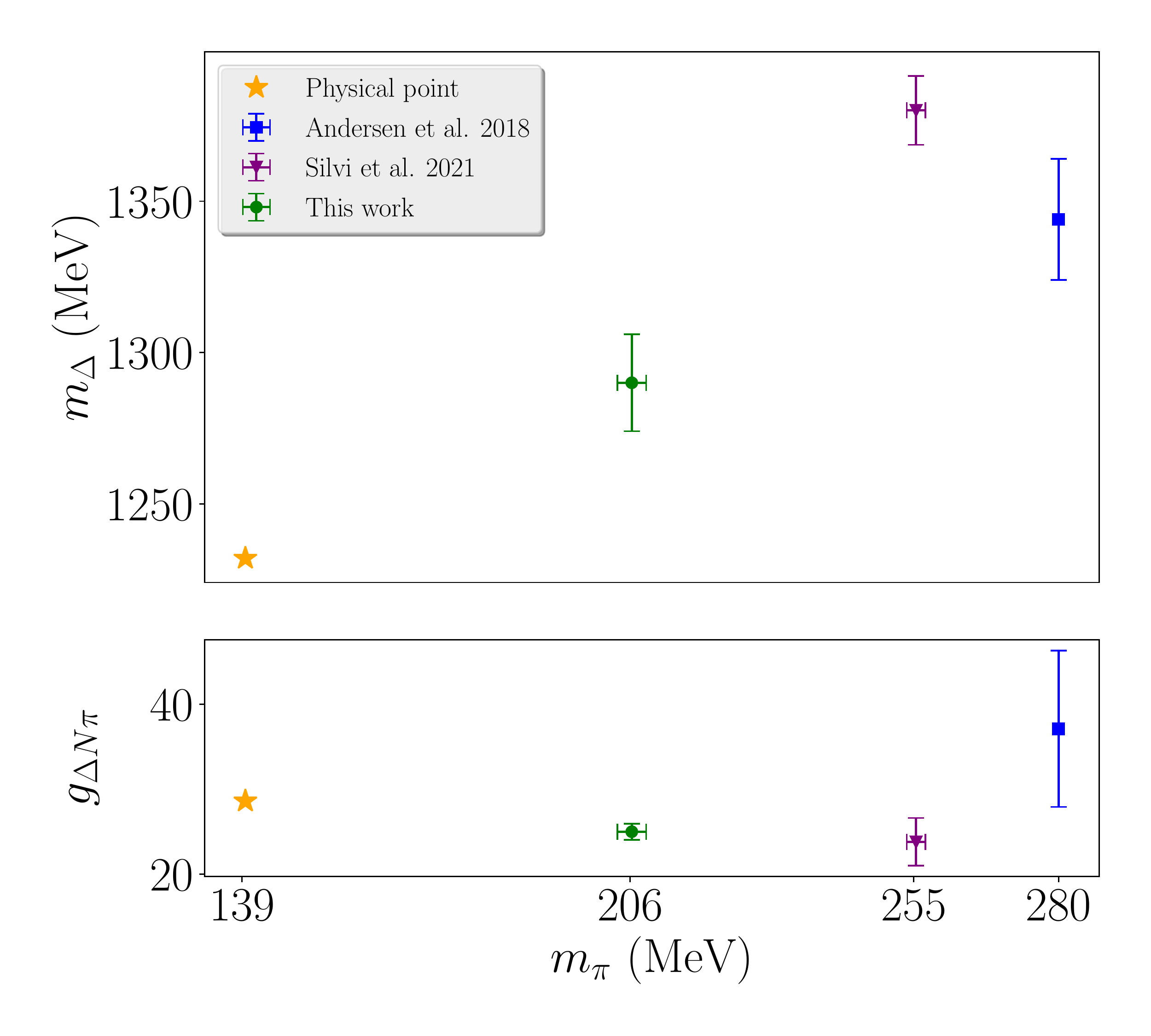}
    \includegraphics[width=0.49\linewidth]{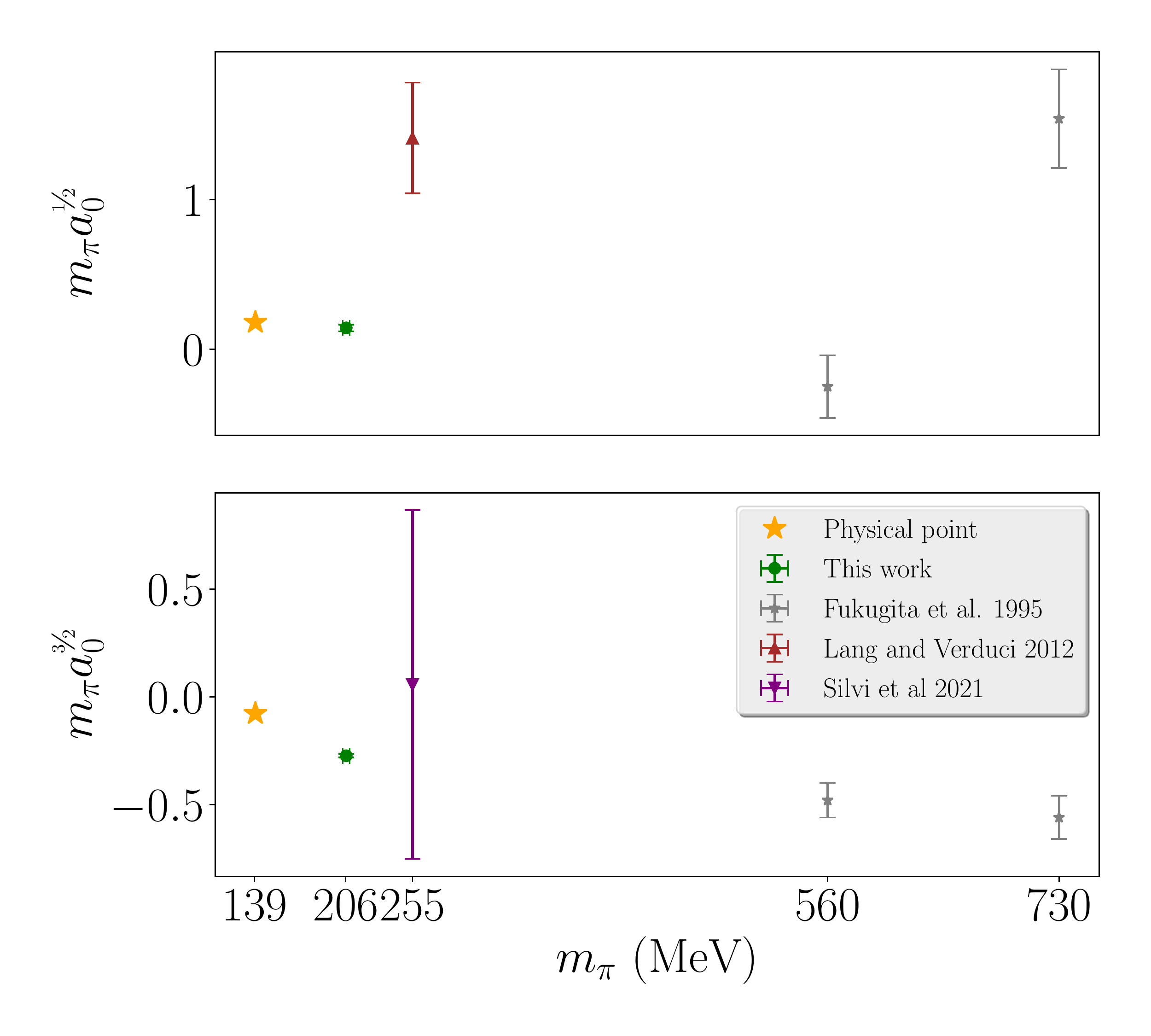}
    \caption{\label{fig:m_delta} Comparison of results from this work to previous lattice calculations.  
    Top left: the Breit-Wigner mass $m_\Delta$. Bottom left: the coupling $\gloeft$ from leading-order 
    effective field theory. Top right: the $N\pi$ isodoublet scattering length $m_\pi a_0^{1/2}$ in 
    terms of the pion mass.  Bottom right: the $N\pi$ isoquartet scattering length $m_\pi a_0^{3/2}$. 
    Prior results are indicated by `Anderson et al. 2018' \cite{Andersen:2017una}, `Silvi et al. 2021' 
    \cite{Silvi:2021}, `Fukugita et al. 1995' \cite{Fukugita:1994ve}, and `Lang and Verduci 2012' 
    \cite{Lang:2012db}. Physical point values are obtained using 
    Refs.~\cite{pdg:2022,Hemmert:1994ky,Hoferichter:2015tha}.
    }
\end{figure}

The final results for the scattering lengths in this work are taken from the determinant residual method 
fit in Table~\ref{t:i32} with $N_{\rm pw} = 2$ for $I=3/2$ and the spectrum method fit for $I=1/2$ 
including all five levels
\begin{align}\label{e:slen2}
 m_{\pi}a_0^{3/2} = -0.2735(81) \, , \qquad m_{\pi}a_0^{1/2} = 0.142(22),
\end{align}
which are already given in Eq.~(\ref{e:slen}). In Fig.~\ref{fig:m_delta}, the results from this work 
for the Breit-Wigner parameters of the $\Delta(1232)$ resonance in the $I=3/2$, $J^P=3/2^+$ partial 
wave are compared to the published numbers in Refs.~\cite{Andersen:2017una} and~\cite{Silvi:2021}
where, as is customary, the definition of the $\gloeft$ coupling from leading-order effective field 
theory is used, as defined in Eq.~(39) of Ref.~\cite{Silvi:2021}.  When considering Fig.~\ref{fig:m_delta}, 
keep in mind that the quark mass trajectory used here and in Ref.~\cite{Andersen:2017una}, which fixes 
the sum of the quark masses, differs from that used in Ref.~\cite{Silvi:2021}, which fixes the strange 
quark mass to its physical value.  A comparison of the scattering lengths determined here to past 
lattice QCD results is also shown in Fig.~\ref{fig:m_delta}.

\subsection{Comparison with phenomenology and chiral perturbation theory}

Although the results here are only at a single pion mass, it is interesting to compare our 
scattering lengths to those extracted phenomenologically~\cite{Baru:2010xn,Baru:2011bw} from 
pionic atoms~\cite{Gotta:2008zza,Strauch:2010vu,Hennebach:2014lsa}, as well as to those obtained 
from chiral perturbation theory.  The proximity of $m_\pi=200$~MeV to its physical value
naively suggests that $SU(2)$ baryon $\chi$PT may be applicable.  While $\chi$PT may be poorly converging 
for $g_A$ and $m_N$, the convergence pattern is an observable dependent issue which has not been explored 
for $N\pi$ scattering. 

Pionic hydrogen ($\pi H$) is sensitive to one combination of the isoscalar ($a^+$) and isovector ($a^-$) 
scattering lengths and pionic deuterium ($\pi D$) is sensitive to a different combination, allowing for 
a percent level determination~\cite{Baru:2010xn,Baru:2011bw}. Phenomenological values extracted 
from a full $\pi N$ partial wave analysis have been reported in Ref.~\cite{Hoferichter:2015hva}.  
Ref.~\cite{Hoferichter:2016ocj} determined the values of these scattering lengths in the 
isospin limit. While it is customary for the lattice community to use $m_\pi=135$~MeV as the 
pion mass in the isospin limit, it is common in the phenomenological estimates to use the charged pion mass 
to determine the QCD quantities in the absence of QED corrections.  In order to be consistent with the 
phenomenological estimates, we similarly use the charged pion mass when quoting results at the physical pion mass.

These scattering lengths are known to fourth order in the baryon chiral 
expansion~\cite{Fettes:1998ud,Fettes:2000xg,Becher:2001hv} and expressed in 
Appendix F of Ref.~\cite{Hoferichter:2015hva} and Ref.~\cite{Hoferichter:2015tha} in a form convenient 
for extrapolating LQCD results.  In terms of the $s$-wave $a_0^\pm$ scattering lengths, the 
isospin 1/2 and 3/2 $\pi N$ scattering lengths are given by
\begin{align}
&a_{0}^{3/2} =  a_0^+ - a_0^-\, ,&
&a_{0}^{1/2} = a_0^+ + 2 a_0^-\, .&
\end{align}
At leading order (LO), the scattering lengths are free of LECs and given by
\begin{align}
&m_\pi a_{0}^{3/2}[{\rm LO}] = -\epi^2 \frac{2\pi}{1+\mu}\, ,&
&m_\pi a_{0}^{1/2}[{\rm LO}] =  \epi^2 \frac{4\pi}{1+\mu}\, ,&
\end{align}
where
\begin{align}
&\epi = \frac{m_\pi}{4\pi F_\pi},&
&\mu = \frac{m_\pi}{m_N}. &
\end{align}
The values of these input parameters on D200 and at the physical (charged) pion mass are
\begin{align}
&\epi^{\rm D200} = 0.1759(12),&
&\mu^{\rm D200} = 0.2102(19),&
\nonumber\\
&\epi^{\rm phys} = 0.12064(74),&
&\mu^{\rm phys} = 0.14875(05)\, .&
\end{align}

\begin{table}
\begin{center}
\begin{tabular}{l@{\hspace*{8mm}}c@{\hspace*{8mm}}l@{\hspace*{8mm}}l}
   & $m_\pi$ (MeV) & $\ \ m_\pi a_{0}^{1/2}$ & $\quad\ m_\pi a_{0}^{3/2}$ \\ \hline
This work  & 200 &  $0.142(22)$ & $-0.2735(81)$ \\
LO $\chi$PT & 200 & $0.321(04)(57)$ & $-0.161(02)(28)$ \\
LO $\chi$PT & 140 & $0.159(02)(19)$ & $-0.080(01)(10)$ \\
Pheno.~(isospin limit)\cite{Hoferichter:2016ocj} & 140 & $0.1788(38)$  & $-0.0775(35)$\\
\end{tabular}
\end{center}
\caption{A comparison of our $N\pi$ scattering length results at $m_\pi=200$~MeV
with phenomenological values in the
isospin limit and predictions from leading order chiral perturbation theory.
For the $\chi$PT predictions, the first error is from uncertainties on the input parameters,
$\epi$ and $\mu$, and the second error is a $\chi$PT truncation uncertainty given by 
$\epi m_\pi a_0^I[\rm LO]$.
\label{tab:LOphenocompare}}
\end{table}

A comparison of our results with the LO $\chi$PT predictions and phenomenological values in the
isospin limit from Ref.~\cite{Hoferichter:2016ocj} is presented in Table~\ref{tab:LOphenocompare}.
Not only do our results disagree with LO $\chi$PT, but we also find the magnitude of $m_\pi a_0^{3/2}$ 
exceeds that of $m_\pi a_0^{1/2}$, in conflict with both LO $\chi$PT and phenomenology.  Note that 
the LO $\chi$PT predictions at the physical point are in reasonable agreement with the phenomenological 
values, lying within one sigma of the estimated $\chi$PT truncation uncertainty.  With only one pion 
mass available in this work, the reasons for the discrepancy of our results with LO $\chi$PT cannot 
be ascertained.  Interestingly, our scattering length results can be described at next-to-leading order 
(NLO) using a single LEC. At NLO, one finds
\begin{align}
m_\pi a_{0}^{3/2}[{\rm NLO}] &= -\epi^2 \frac{2\pi}{1+\mu}\left\{ 1
    +\frac{\epi}{2} \frac{\Lchi}{m_N}(g_A^2 +8C)
    \right\}\, ,
\nonumber\\
m_\pi a_{0}^{1/2}[{\rm NLO}] &= \phantom{-}\epi^2 \frac{2\pi}{1+\mu}\left\{ 1
    -\frac{\epi}{4} \frac{\Lchi}{m_N}(g_A^2 +8C)
    \right\}\, ,
\end{align}
where $\Lchi = 4\pi F_\pi$ and we have defined the dimensionless LEC
\begin{equation}
C = m_N (2c_1 - c_2 - c_3)\, ,
\end{equation}
in terms of the $c_i$ LECs in the baryon chiral Lagrangian~\cite{Fettes:2000gb}.
The scattering lengths in this work can be described by these NLO formulae if $C$ is in the range 0.6-0.7.
The NLO phenomenological determination finds a value of $C\approx 0.3$, which is not significantly 
different from that needed to describe our results. However, the phenomenological extraction
of the LECs in Ref.~\cite{Hoferichter:2015hva} is clouded by issues related to the $\Delta$ 
degrees of freedom~\cite{Siemens:2016jwj} and is not stable until at least next-to-next-to-next-to-leading 
order (N$^3$LO)~\cite{Hoferichter:2015hva}. When results at additional pion masses, particularly lighter 
ones, become available, a more thorough understanding of the pion mass dependence of the scattering 
lengths can be achieved and a more quantitative comparison with the results from the phenomenological 
analysis and $\chi$PT can be performed.

\section{Conclusion}
\label{s:conc}

This work presents a computation of the lowest partial waves for the elastic nucleon-pion scattering
amplitude on a single ensemble of gauge configurations with $m_\pi = 200~{\rm MeV}$. The $s$-wave 
scattering lengths are determined for both isospins $I=1/2$ and $I=3/2$ and compared to determinations 
from LO $\chi$PT and a Roy-Steiner analysis\cite{Hoferichter:2015hva}.  To our knowledge, this is the 
first (unquenched) lattice QCD determination of both nucleon-pion scattering lengths for $m_\pi<250$~MeV.
The Breit-Wigner resonance parameters of the $\Delta(1232)$ in the $J^P=3/2^+$ partial wave with 
$I=3/2$ are determined as well. 

A comparison of two different methods, the spectrum method and the determinant residual method, of 
extracting $K$-matrix information from finite-volume spectra is also performed.  Although the 
determinant residual method avoids awkward root-finding, it was found to be less sensitive to 
higher partial wave contributions. Nonetheless, the consistency between these two different 
fitting procedures is reassuring. 

These results suggest that the methods used here will prove useful for future work at the physical 
values of the quark masses and for other lattice spacings. Larger volumes needed at smaller 
quark masses will require an increase of $N_{\rm ev}$, the dimension of the LapH subspace discussed 
in Sec.~\ref{s:laph}, but not the number of Dirac matrix inversions in the stochastic-LapH 
algorithm for all-to-all quark propagators.  Nevertheless, the increasingly severe signal-to-noise 
problem will likely require more configurations and source times to achieve a similar statistical 
precision.

This work is part of a larger effort to compute baryon scattering amplitudes on lattice QCD gauge 
field ensembles at quark masses in the chiral regime $m_{\pi} \lesssim 300~{\rm MeV}$ where effective 
theories may be applicable. As discussed in Sec.~\ref{s:laph}, the stochastic LapH approach to quark 
propagation enables considerable re-use of the hadron tensors in multiple multi-hadron correlation 
functions on the D200 ensemble employed here. Analyses are currently underway to compute the 
analogous amplitudes for the $N\Lambda-N\Sigma$ and $NN$ systems. Hopefully this exploratory 
computation has sufficient statistical precision to impact chiral effective theories for these 
baryon-baryon channels as well.

\noindent 
\section*{Acknowledgments}
Helpful discussions are acknowledged with D. Mohler, J. Ruiz de Elvira, M. Hoferichter and M. Lutz. 
Computations were carried out on Frontera~\cite{frontera} at the Texas Advanced Computing 
Center (TACC) under award PHY20009, and at the 
National Energy Research Scientific Computing Center (NERSC), a U.S. Department of Energy 
Office of Science User Facility located at Lawrence Berkeley National Laboratory, operated under 
Contract No. DE-AC02-05CH11231 using NERSC awards NP-ERCAP0005287, NP-ERCAP0010836 and NP-ERCAP0015497.
CJM and SS acknowledge support from the 
U.S.~NSF under awards PHY-1913158 and PHY-2209167.  
FRL has been supported by the U.S.~Department of Energy (DOE), Office
of Science, Office of Nuclear Physics, under grant Contract Numbers DE-SC0011090 and DE-SC0021006.  
The work of ADH is supported by: (i) The U.S. DOE, Office of Science, Office of Nuclear Physics through 
Contract No. DE-SC0012704 (S.M.); (ii) The U.S. DOE, Office of Science, Office of Nuclear Physics and 
Office of Advanced Scientific Computing Research within the framework of Scientific Discovery through 
Advanced Computing (SciDAC) award Computing the Properties of Matter with Leadership Computing Resources.
AN is supported by the National Science Foundation CAREER award program under award PHY-2047185.  
The work of PV was supported 
in part by the U.S. DOE, Office of Science through contract No. DE-AC52-07NA27344 under which LLNL is operated.
The work of AWL was supported in part by the U.S. DOE, Office of Science through contract 
No.~DE-AC02-05CH11231 under which the Regents of the University of California manage LBNL.

\newlength\apwidth
\setlength{\apwidth}{0.32\textwidth}
\vspace{4mm}

\appendix
\section{Systematic errors from correlation matrix rotation}
\label{s:appendix_gevp}

\begin{figure}[tp]
\centering
\begin{subfigure}{\linewidth}
    \centering
    \includegraphics[width=0.8\linewidth]{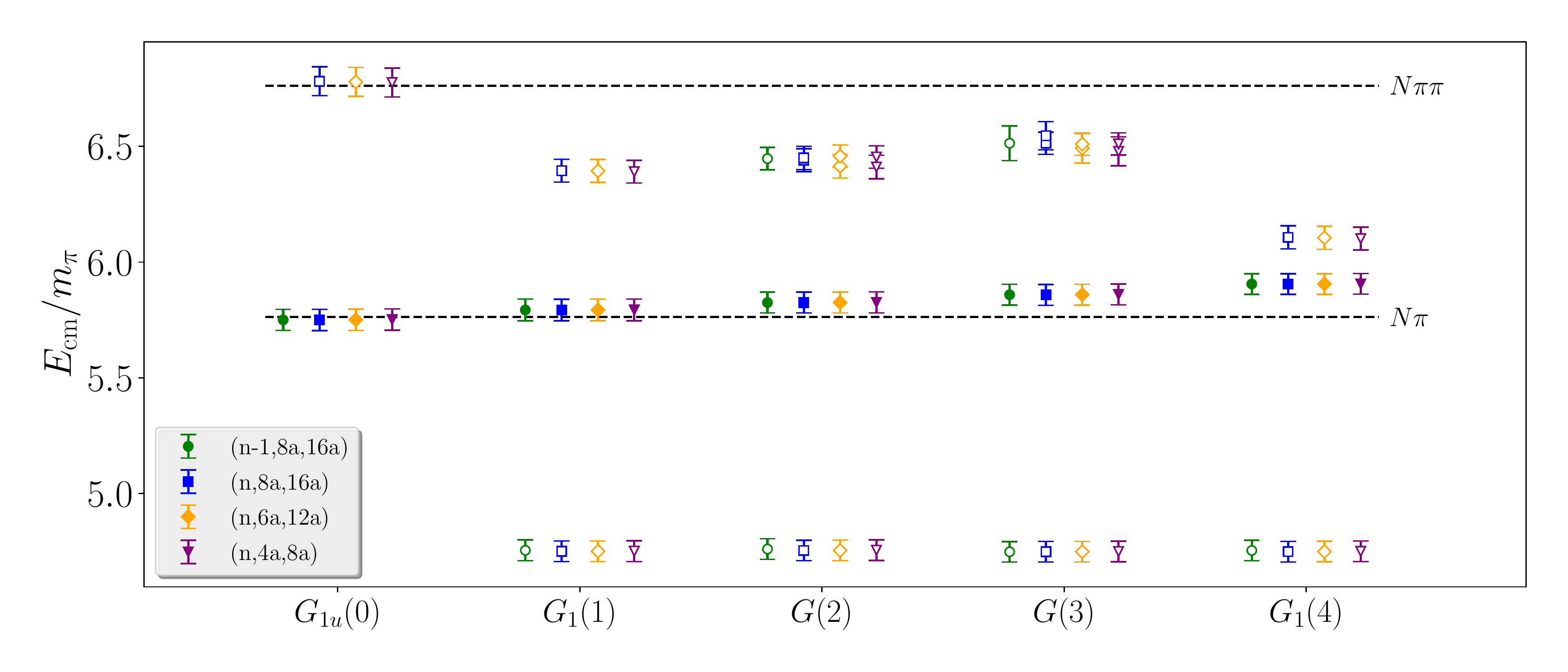}
    \caption{\label{fig:isodoublet_gevp_stability} the $I=1/2$ spectrum.}
\end{subfigure}
\begin{subfigure}{\linewidth}
    \centering
    \includegraphics[width=0.8\linewidth]{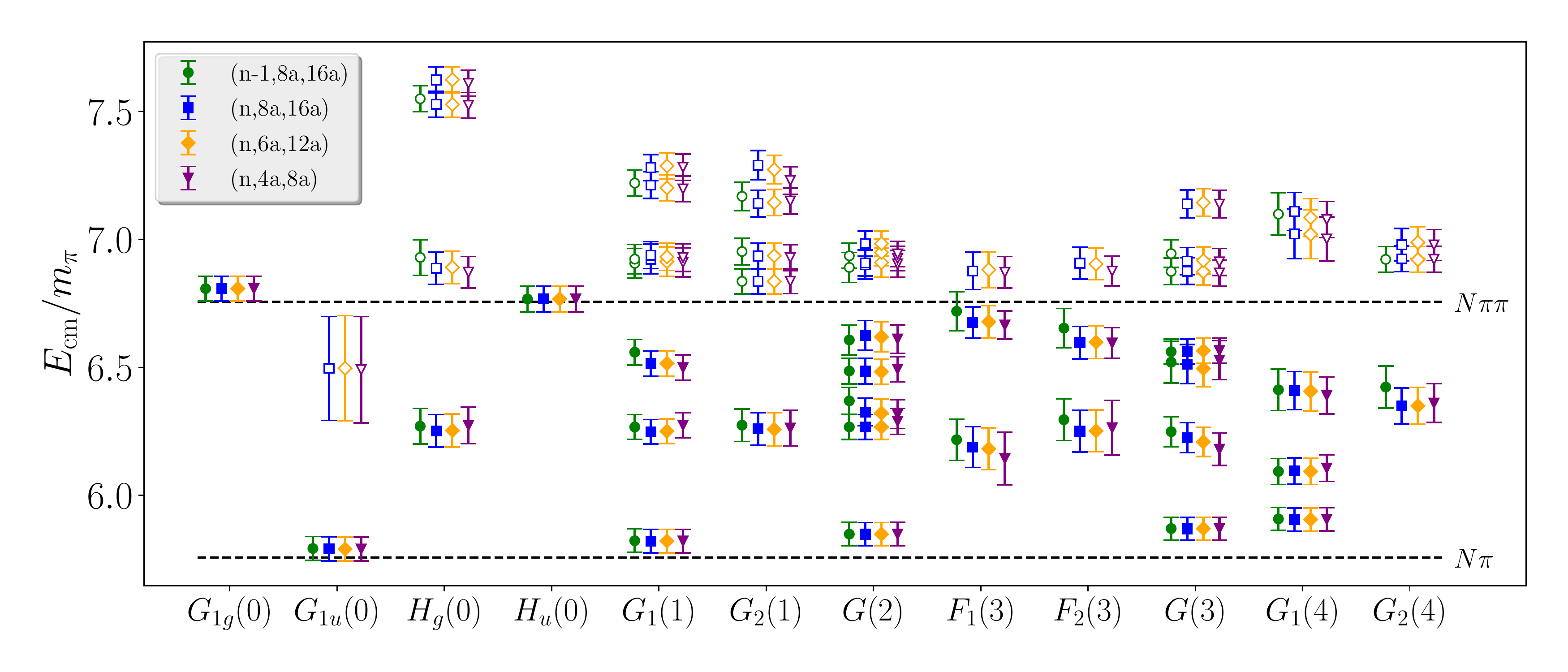}
    \caption{\label{fig:isoquartet_gevp_stability}the $I=3/2$ spectrum.}
\end{subfigure}
\caption{\label{fig:gevp_stability} Stability of the finite-volume spectra under variation of the GEVP as 
discussed in Sec.~\ref{s:fit}. Multiple values of $(N_{\rm op}, t_0, t_{\rm d})$ for each level are shown 
with a horizontal displacement for clarity. Each irrep is shown in a column in the same manner as 
Fig.~\ref{fig:spectrums} with $n$ denoting the maximum $N_\textup{op}$.} 
\end{figure}

As discussed in Sec.~\ref{s:fit}, the optimized diagonal correlation functions $D_n(t)$ are obtained in this work 
from the GEVP using a single-pivot approach which uses one choice of $(t_0,t_{\rm d})$.  The systematic error 
associated with this approach is estimated for each energy level by fixing the fit range $[t_{\rm min}, t_{\rm max}]$ 
and varying the GEVP metric and diagonalization times $(t_0,t_{\rm d})$ defined in Eq.~(\ref{e:gevp}), as well as the 
dimension of the input correlation matrix $N_{\rm op}$.   Taking both GEVP stability and statistical precision into 
account, the parameters $(t_0, t_{\rm d}) = (8a, 16a)$ are found to work well for all energies presented here. As 
shown in Fig.~\ref{fig:gevp_stability}, the spectrum is rather insensitive to variations in $(t_0, t_{\rm d})$ and 
$N_{\rm op}$.

\section{Systematic errors from varying fit forms and time ranges}
\label{s:appendix_tmin}
\begin{figure}[t]
\centering
\begin{subfigure}{\apwidth}
    \centering
    \includegraphics[width=1.0\linewidth]{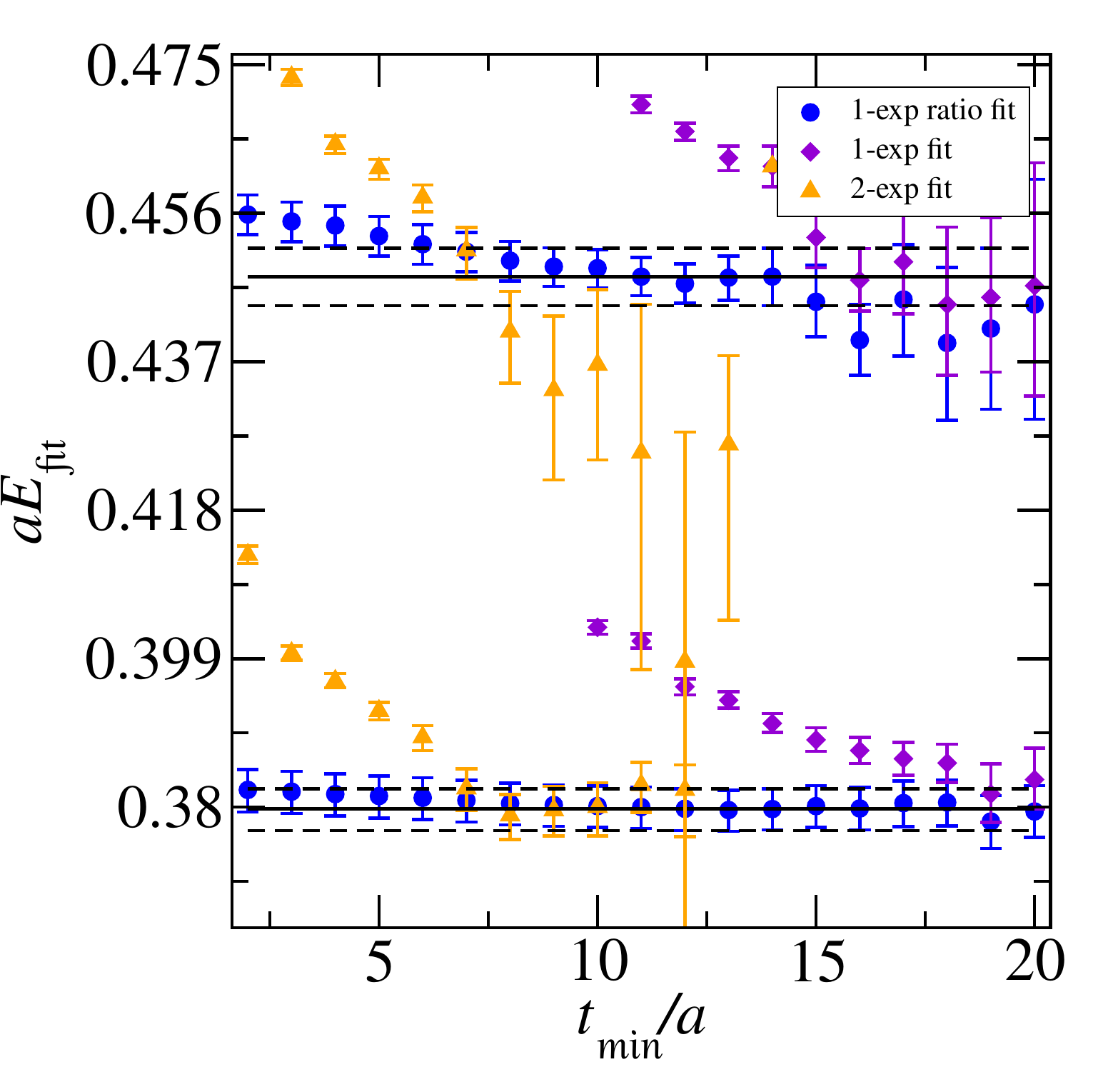}
    \caption{\label{fig:isodoublet_G1uP0_tmin}$G_{1u}(0)$}
\end{subfigure}
\hfill
\begin{subfigure}{\apwidth}
    \centering
    \includegraphics[width=1.0\linewidth]{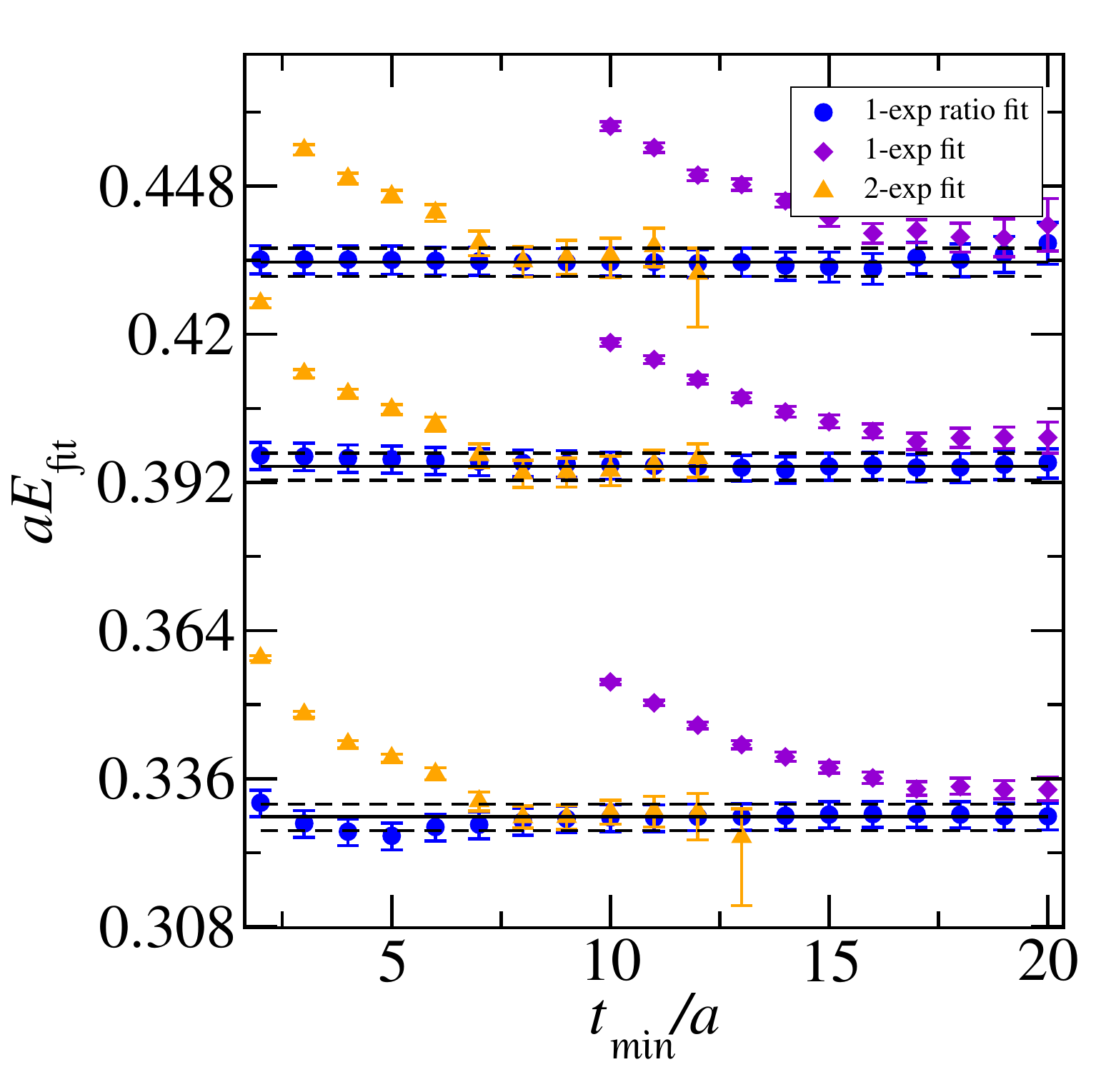}
    \caption{\label{fig:isodoublet_G1P1_tmin}$G_{1}(1)$}
\end{subfigure}
\hfill
\begin{subfigure}{\apwidth}
    \centering
    \includegraphics[width=1.0\linewidth]{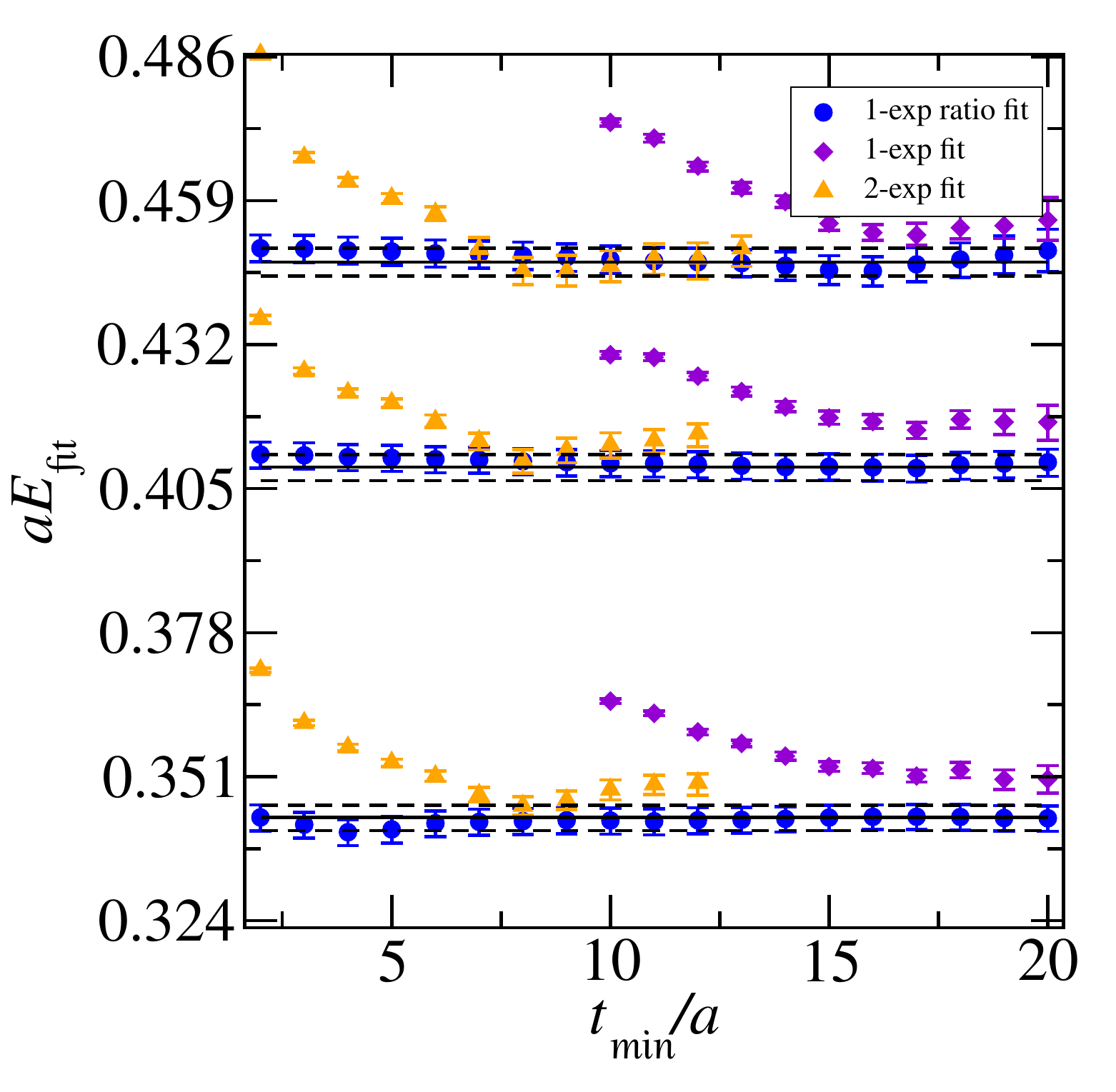}
    \caption{\label{fig:isodoublet_GP2_tmin}$G(2)$}
\end{subfigure}
\begin{subfigure}{\apwidth}
    \centering
    \includegraphics[width=1.0\linewidth]{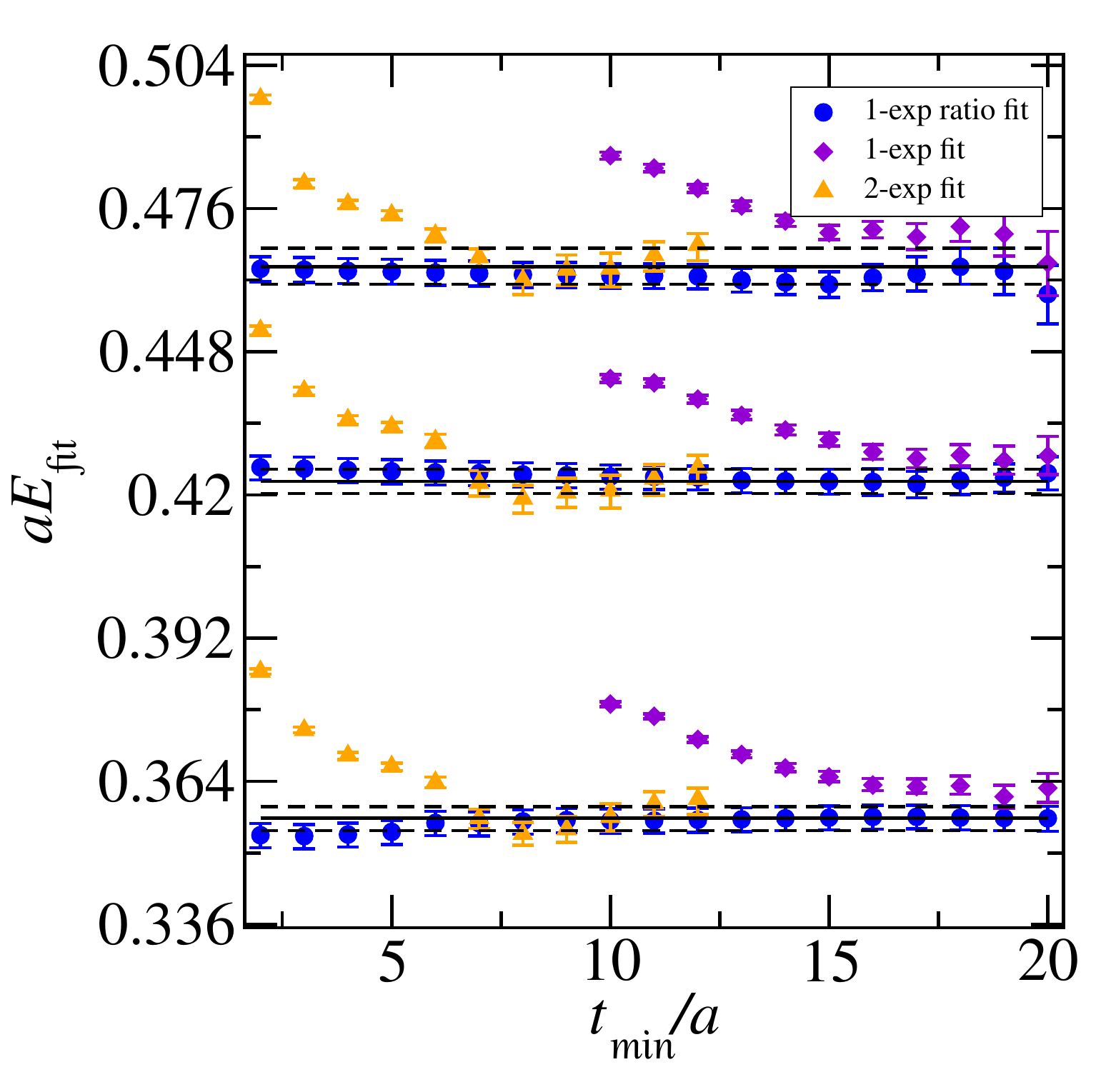}
    \caption{\label{fig:isodoublet_GP3_tmin}$G(3)$}
\end{subfigure}
\begin{subfigure}{\apwidth}
    \centering
    \includegraphics[width=1.0\linewidth]{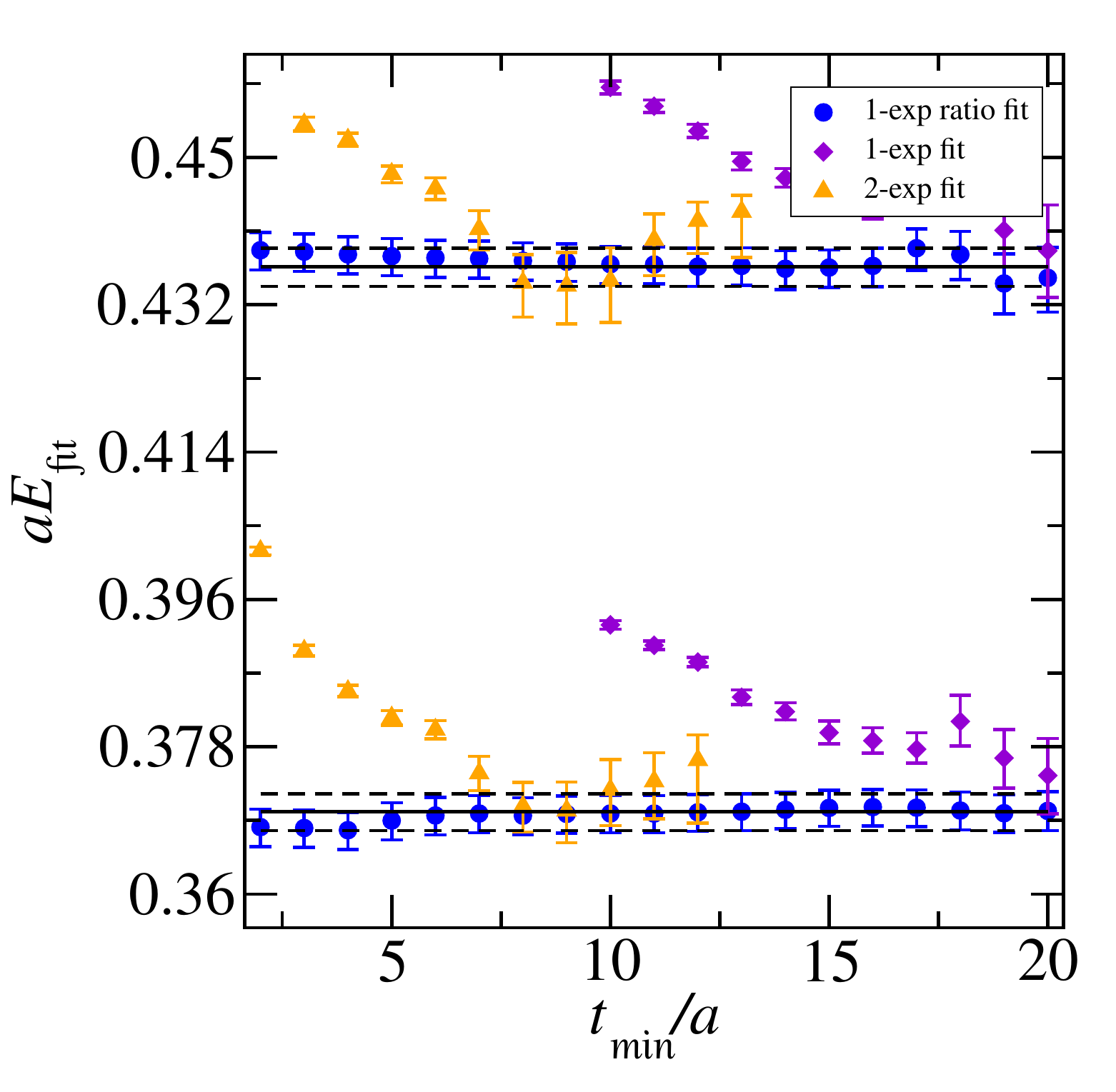}
    \caption{\label{fig:isodoublet_G1P4_tmin}$G_{1}(4)$}
\end{subfigure}
\caption{\label{fig:isodoublet_tmin} Stability of the $I=1/2$ spectrum illustrated by varying the fit range and 
fit form. The chosen fit for each level is indicated by the solid black line and the corresponding errors are 
indicated by dotted lines. Each subplot contains the spectrum for a single irrep labeled in the same manner 
as Fig.~\ref{fig:spectrums}. The chosen values are taken from ratio fits and compared to both single- and 
double-exponential fits over a range of $t_\textup{min}$ with $t_{\rm max}= 25a$.} 
\end{figure}

\begin{figure}[p]
    \centering
    \begin{subfigure}{\textwidth}
        \centering
        \begin{subfigure}{\apwidth}
            \centering
            \includegraphics[width=1.0\linewidth]{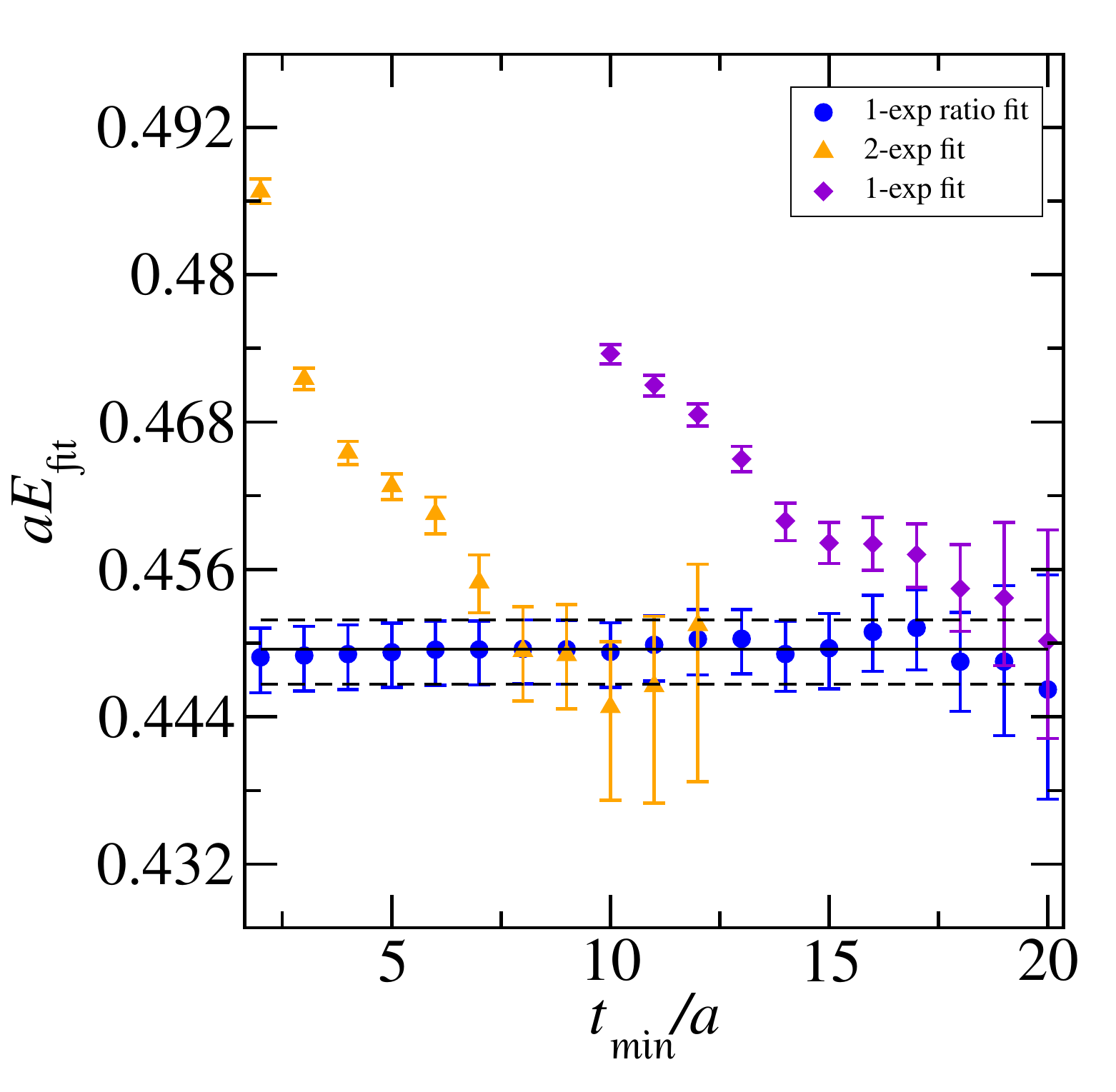}
            \caption{\label{fig:isoquartet_G1gP0_0_tmin}$G_{1g}(0)$ level 0}
        \end{subfigure}
        \begin{subfigure}{\apwidth}
            \centering
            \includegraphics[width=1.0\linewidth]{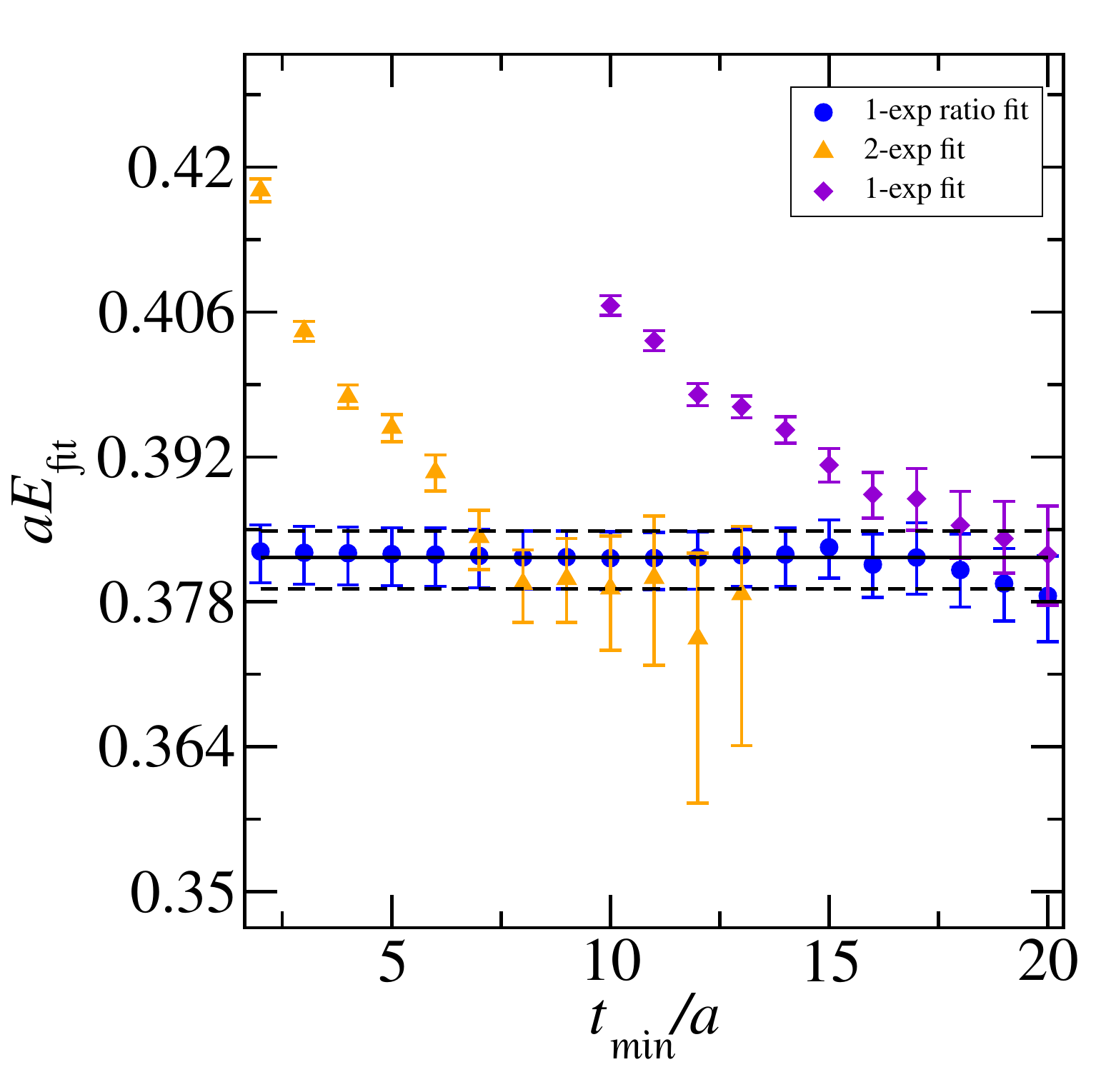}
            \caption{\label{fig:isoquartet_G1uP0_0_tmin}$G_{1u}(0)$ level 0}
        \end{subfigure}
        \begin{subfigure}{\apwidth}
            \centering
            \includegraphics[width=1.0\linewidth]{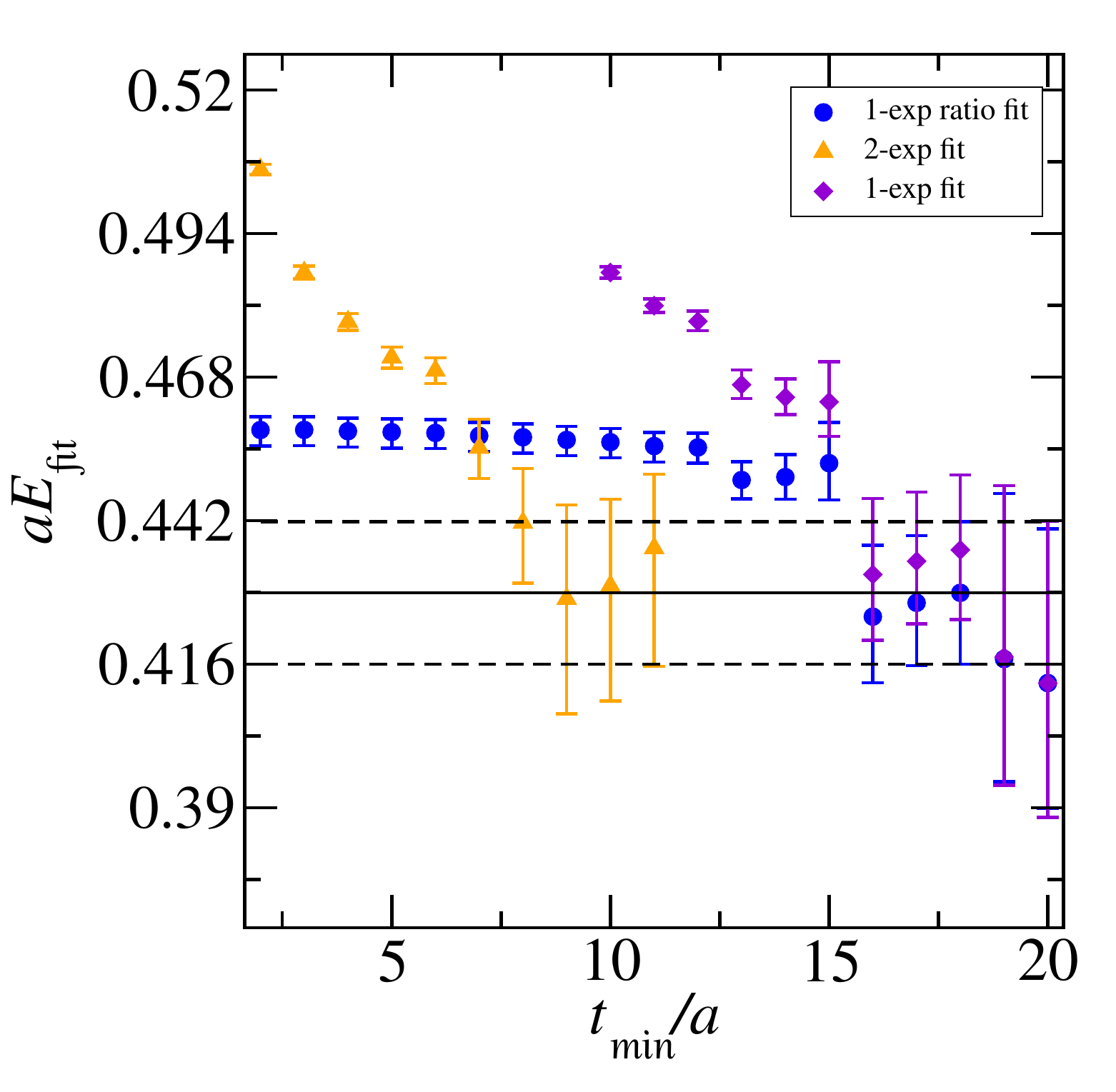}
            \caption{\label{fig:isoquartet_G1uP0_1_tmin}$G_{1g}(0)$ level 1}
        \end{subfigure}
        \begin{subfigure}{\apwidth}
            \centering
            \includegraphics[width=1.0\linewidth]{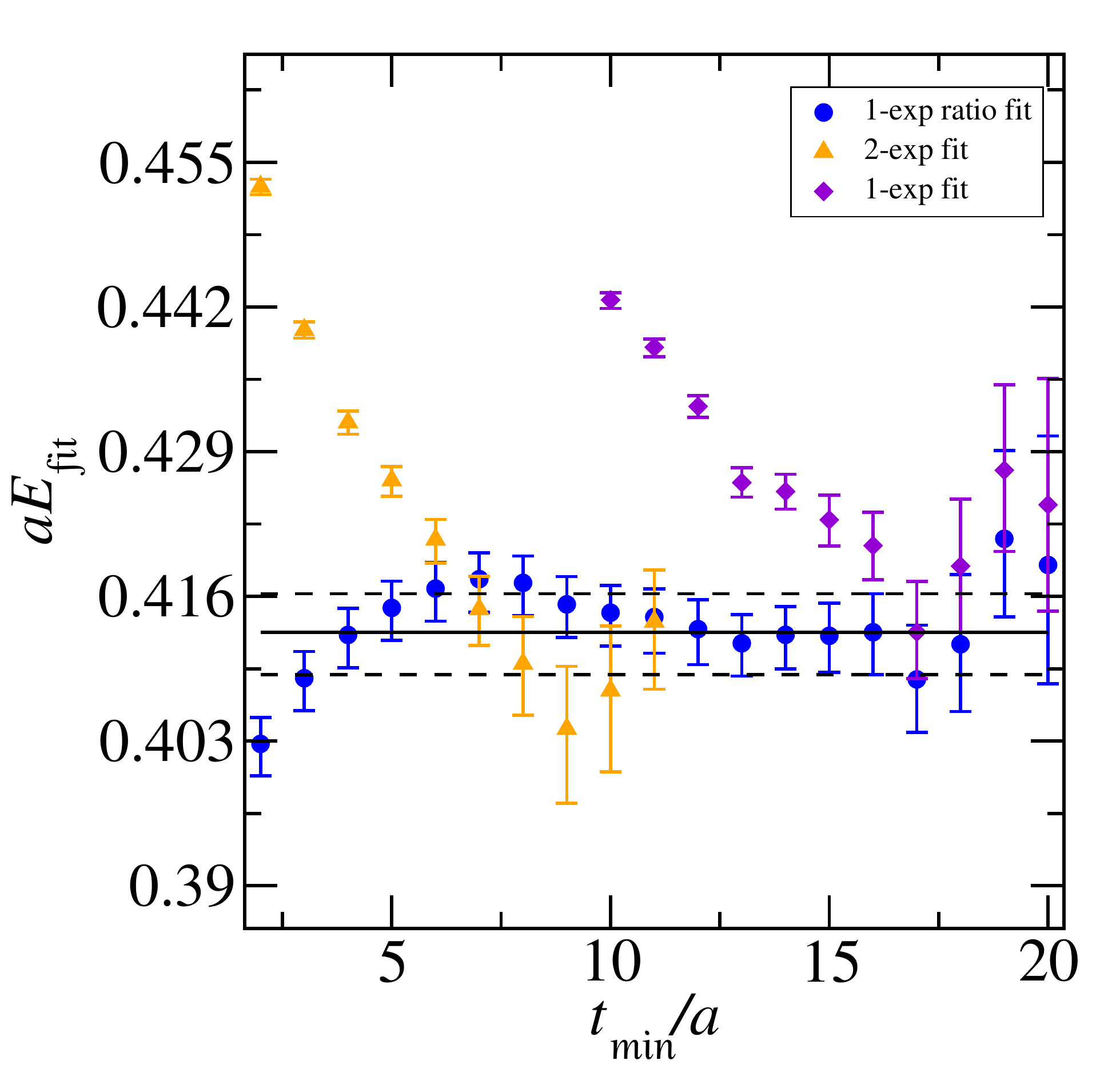}
            \caption{\label{fig:isoquartet_HgP0_0_tmin}$H_g(0)$ level 0}
        \end{subfigure}
        \begin{subfigure}{\apwidth}
            \centering
            \includegraphics[width=1.0\linewidth]{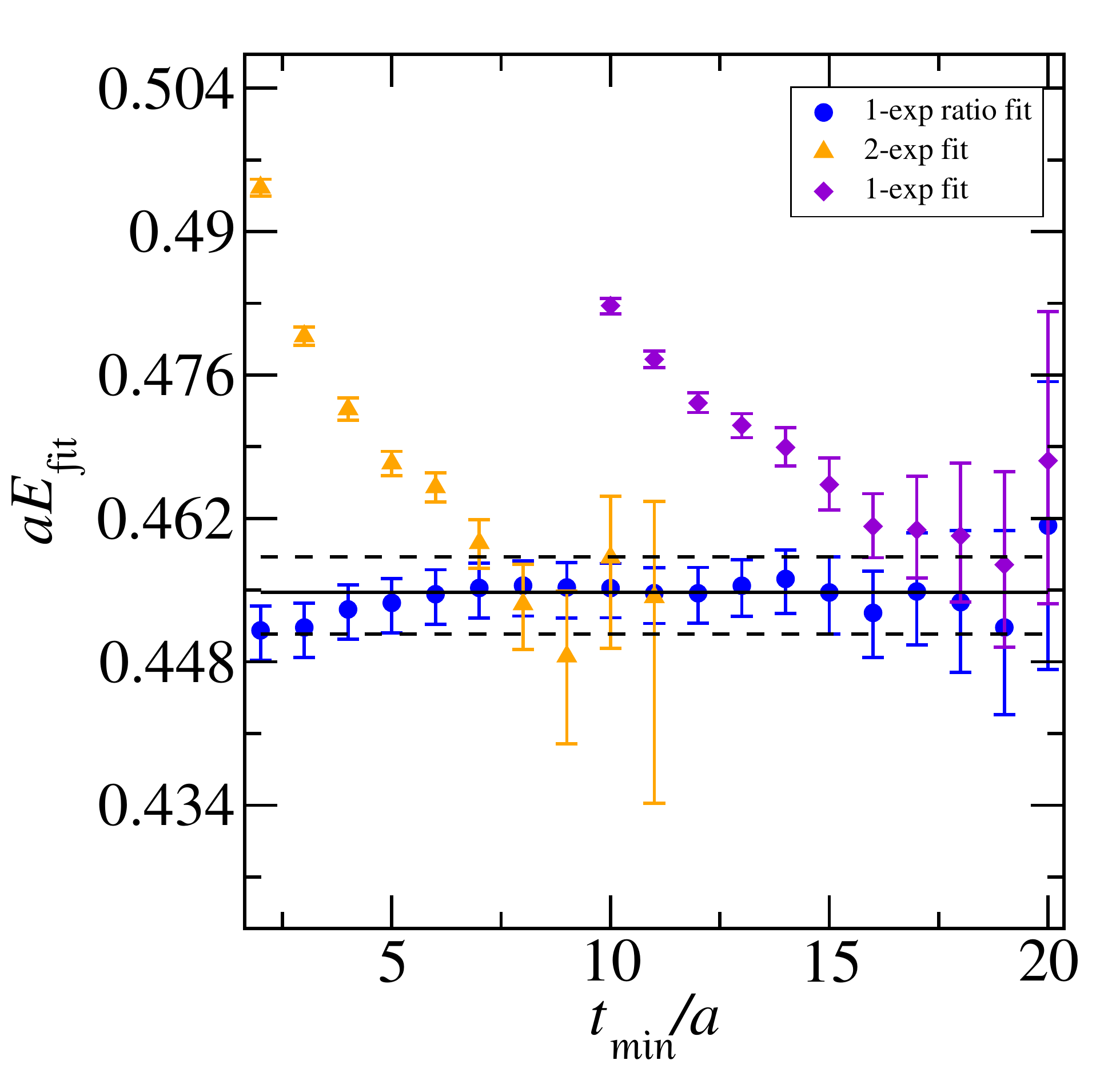}
            \caption{\label{fig:isoquartet_HgP0_1_tmin}$H_g(0)$ level 1}
        \end{subfigure}
        \begin{subfigure}{\apwidth}
            \centering
            \includegraphics[width=1.0\linewidth]{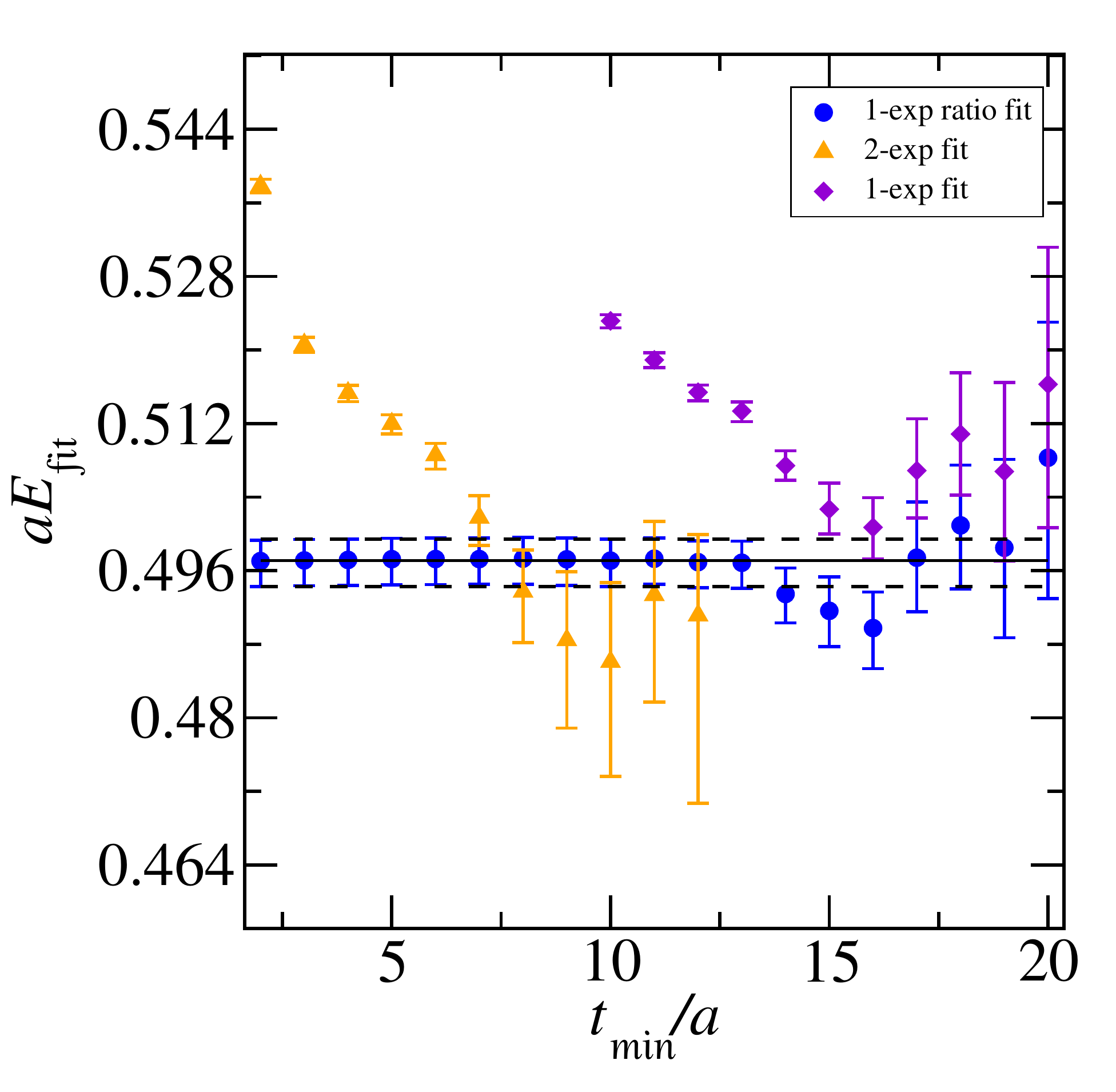}
            \caption{\label{fig:isoquartet_HgP0_2_tmin}$H_g(0)$ level 2}
        \end{subfigure}
        \begin{subfigure}{\apwidth}
            \centering
            \includegraphics[width=1.0\linewidth]{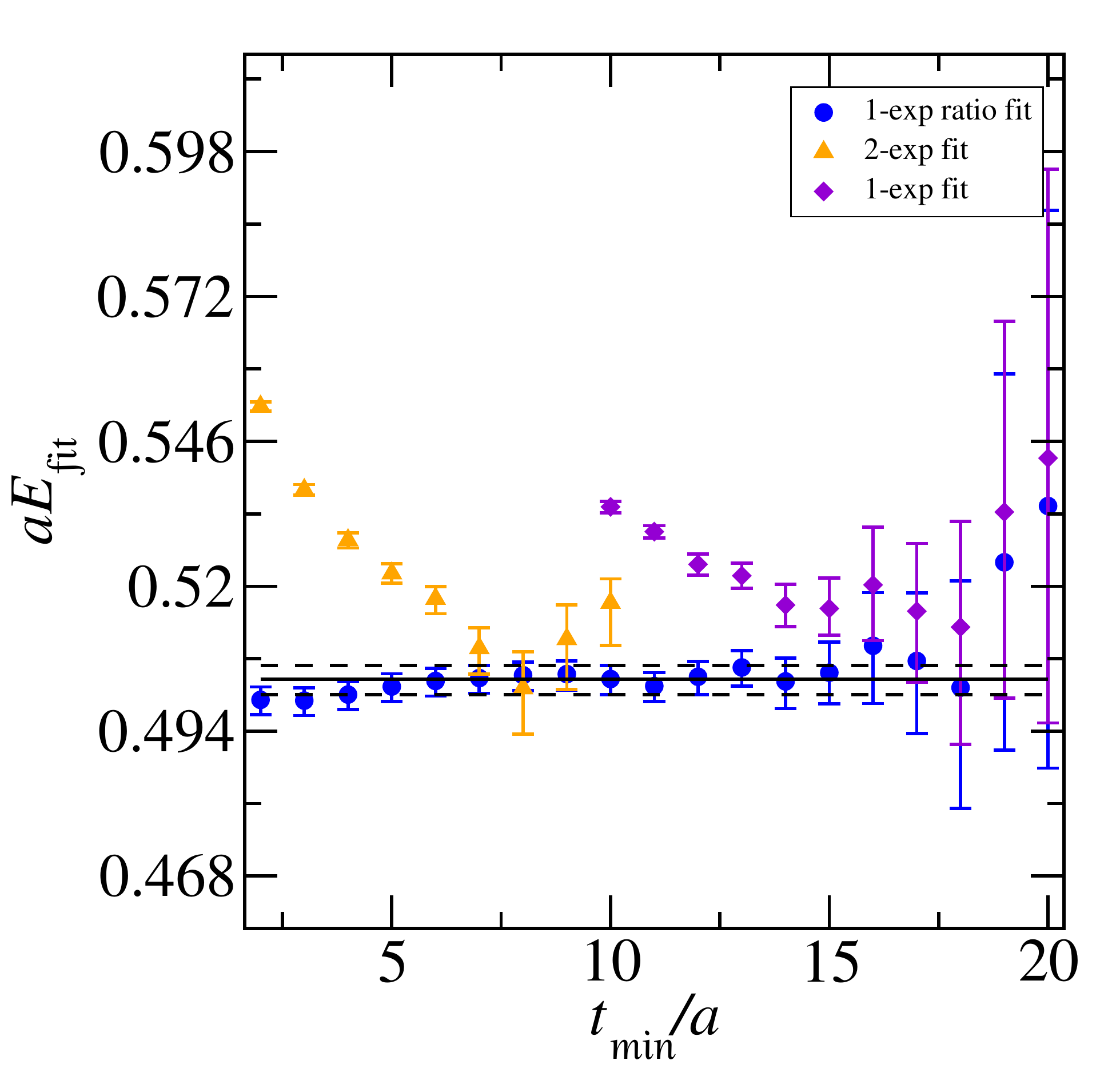}
            \caption{\label{fig:isoquartet_HgP0_3_tmin}$H_g(0)$ level 3}
        \end{subfigure}
        \begin{subfigure}{\apwidth}
            \centering
            \includegraphics[width=1.0\linewidth]{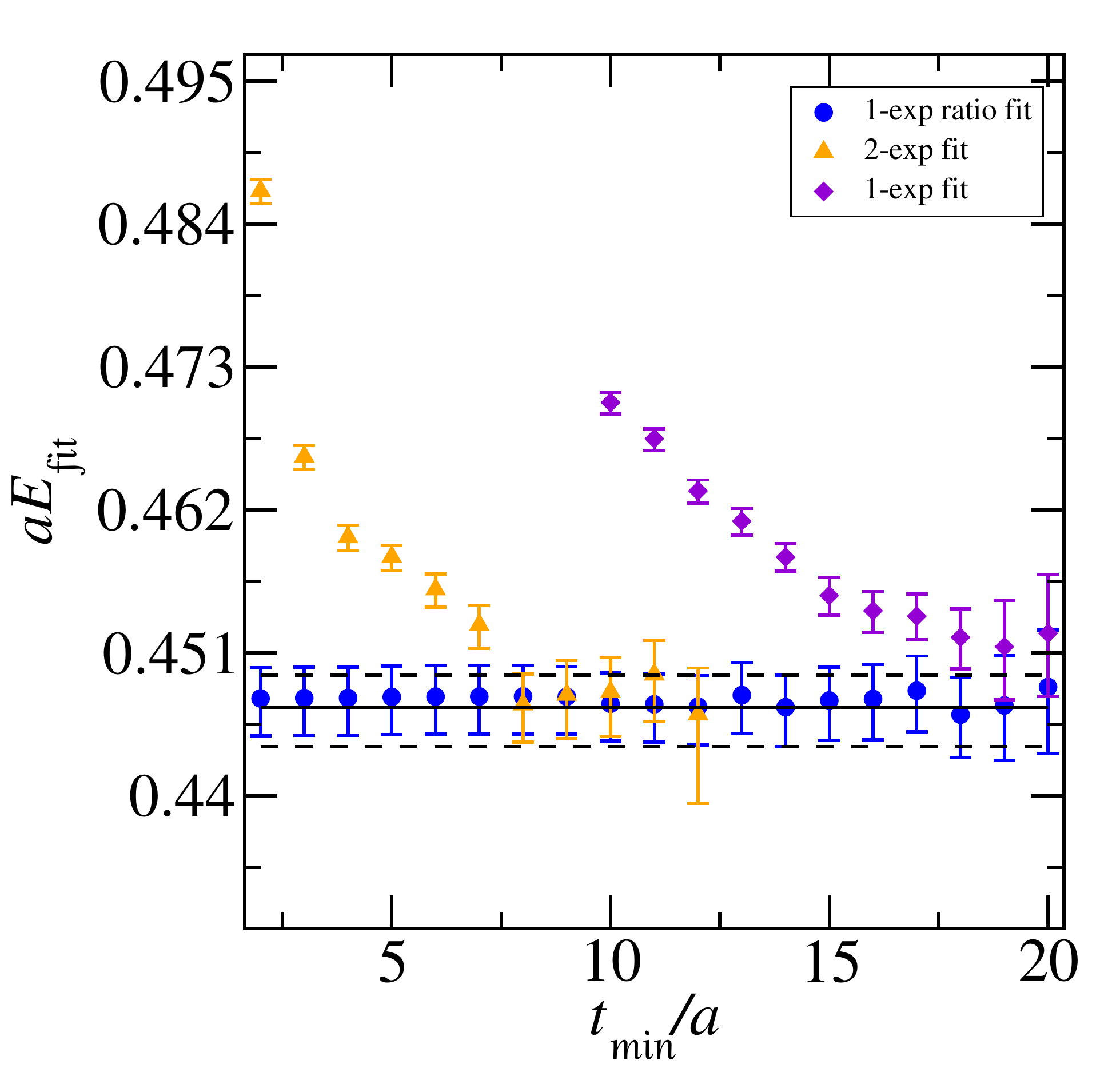}
            \caption{\label{fig:isoquartet_HuP0_0_tmin}$H_u(0)$ level 0}
        \end{subfigure}
        \setcounter{subfigure}{0}
        \renewcommand{\thesubfigure}{\Roman{subfigure}}
    \end{subfigure}
\caption{\label{fig:isoquartet_P0_tmin} Stability of fits to determine the $I=3/2$ spectrum for total 
momentum having $\boldsymbol{d}^2=0$. As in Fig.~\ref{fig:isodoublet_tmin}, a variety of fit ranges and 
fit forms is compared for each level. Each plot contains all fits for a single level in a particular irrep. 
Indexing for the levels begins at zero for the lowest and increases with increasing energy.}
\end{figure}

\begin{figure}[p]
    \begin{subfigure}{\textwidth}
        \setcounter{subfigure}{0}
        \renewcommand{\thesubfigure}{\alph{subfigure}}
        \centering
        \begin{subfigure}{\apwidth}
            \centering
            \includegraphics[width=1.0\linewidth]{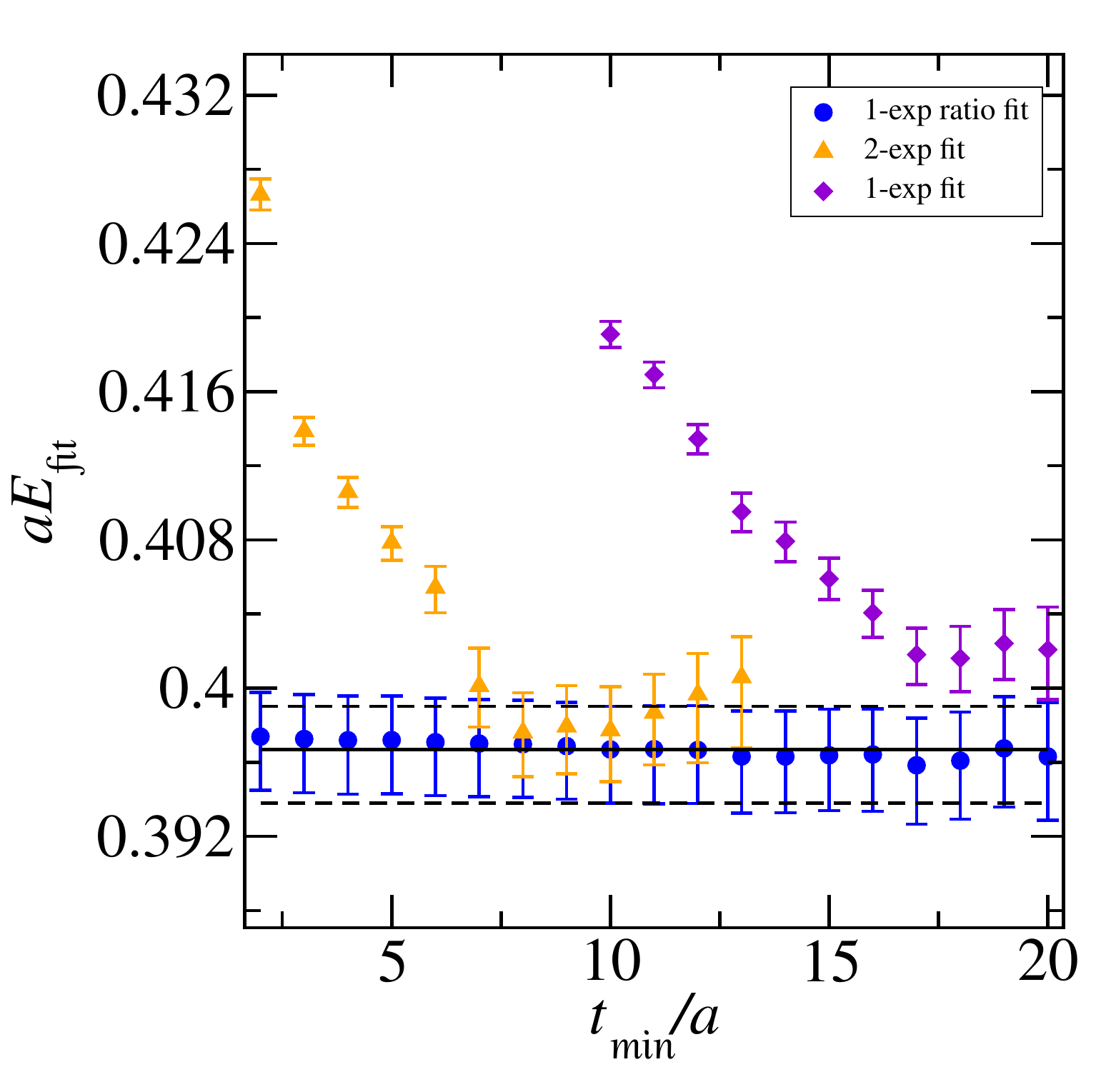}
            \caption{$G_1(1)$ level 0}  
        \end{subfigure}
        \begin{subfigure}{\apwidth}
            \centering
            \includegraphics[width=1.0\linewidth]{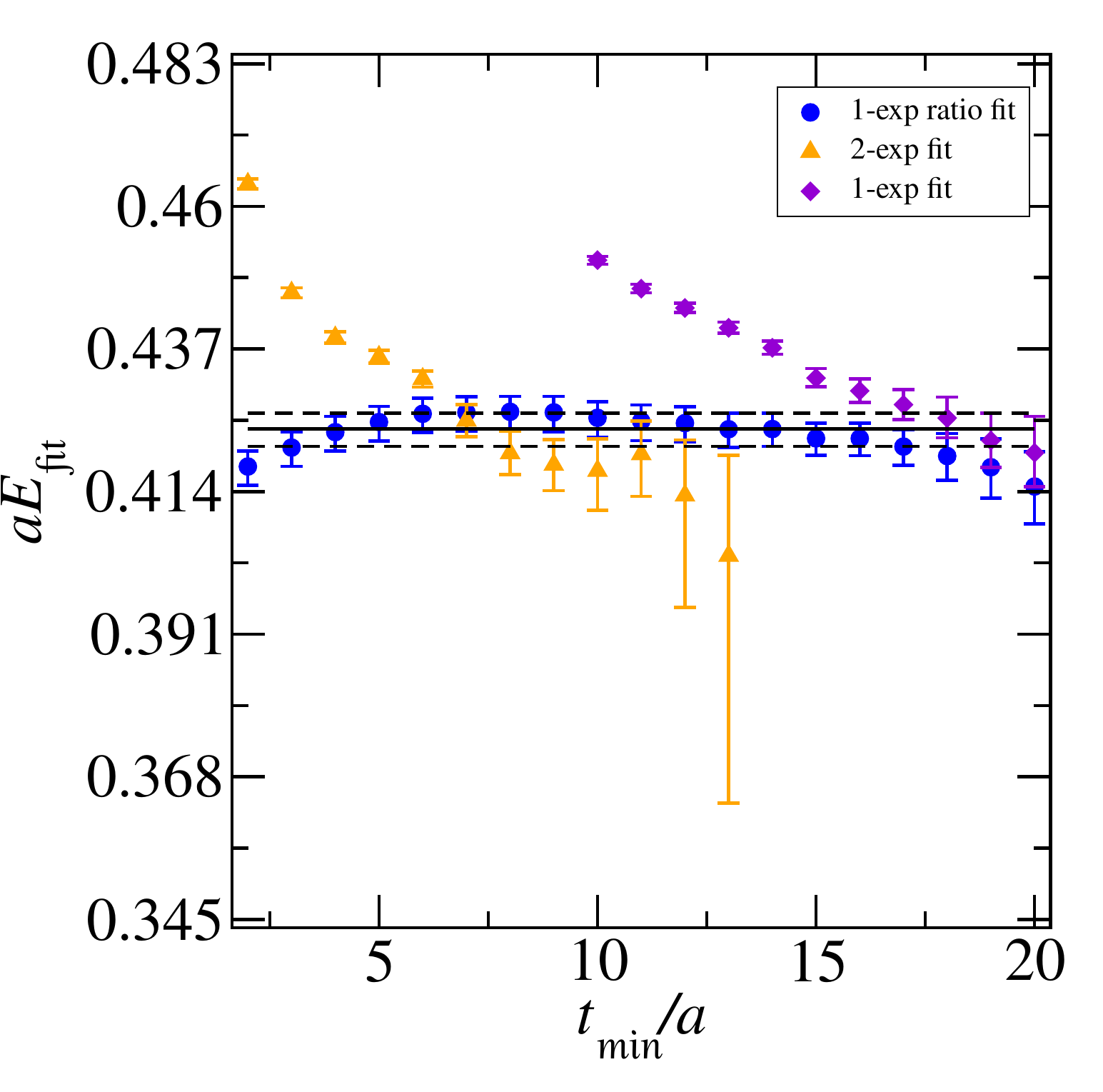}
            \caption{$G_1(1)$ level 1}  
        \end{subfigure}
        \begin{subfigure}{\apwidth}
            \centering
            \includegraphics[width=1.0\linewidth]{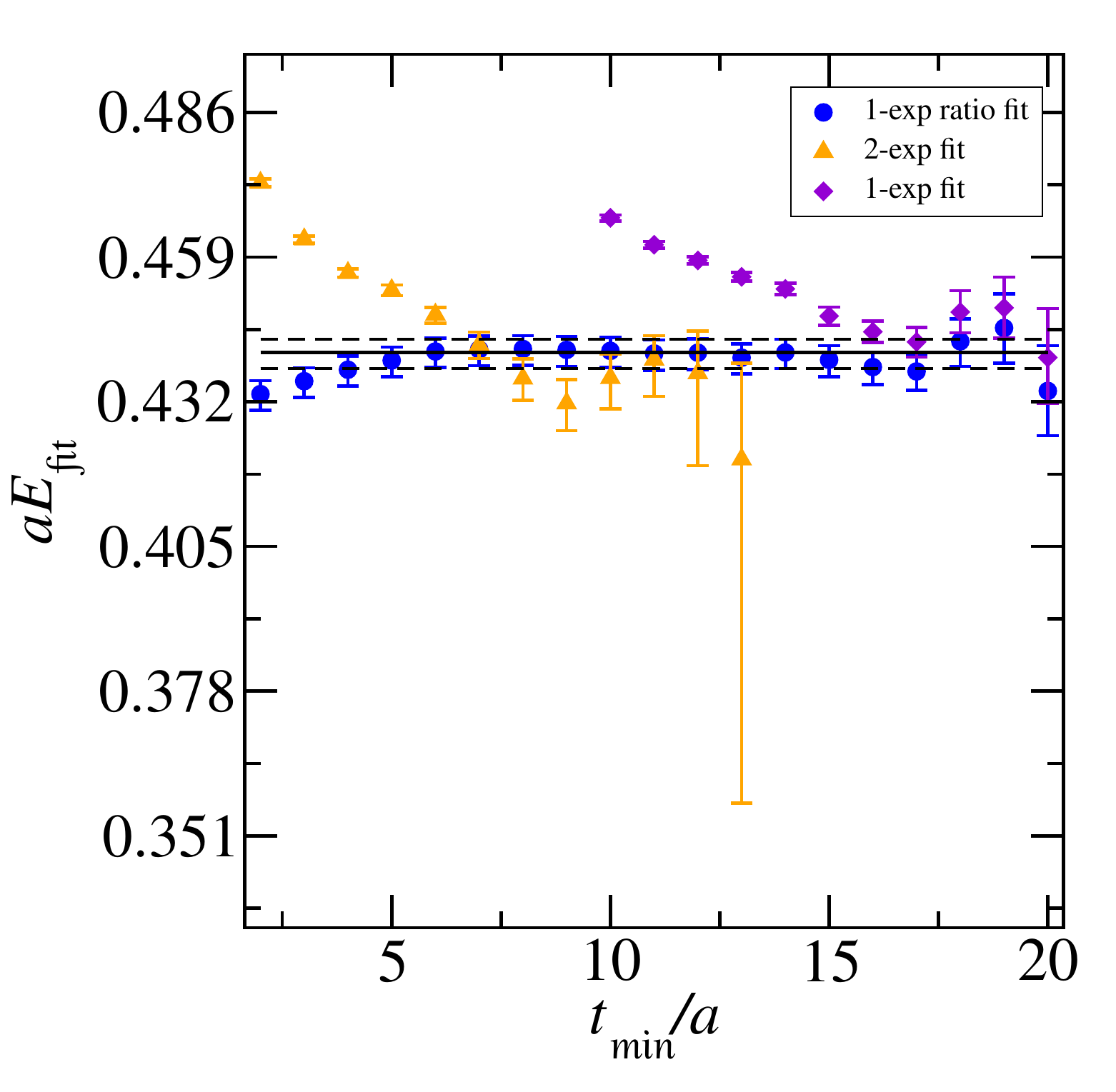}
            \caption{$G_1(1)$ level 2}  
        \end{subfigure}
        \begin{subfigure}{\apwidth}
            \centering
            \includegraphics[width=1.0\linewidth]{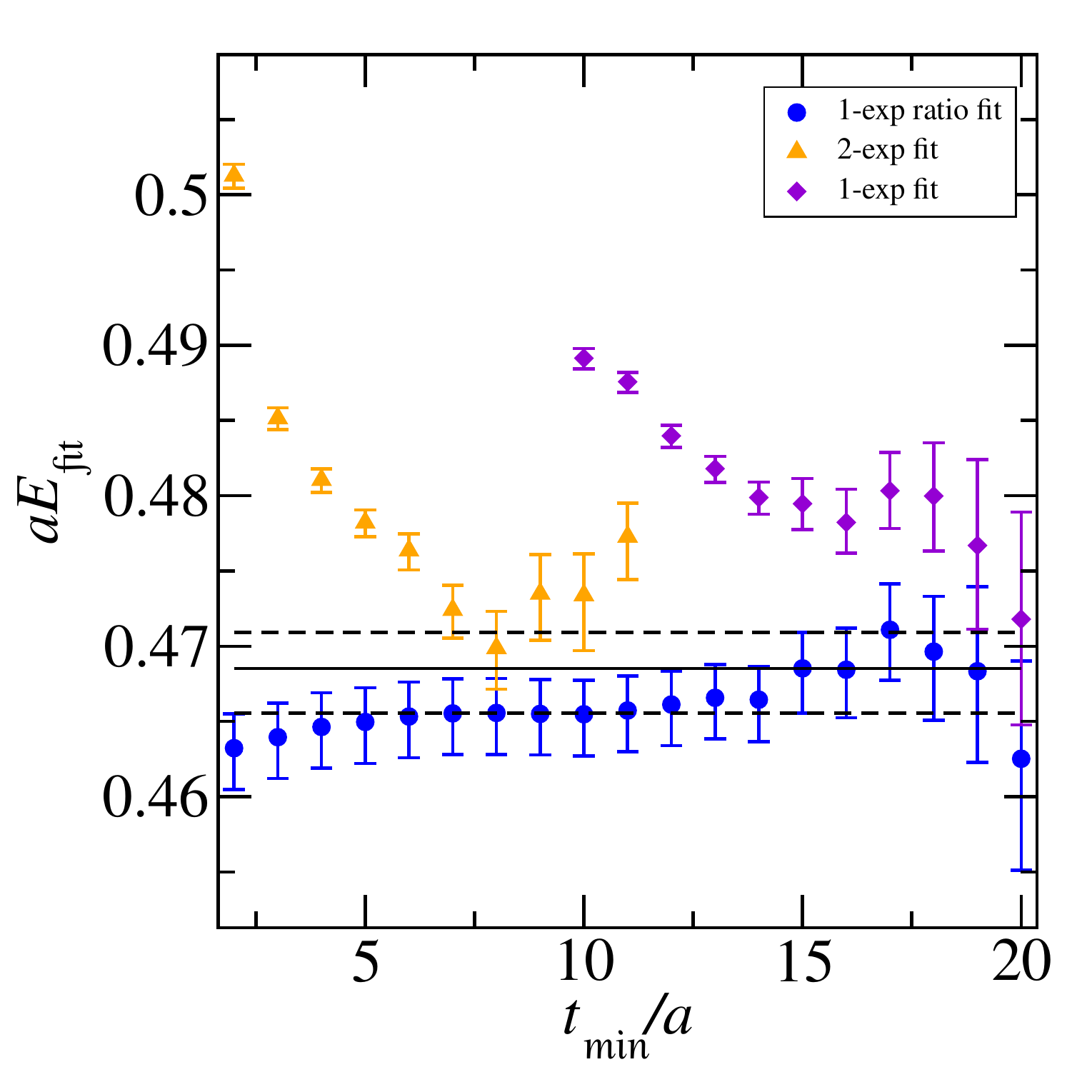}
            \caption{$G_1(1)$ level 3}  
        \end{subfigure}
        \begin{subfigure}{\apwidth}
            \centering
            \includegraphics[width=1.0\linewidth]{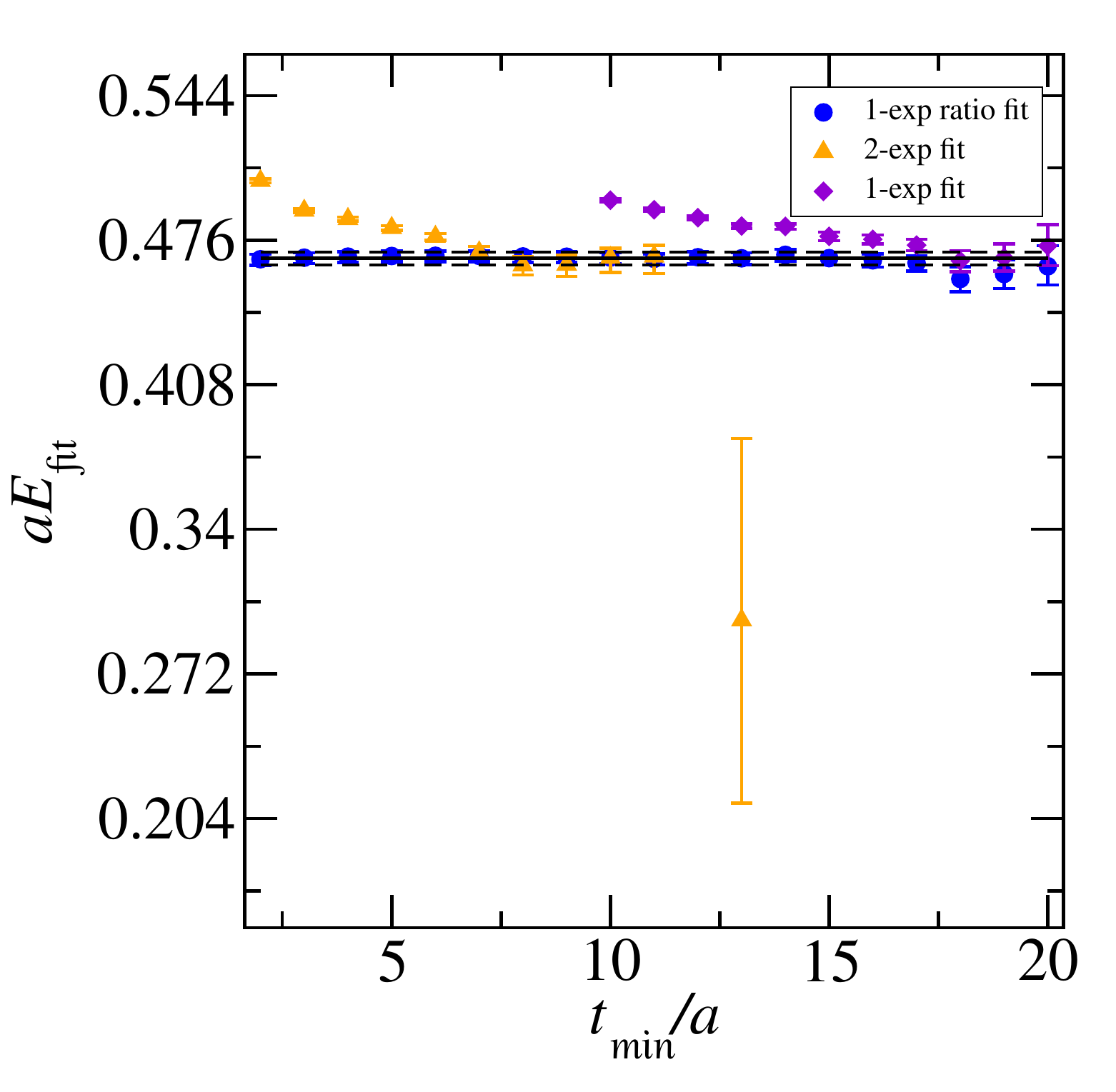}
            \caption{$G_1(1)$ level 4}  
        \end{subfigure}
        \begin{subfigure}{\apwidth}
            \centering
            \includegraphics[width=1.0\linewidth]{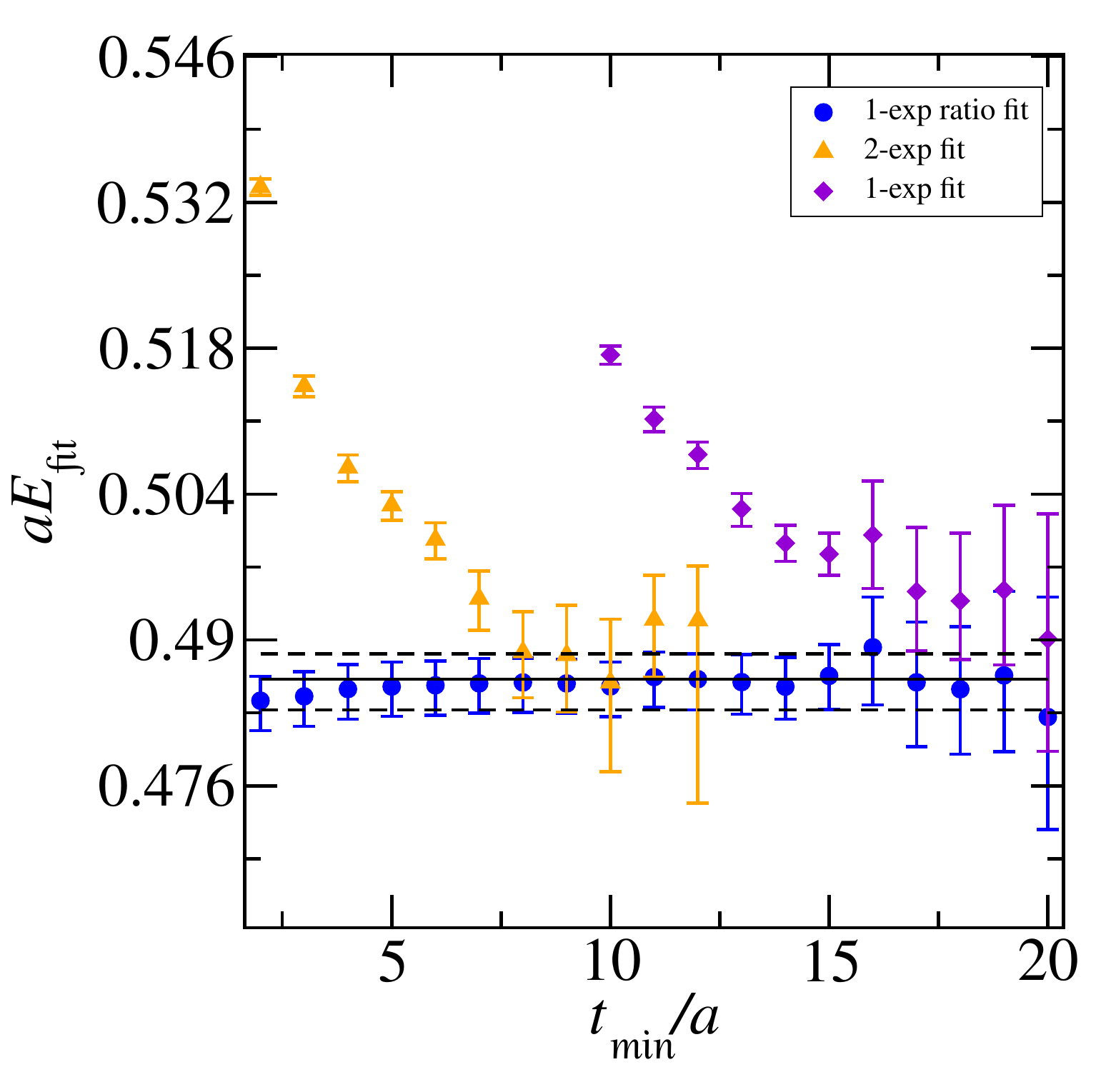}
            \caption{$G_1(1)$ level 5}  
        \end{subfigure}
        \begin{subfigure}{\apwidth}
            \centering
            \includegraphics[width=1.0\linewidth]{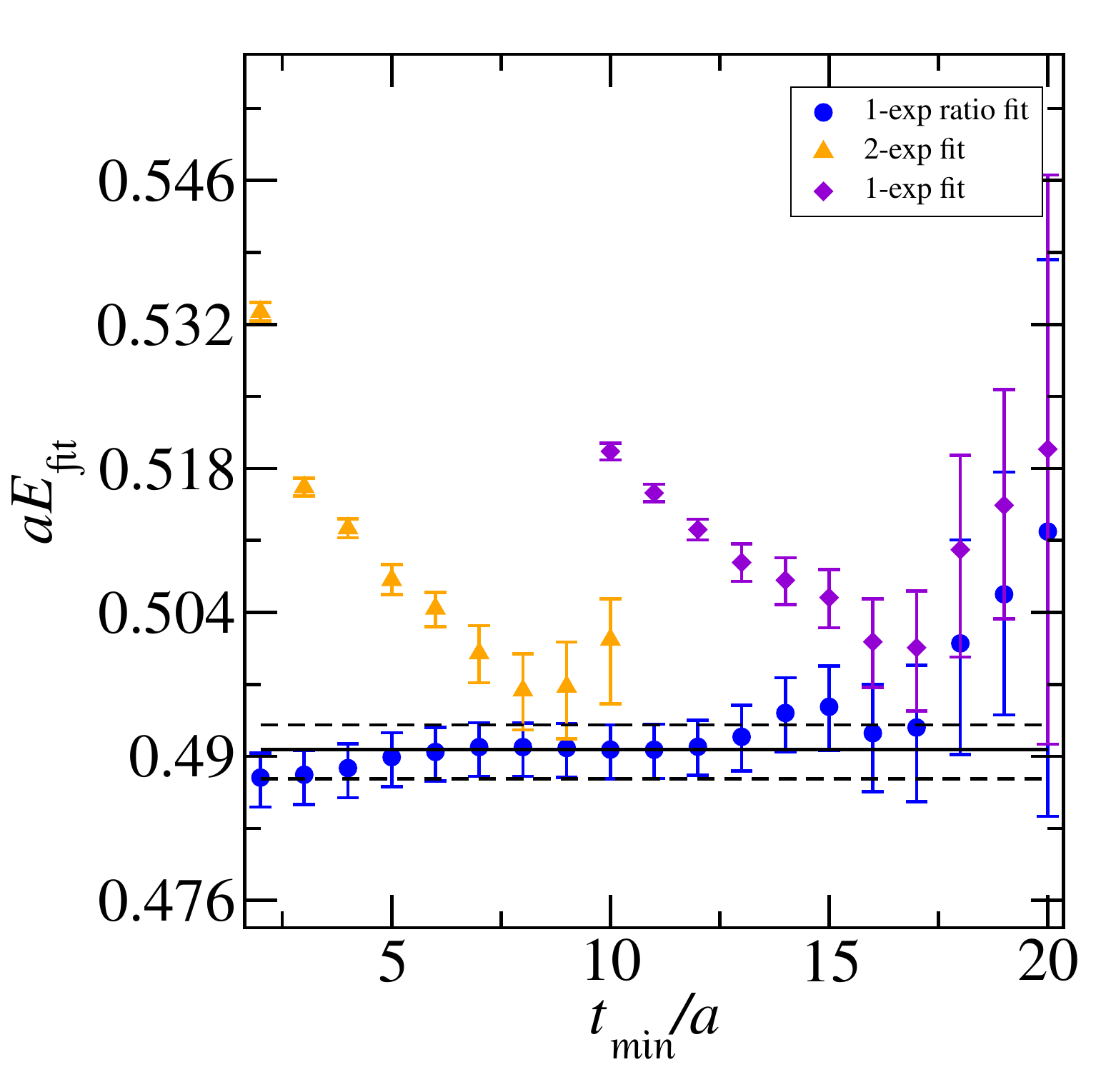}
            \caption{$G_1(1)$ level 6}  
        \end{subfigure}
        \begin{subfigure}{\apwidth}
            \centering
            \includegraphics[width=1.0\linewidth]{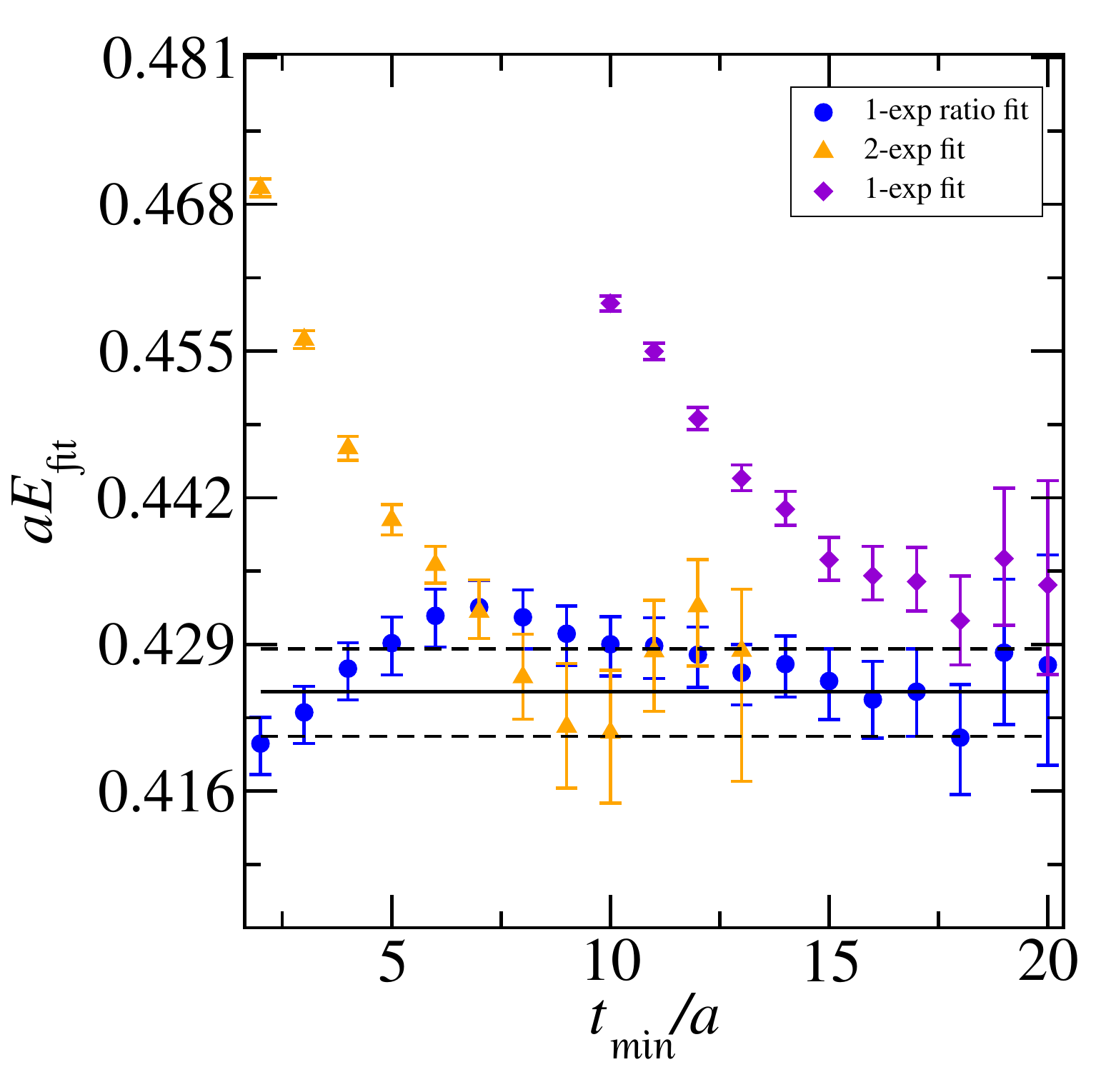}
            \caption{$G_2(1)$ level 0}
        \end{subfigure}
        \begin{subfigure}{\apwidth}
            \centering
            \includegraphics[width=1.0\linewidth]{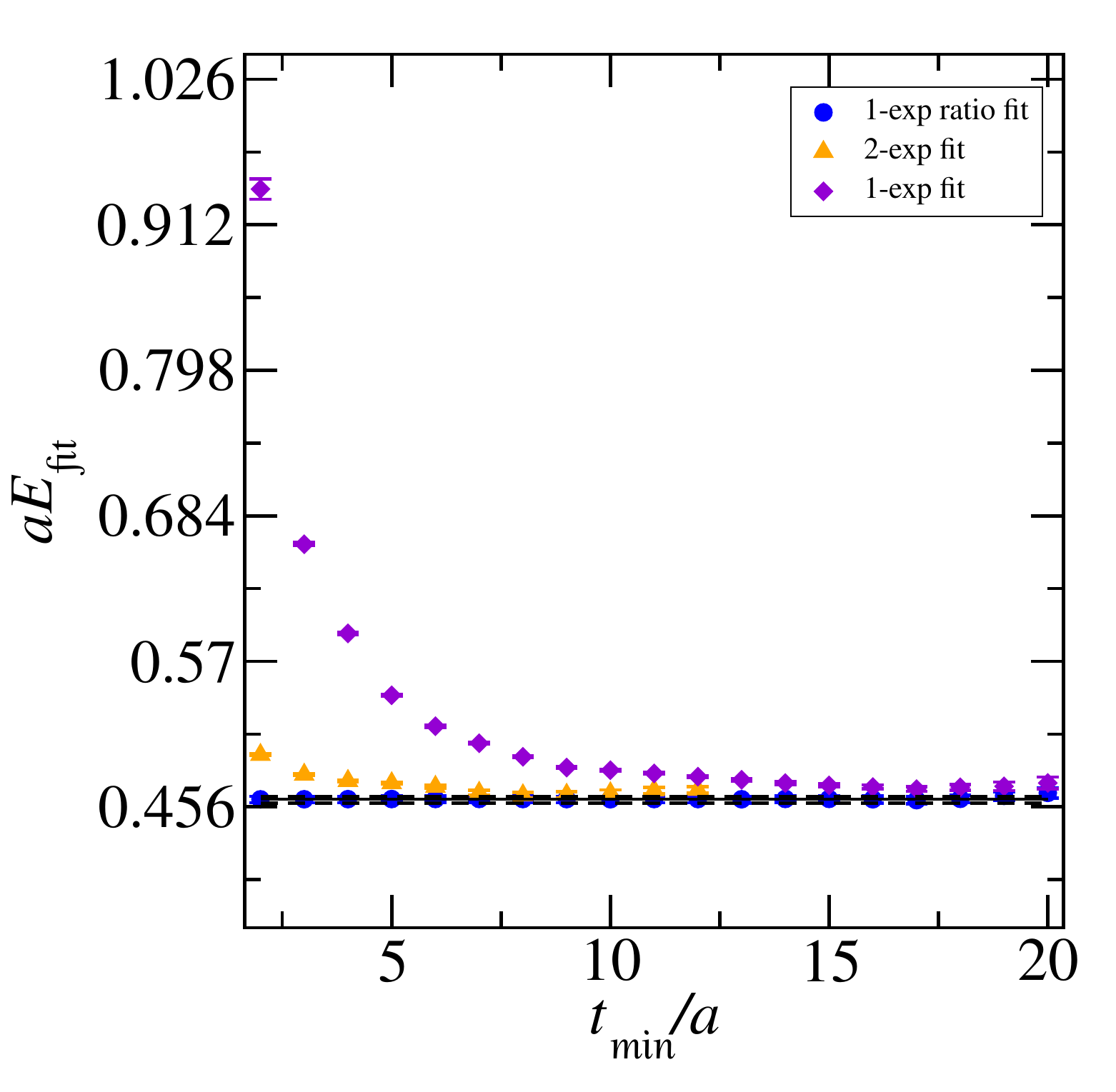}
            \caption{$G_2(1)$ level 1}
        \end{subfigure}
        \begin{subfigure}{\apwidth}
            \centering
            \includegraphics[width=1.0\linewidth]{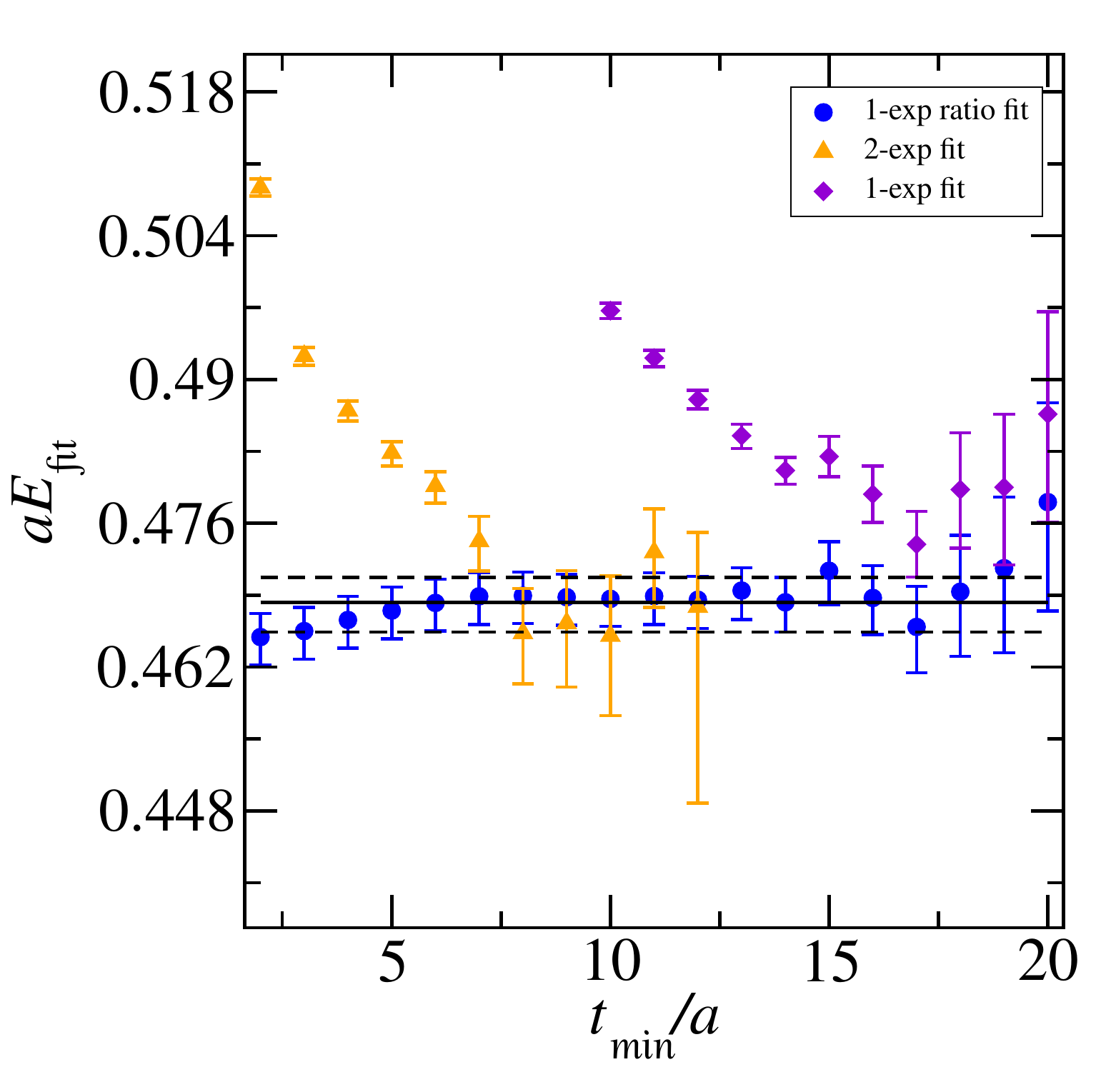}
            \caption{$G_2(1)$ level 2}
        \end{subfigure}
        \begin{subfigure}{\apwidth}
            \centering
            \includegraphics[width=1.0\linewidth]{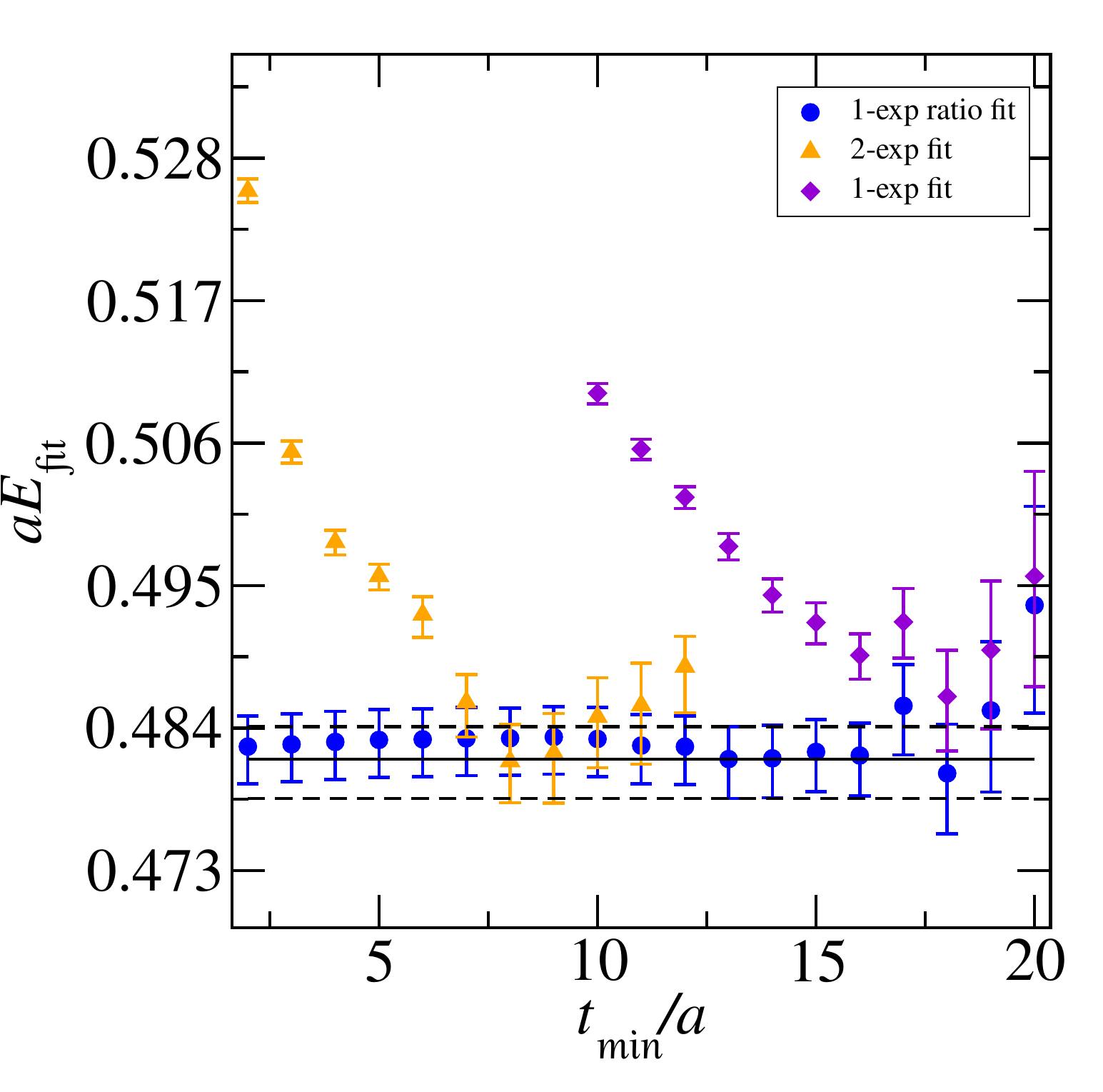}
            \caption{$G_2(1)$ level 3}
        \end{subfigure}
        \begin{subfigure}{\apwidth}
            \centering
            \includegraphics[width=1.0\linewidth]{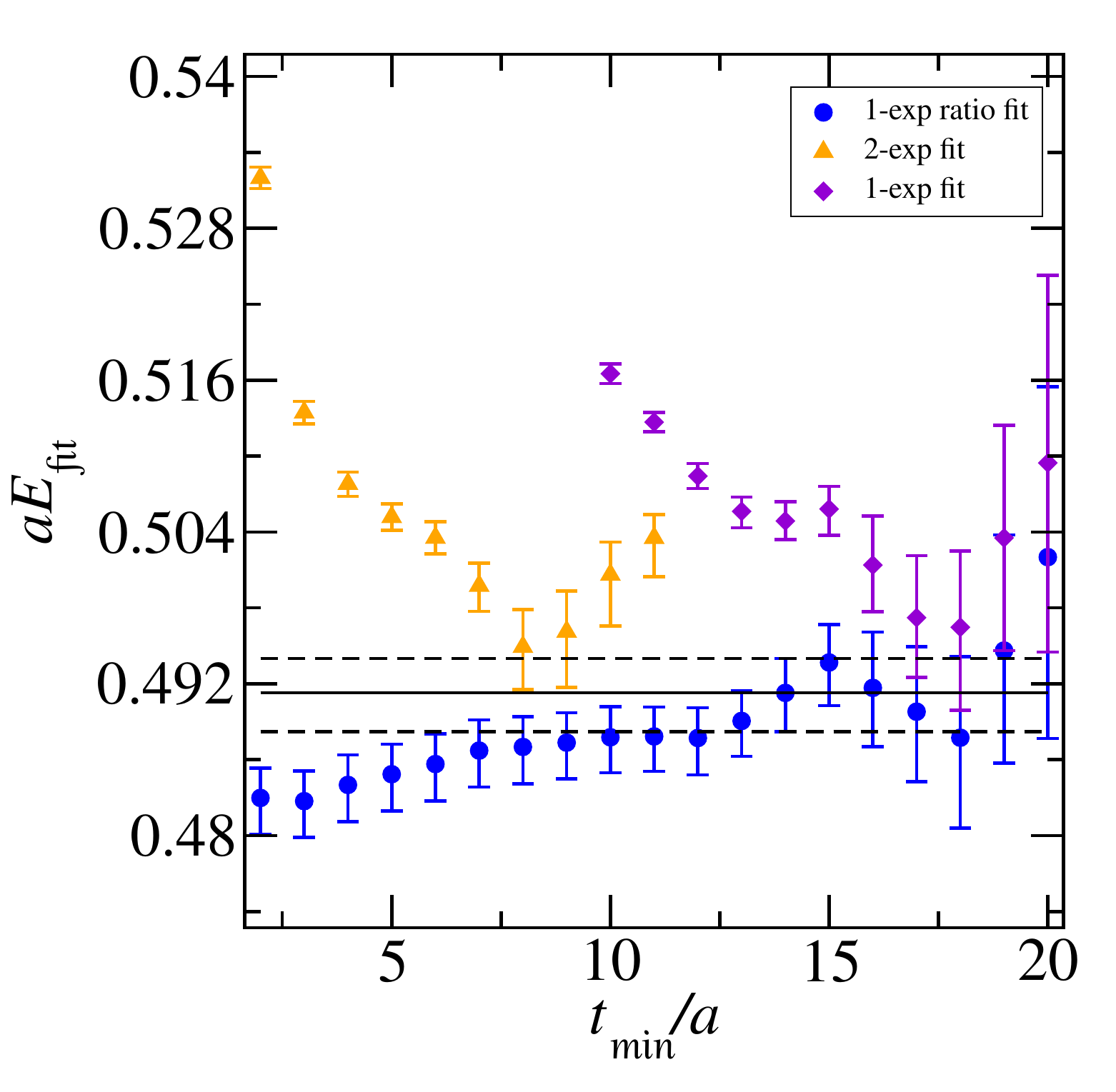}
            \caption{$G_2(1)$ level 4}
        \end{subfigure}
        \setcounter{subfigure}{1}
        \renewcommand{\thesubfigure}{\Roman{subfigure}}
    \end{subfigure}
 \caption{\label{fig:isoquartet_P1_tmin} Same as Fig.~\ref{fig:isoquartet_P0_tmin} for $I=3/2$ except 
that $\boldsymbol{d}^2=1$.} 
\end{figure}

\begin{figure}[p]
    \begin{subfigure}{\textwidth}
        \setcounter{subfigure}{0}
        \renewcommand{\thesubfigure}{\alph{subfigure}}
        \centering
        \begin{subfigure}{\apwidth}
            \centering
            \includegraphics[width=1.0\linewidth]{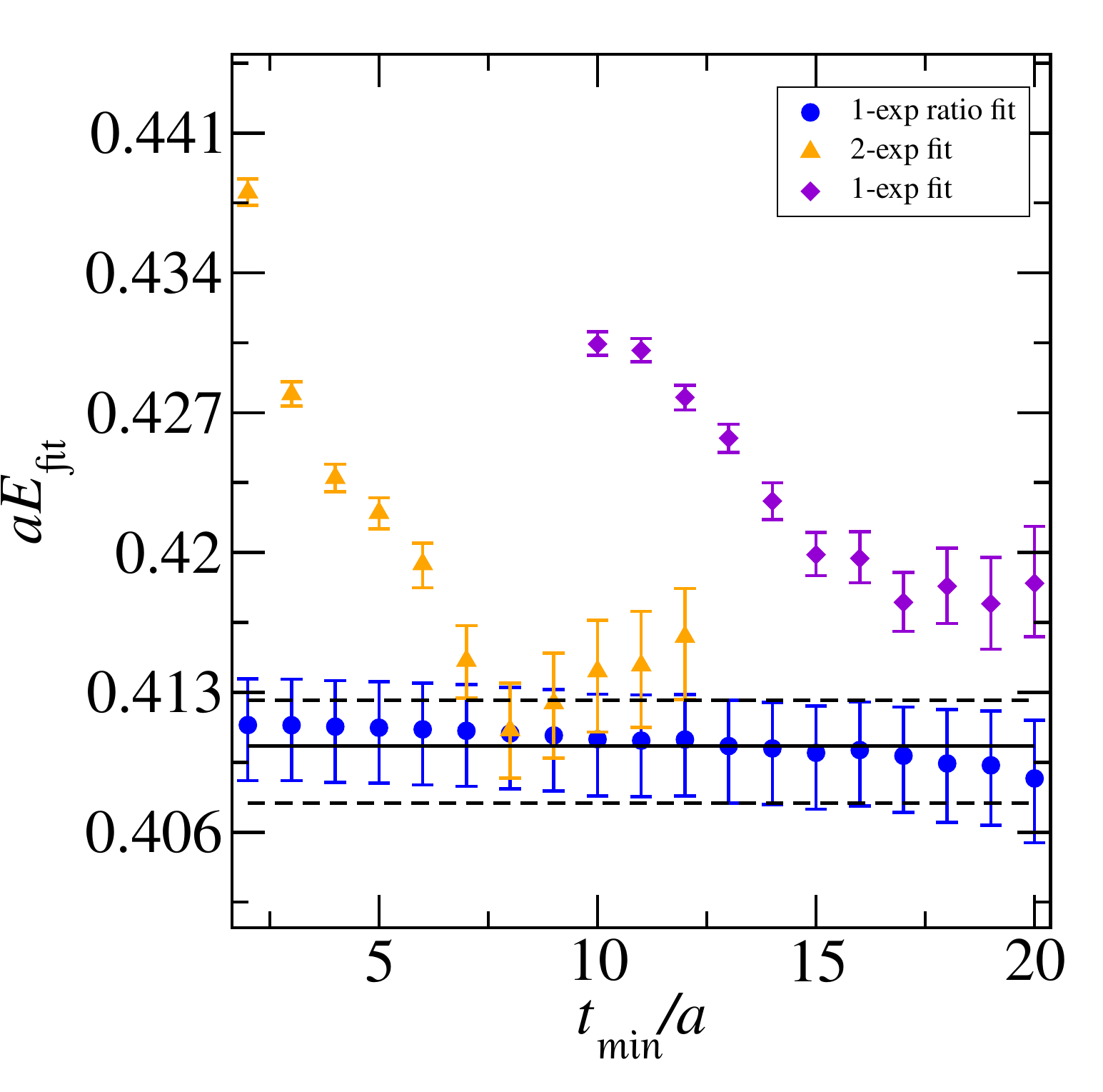}
            \caption{$G(2)$ level 0}
        \end{subfigure}
        \begin{subfigure}{\apwidth}
            \centering
            \includegraphics[width=1.0\linewidth]{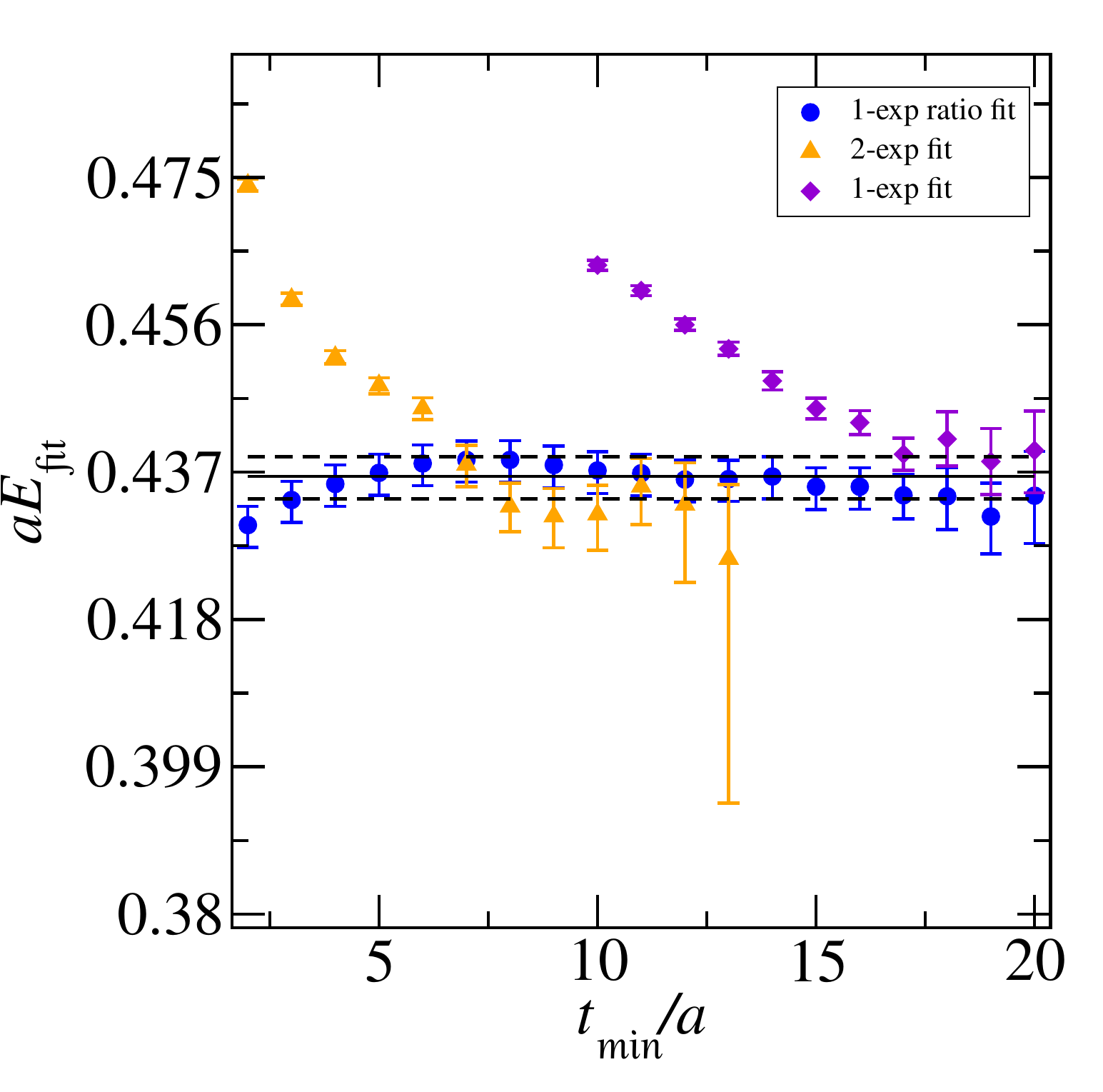}
            \caption{$G(2)$ level 1}
        \end{subfigure}
        \begin{subfigure}{\apwidth}
            \centering
            \includegraphics[width=1.0\linewidth]{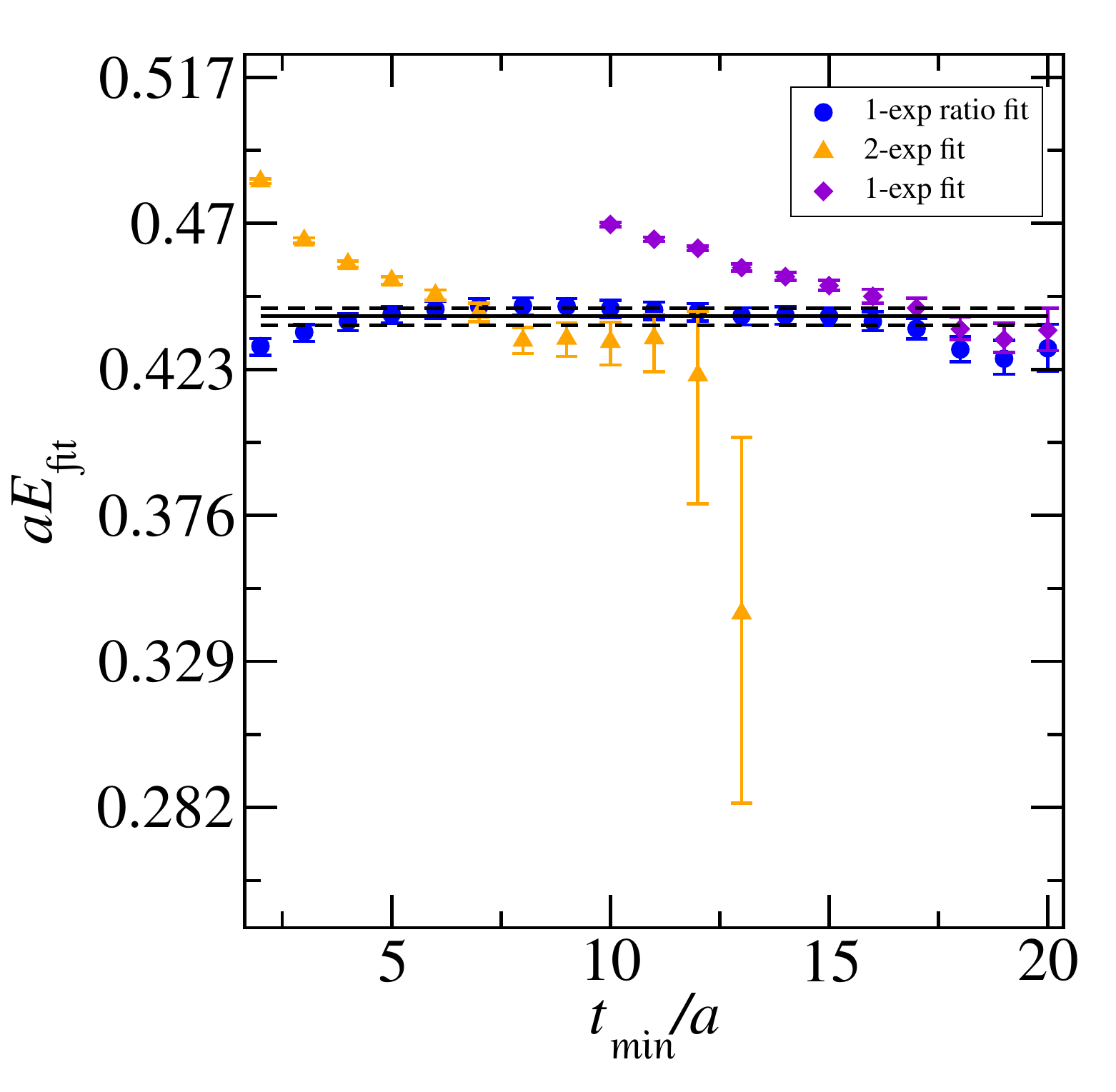}
            \caption{$G(2)$ level 2}
        \end{subfigure}
        \begin{subfigure}{\apwidth}
            \centering
            \includegraphics[width=1.0\linewidth]{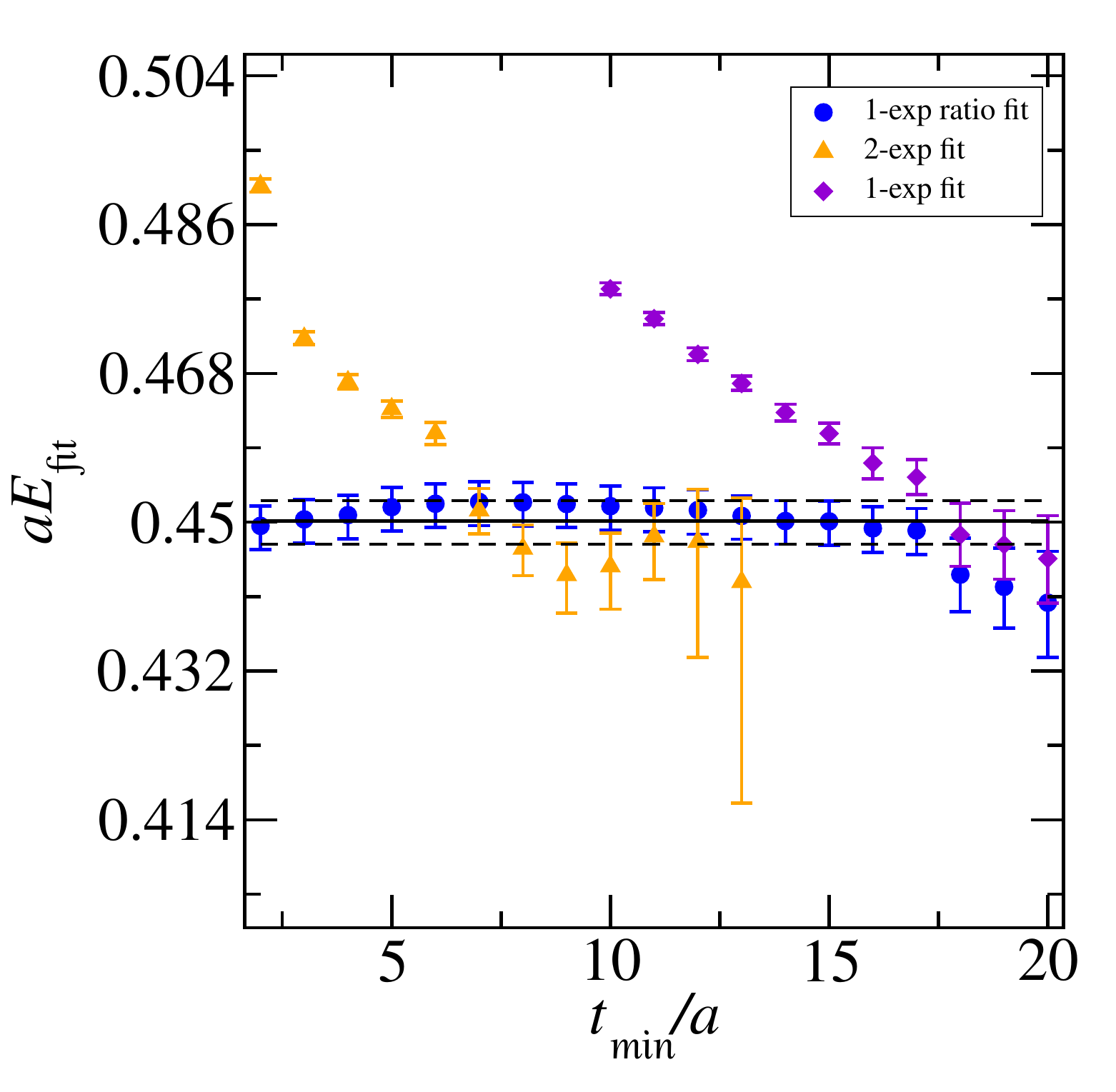}
            \caption{$G(2)$ level 3}
        \end{subfigure}
        \begin{subfigure}{\apwidth}
            \centering
            \includegraphics[width=1.0\linewidth]{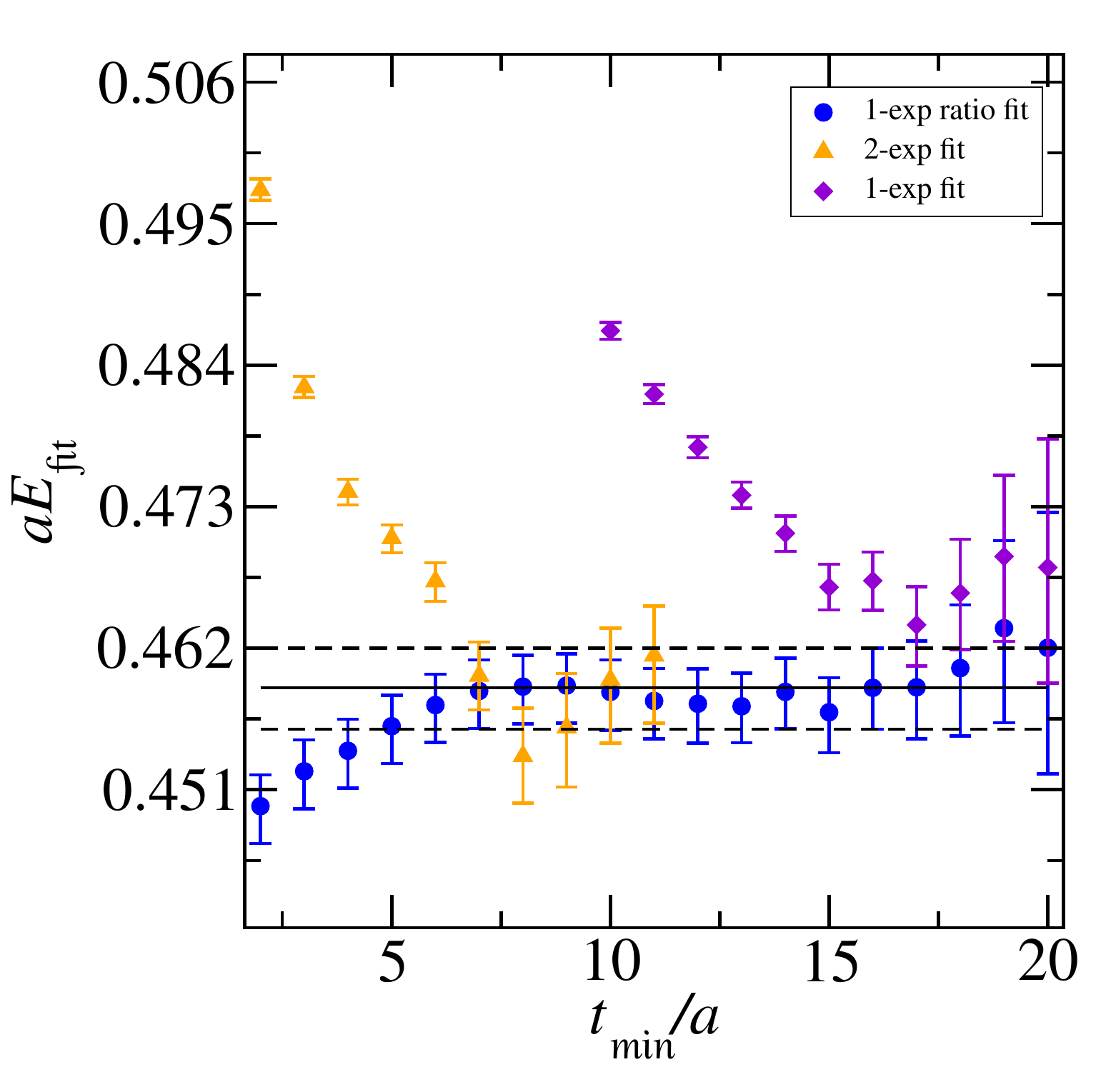}
            \caption{$G(2)$ level 4}
        \end{subfigure}
        \begin{subfigure}{\apwidth}
            \centering
            \includegraphics[width=1.0\linewidth]{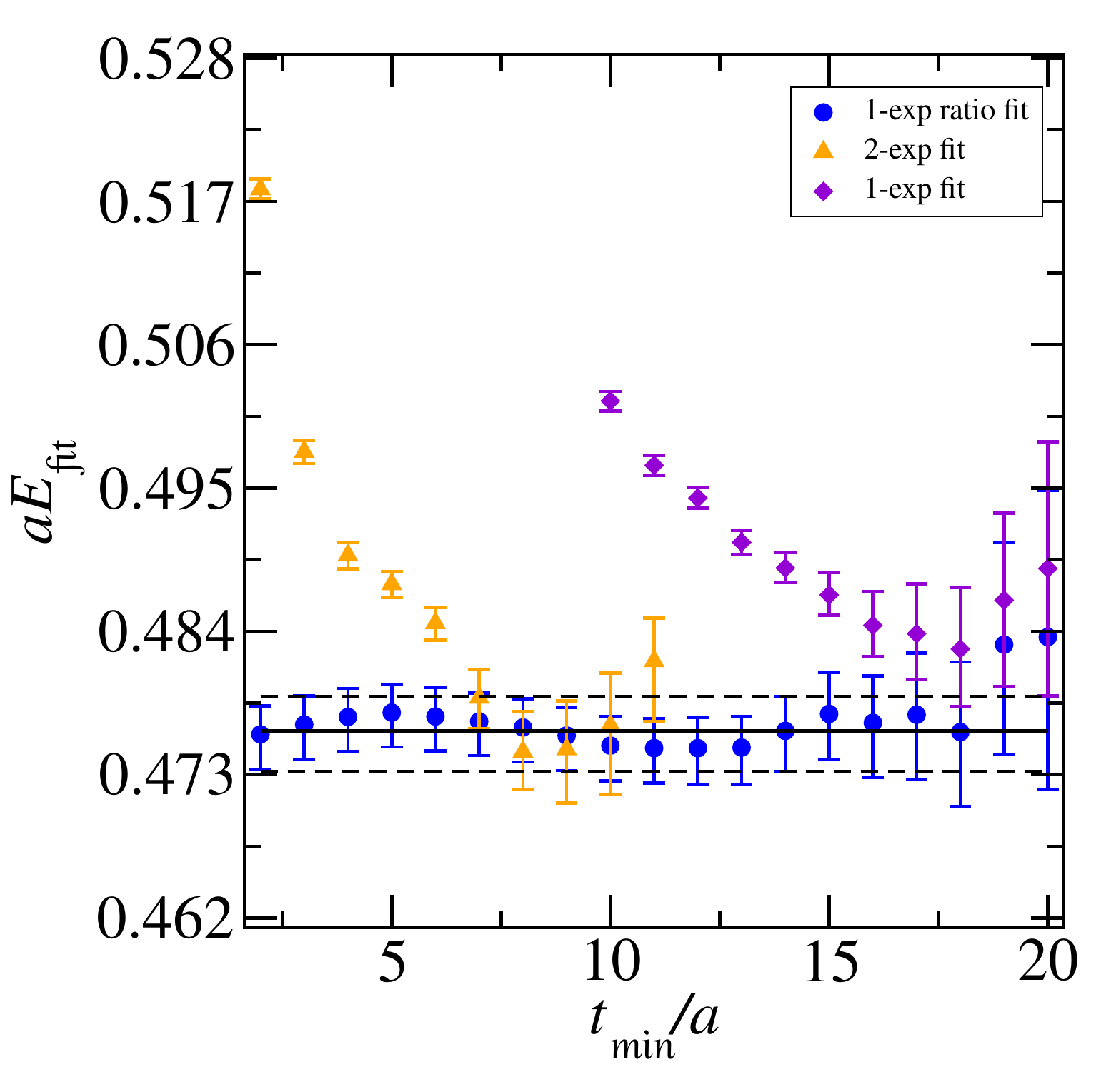}
            \caption{$G(2)$ level 5}
        \end{subfigure}
        \begin{subfigure}{\apwidth}
            \centering
            \includegraphics[width=1.0\linewidth]{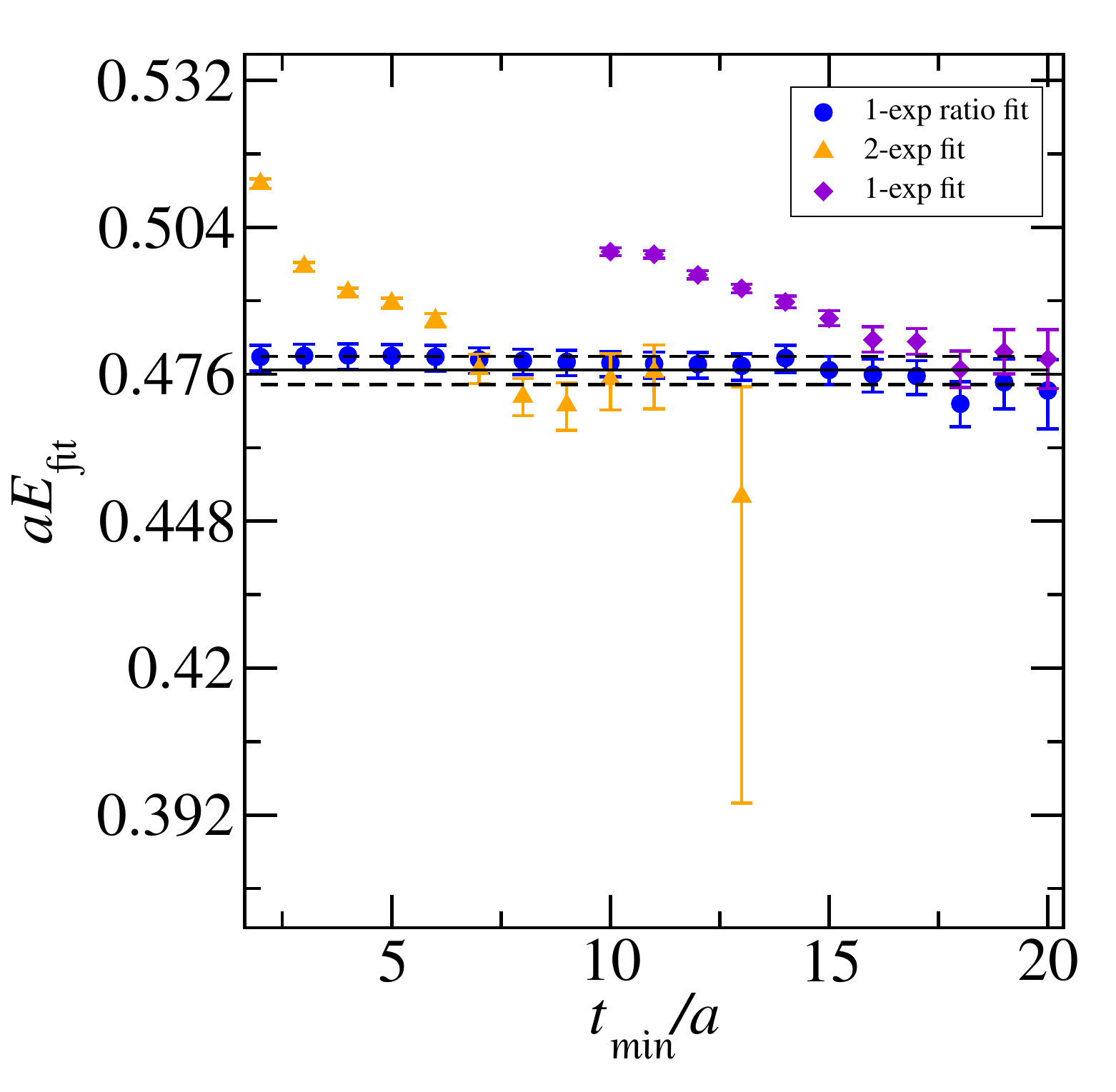}
            \caption{$G(2)$ level 6}
        \end{subfigure}
        \begin{subfigure}{\apwidth}
            \centering
            \includegraphics[width=1.0\linewidth]{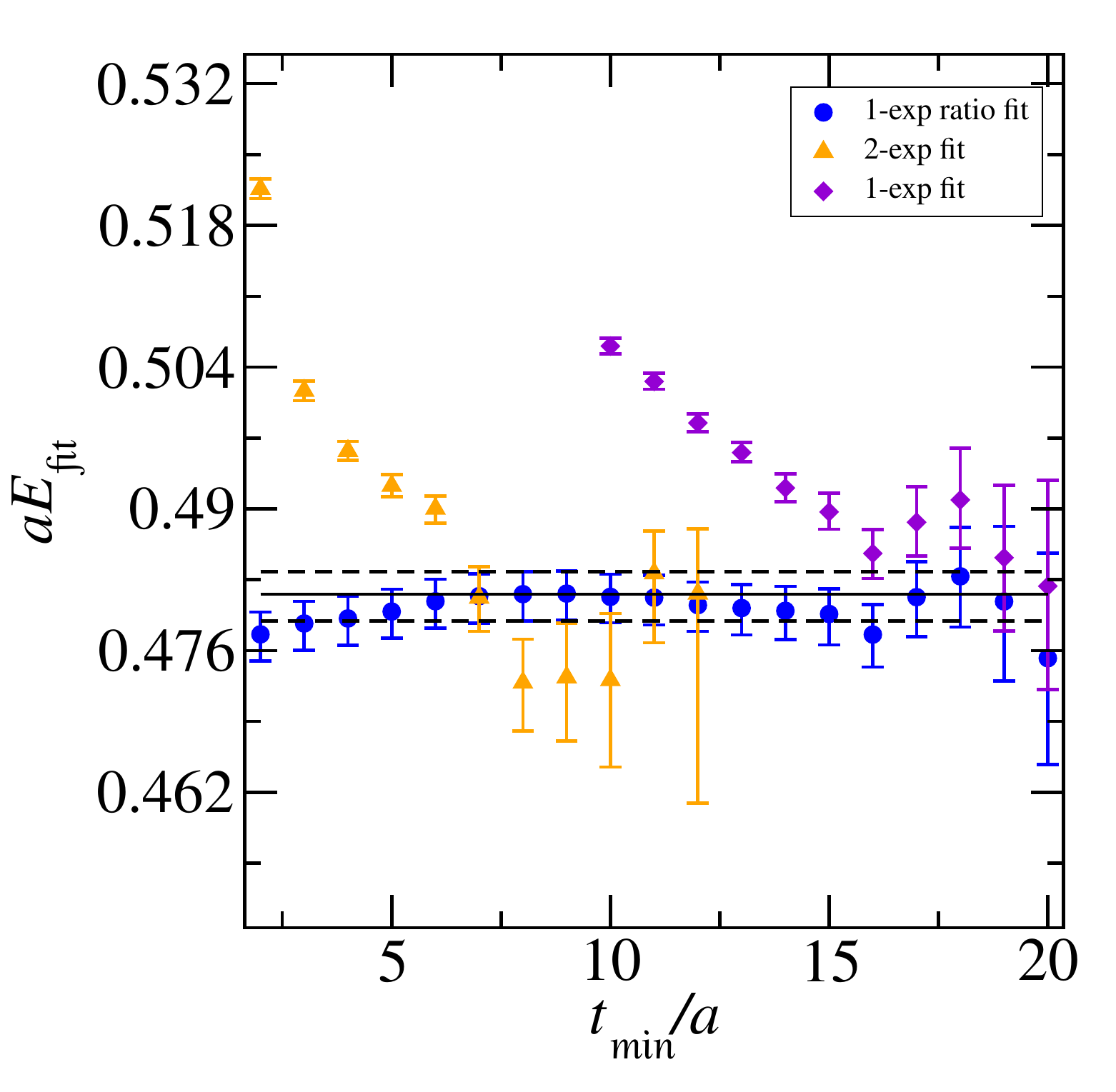}
            \caption{$G(2)$ level 7}
        \end{subfigure}
        \setcounter{subfigure}{2}
        \renewcommand{\thesubfigure}{\Roman{subfigure}}
    \end{subfigure}
\caption{\label{fig:isoquartet_P2_tmin} Same as Fig.~\ref{fig:isoquartet_P0_tmin} for $I=3/2$ 
        except that $\boldsymbol{d}^2=2$.} 
\end{figure}

\begin{figure}[p]
    \begin{subfigure}{\textwidth}
        \setcounter{subfigure}{0}
        \renewcommand{\thesubfigure}{\alph{subfigure}}
        \centering
        \begin{subfigure}{\apwidth}
            \centering
            \includegraphics[width=1.0\linewidth]{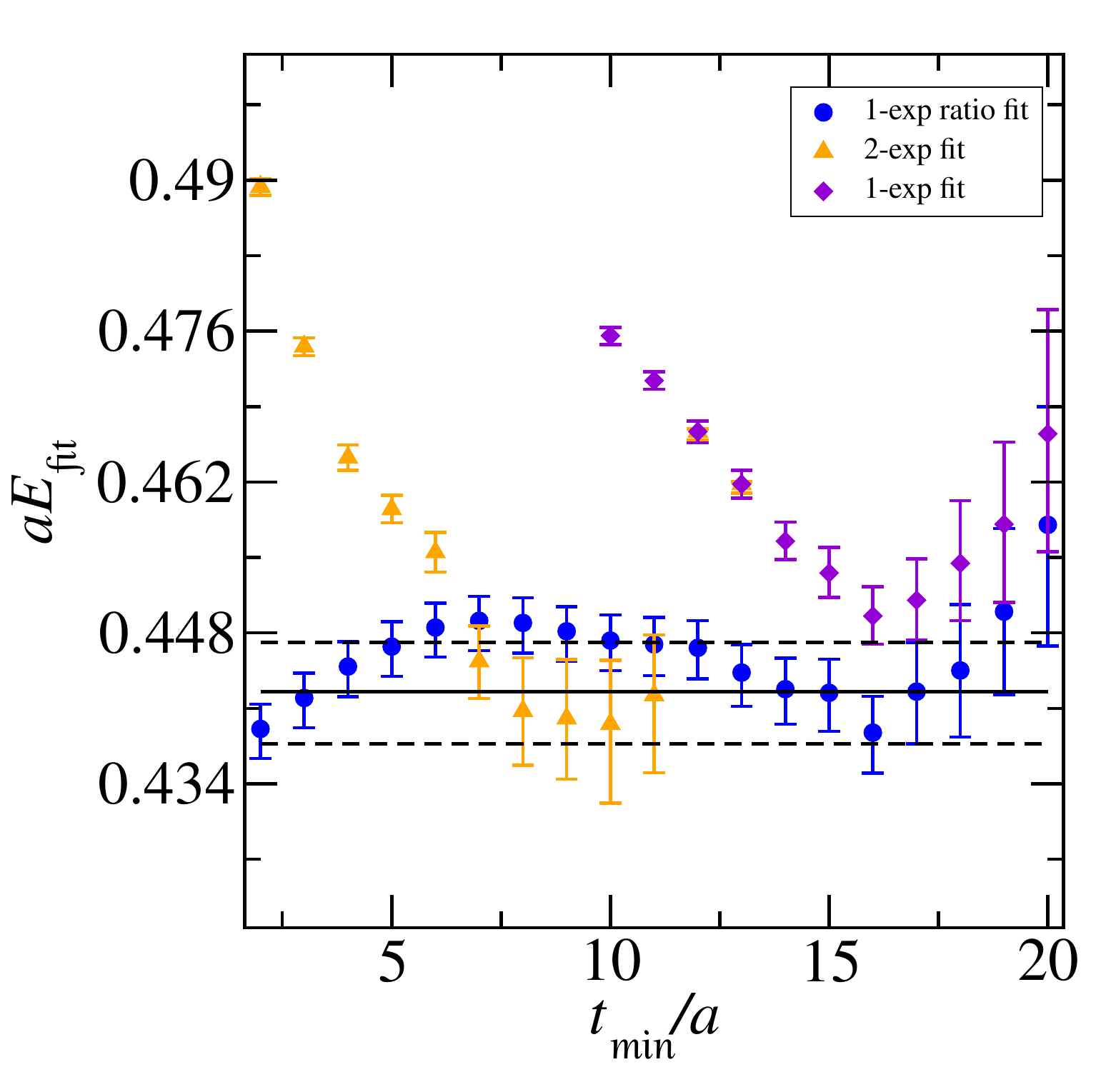}
            \caption{$F_1(3)$ level 0}
        \end{subfigure}
        \begin{subfigure}{\apwidth}
            \centering
            \includegraphics[width=1.0\linewidth]{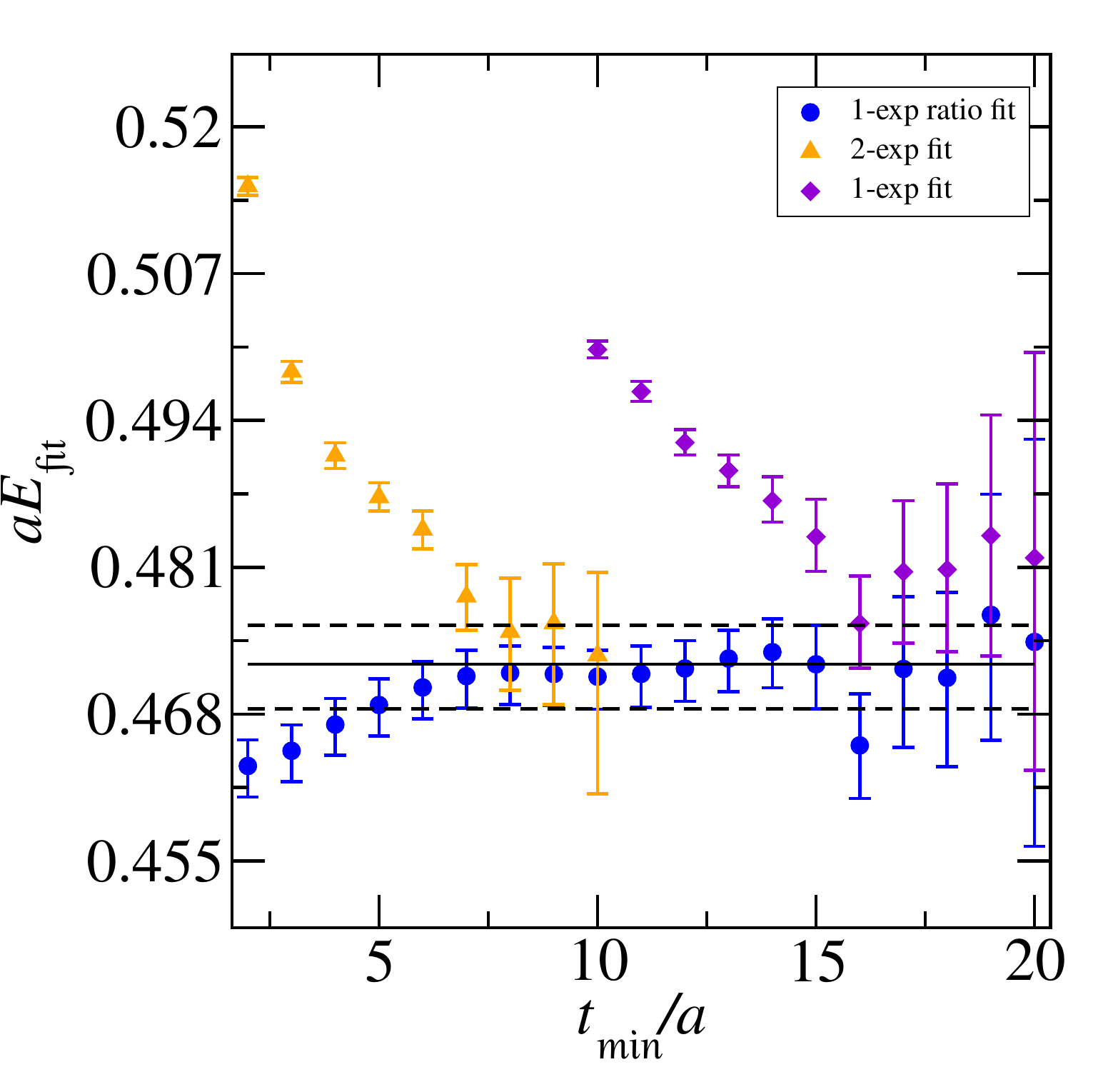}
            \caption{$F_1(3)$ level 1}
        \end{subfigure}
        \begin{subfigure}{\apwidth}
            \centering
            \includegraphics[width=1.0\linewidth]{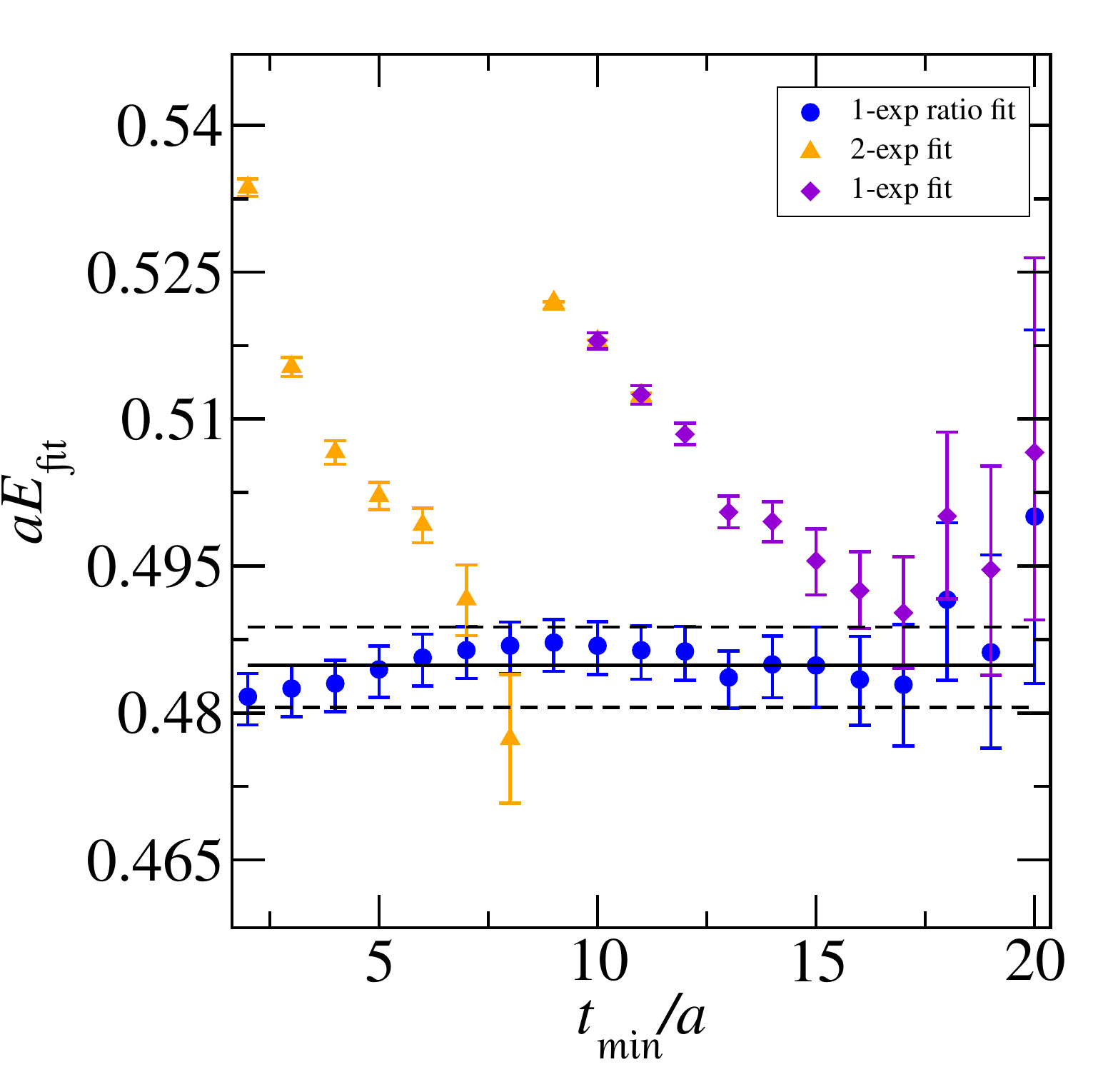}
            \caption{$F_1(3)$ level 2}
        \end{subfigure}
        \begin{subfigure}{\apwidth}
            \centering
            \includegraphics[width=1.0\linewidth]{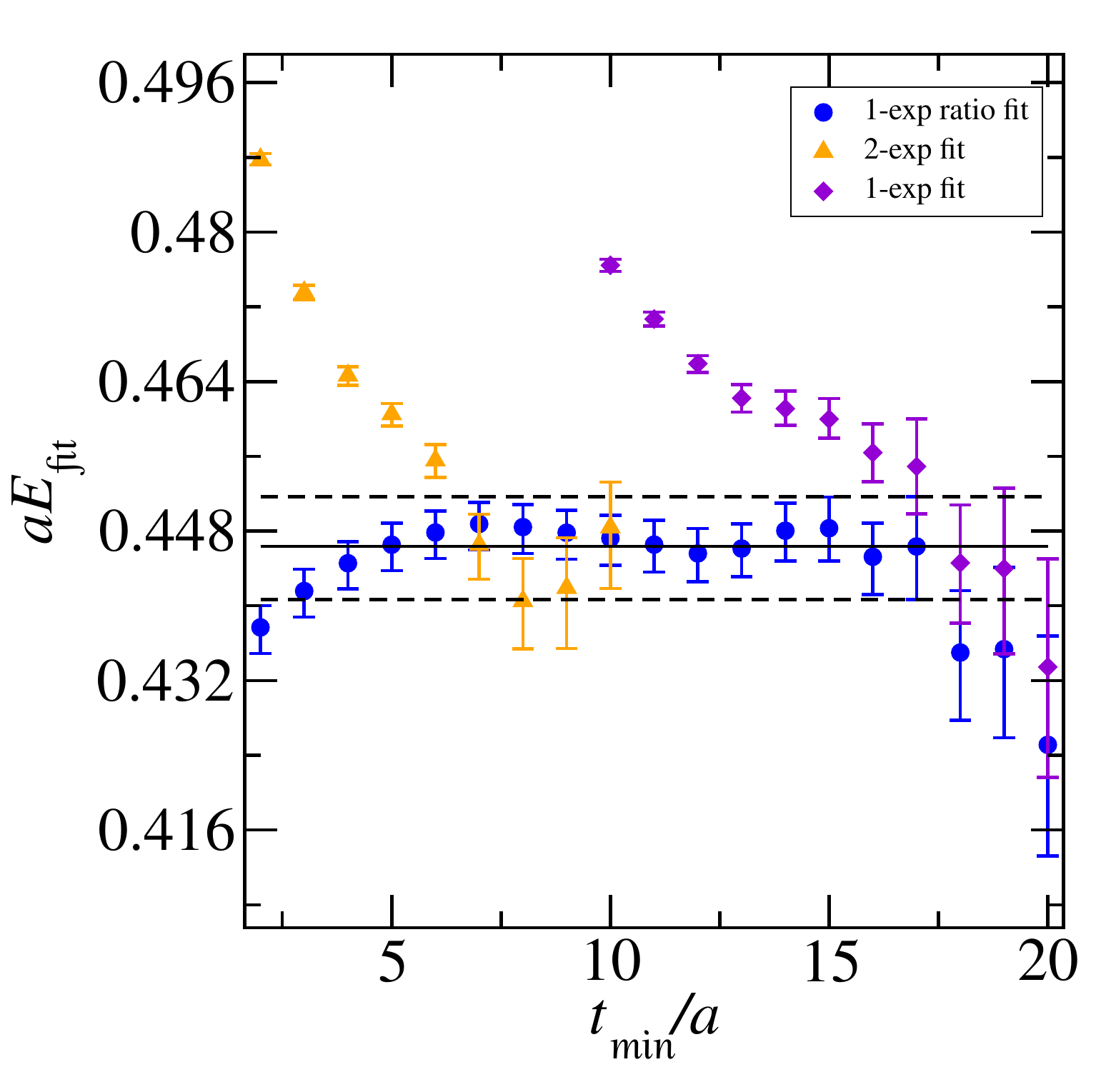}
            \caption{$F_2(3)$ level 0}
        \end{subfigure}
        \begin{subfigure}{\apwidth}
            \centering
            \includegraphics[width=1.0\linewidth]{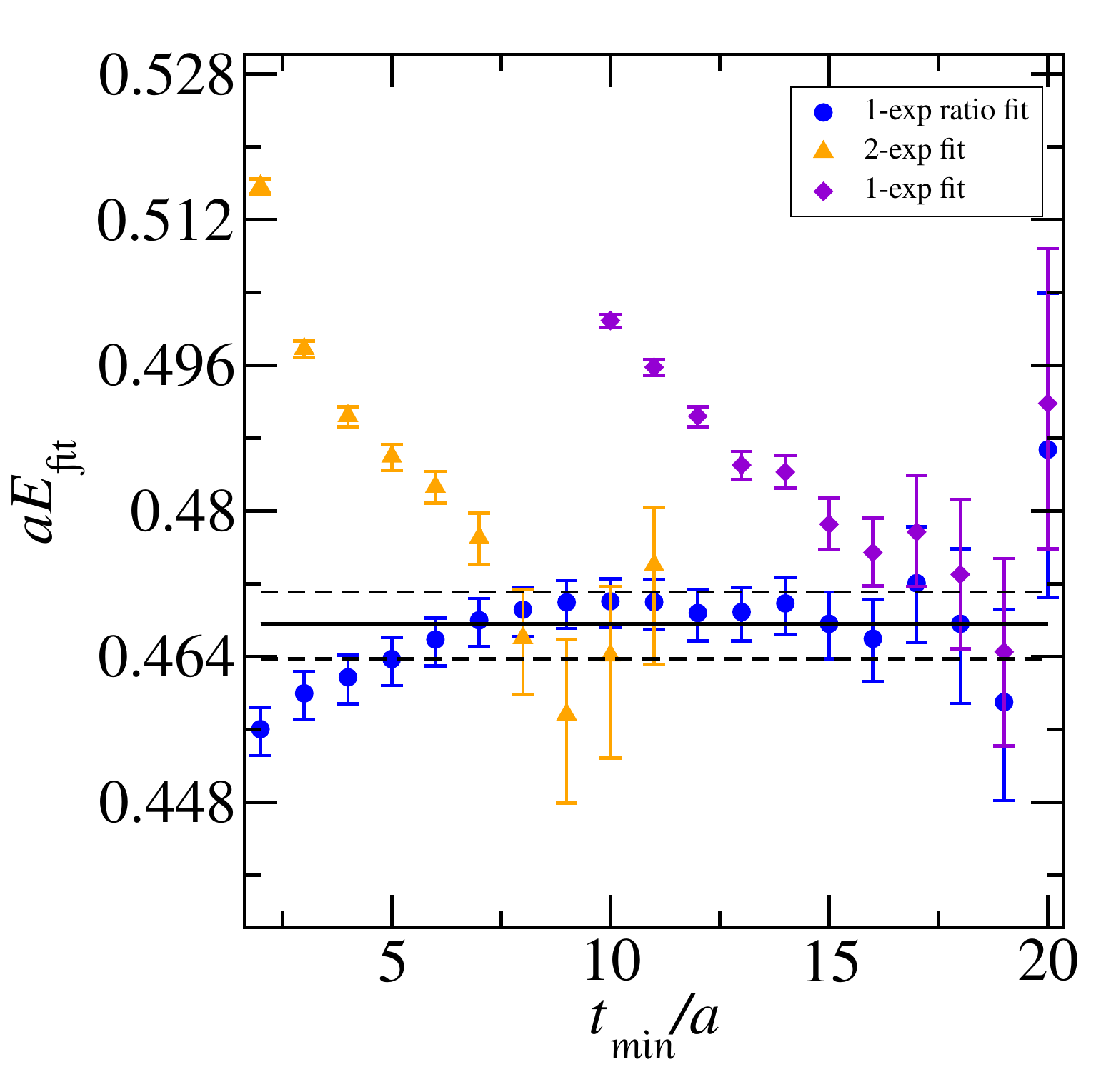}
            \caption{$F_2(3)$ level 1}
        \end{subfigure}
        \begin{subfigure}{\apwidth}
            \centering
            \includegraphics[width=1.0\linewidth]{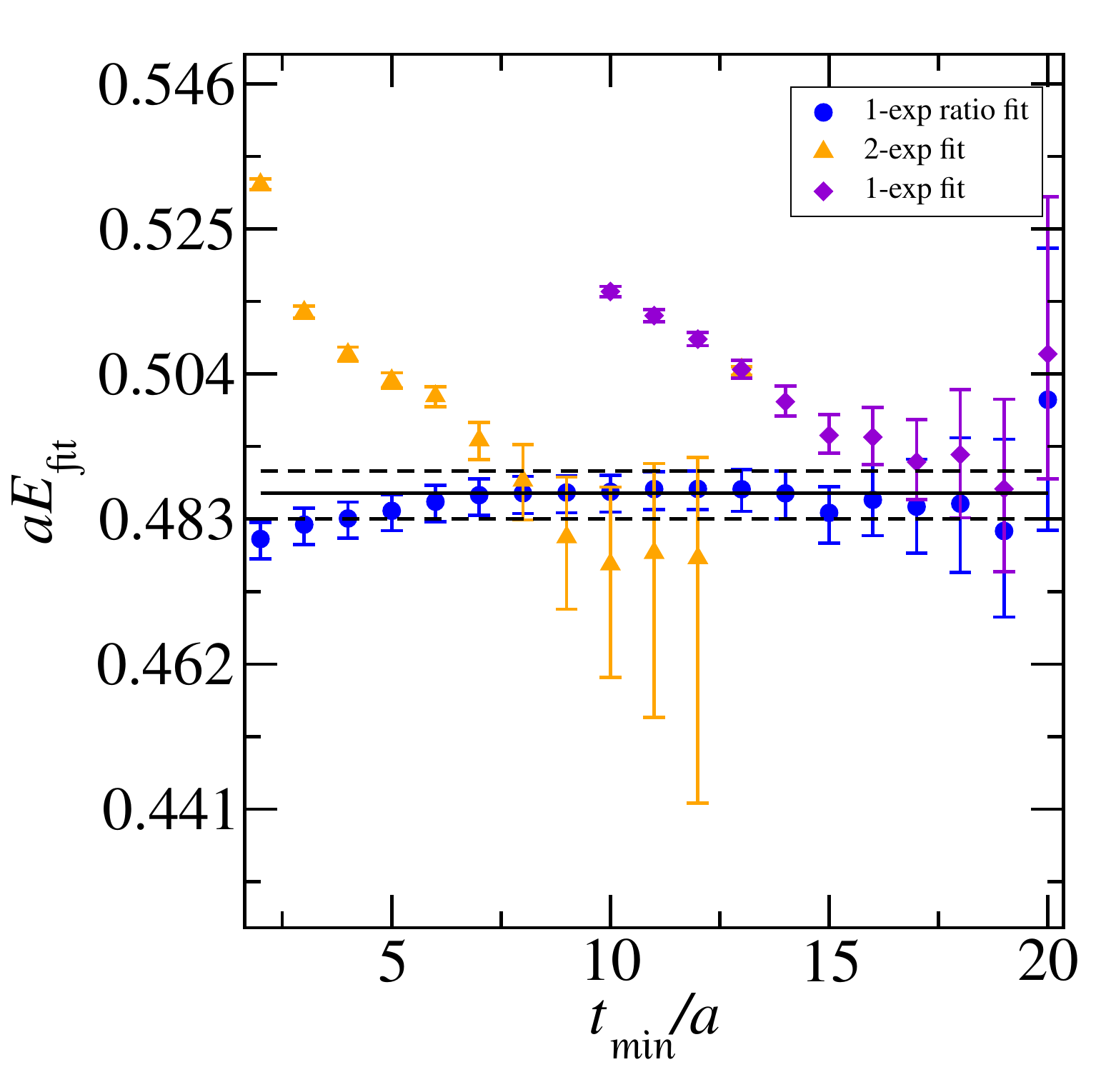}
            \caption{$F_2(3)$ level 2}
        \end{subfigure}
        \begin{subfigure}{\apwidth}
            \centering
            \includegraphics[width=1.0\linewidth]{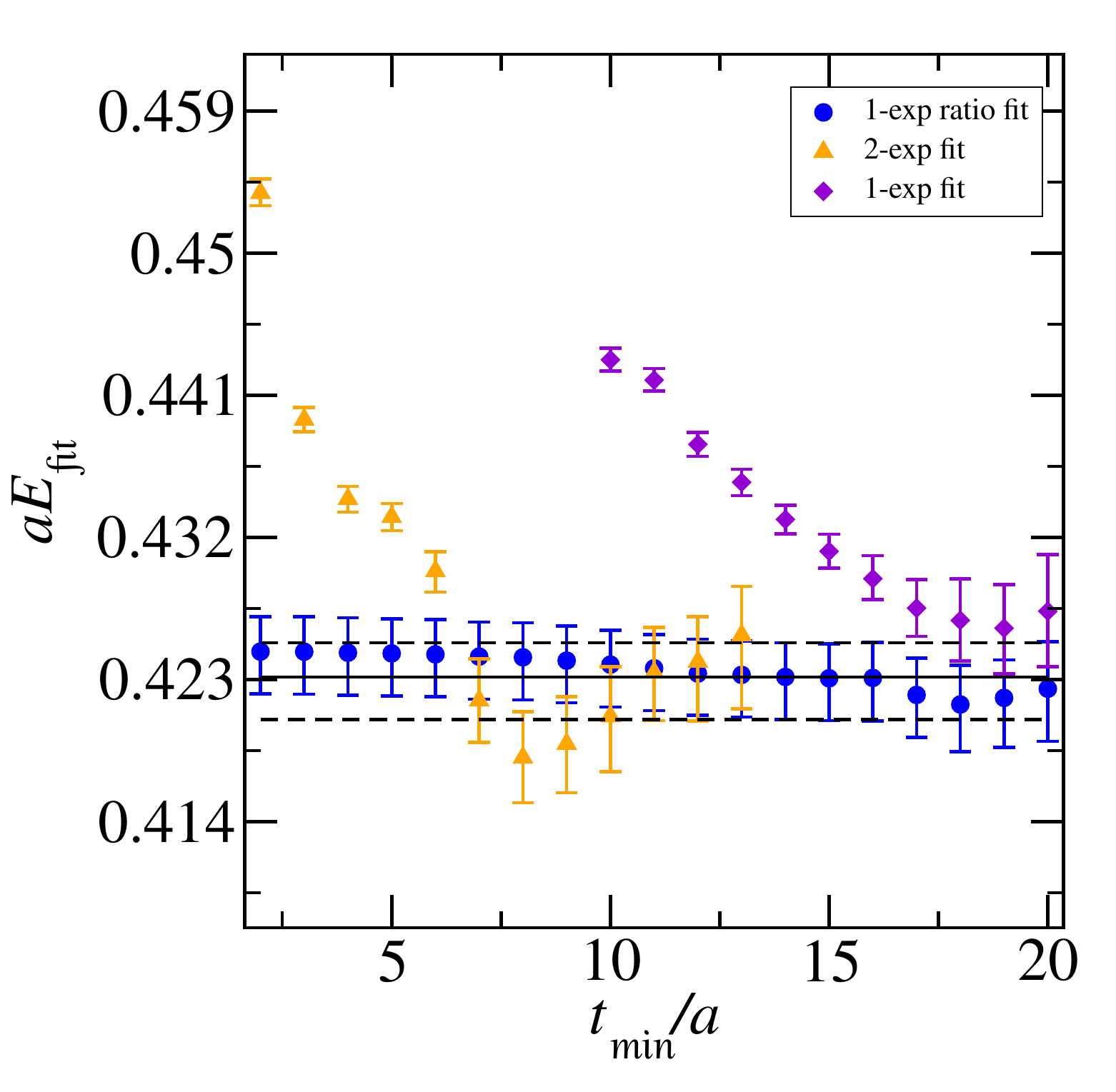}
            \caption{$G(3)$ level 0}
        \end{subfigure}
        \begin{subfigure}{\apwidth}
            \centering
            \includegraphics[width=1.0\linewidth]{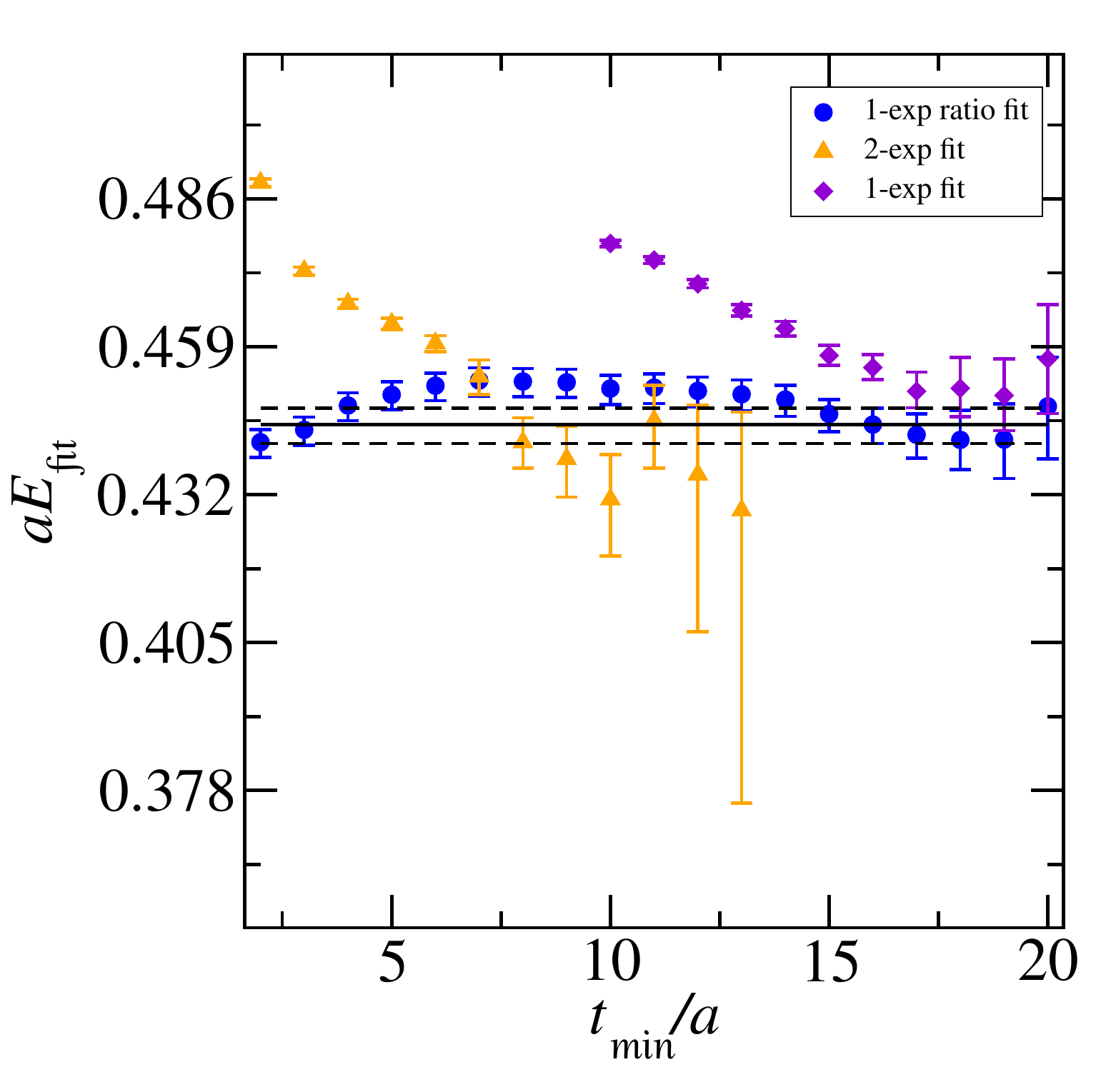}
            \caption{$G(3)$ level 1}
        \end{subfigure}
        \begin{subfigure}{\apwidth}
            \centering
            \includegraphics[width=1.0\linewidth]{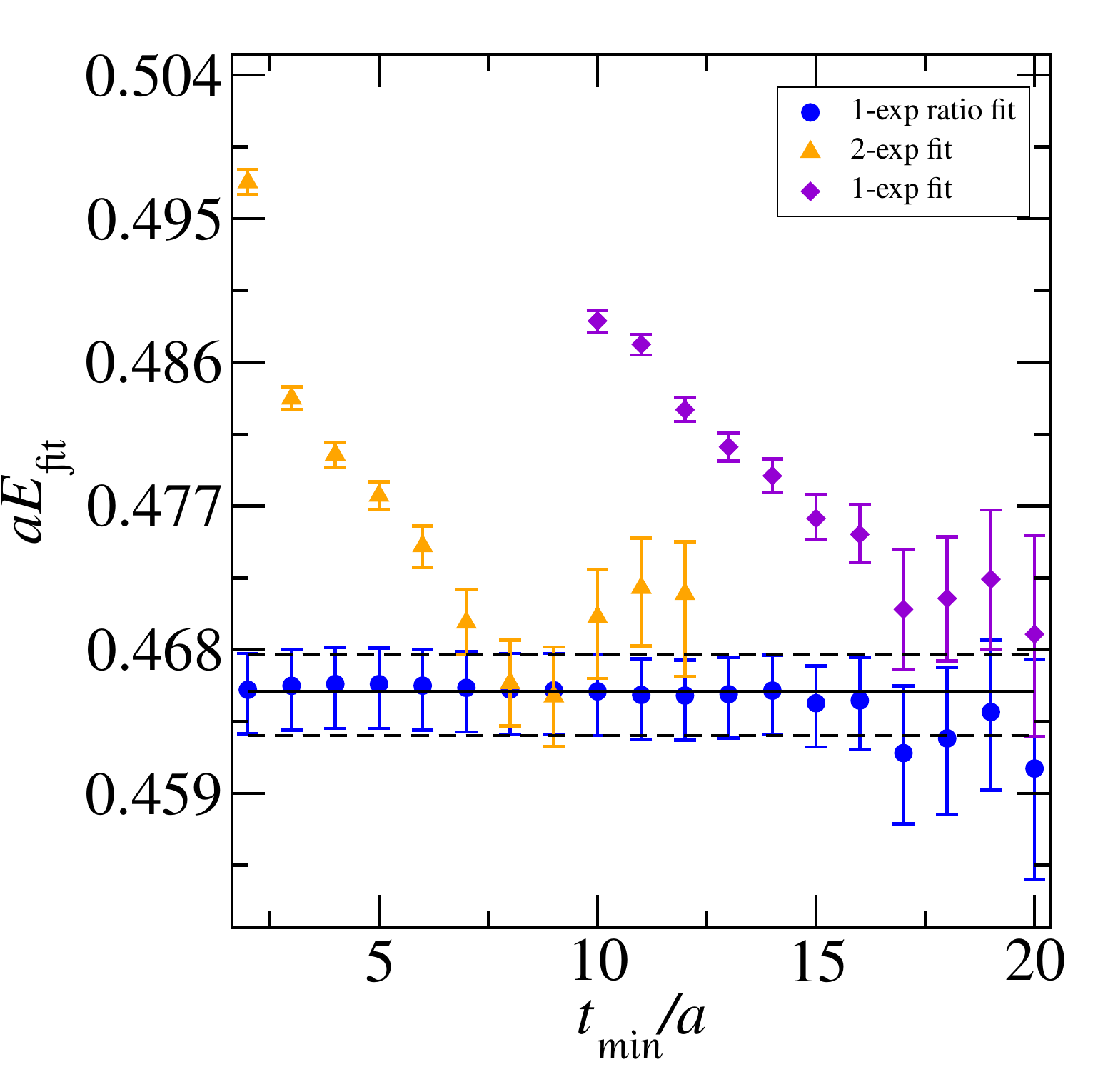}
            \caption{$G(3)$ level 2}
        \end{subfigure}
        \begin{subfigure}{\apwidth}
            \centering
            \includegraphics[width=1.0\linewidth]{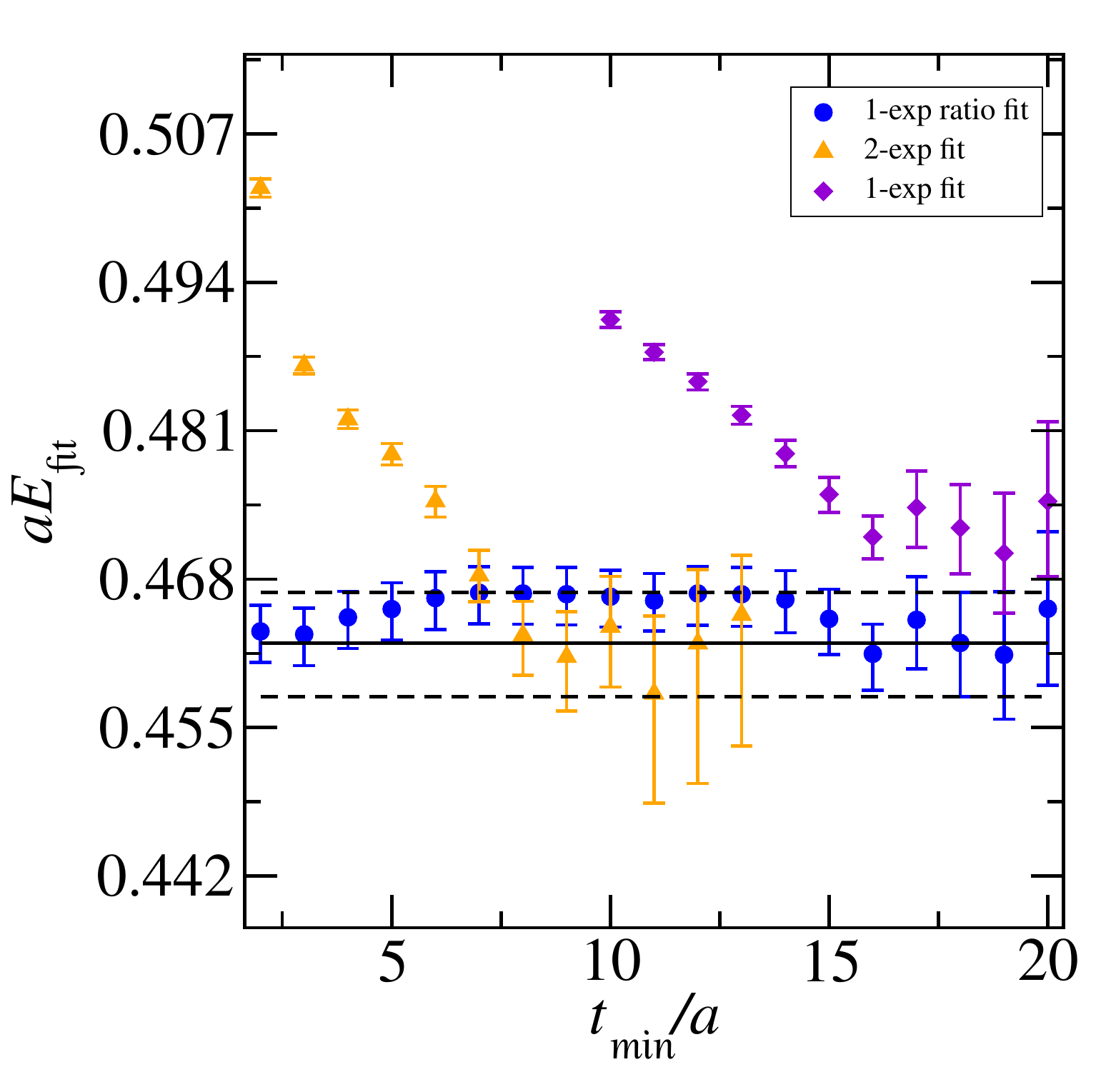}
            \caption{$G(3)$ level 3}
        \end{subfigure}
        \begin{subfigure}{\apwidth}
            \centering
            \includegraphics[width=1.0\linewidth]{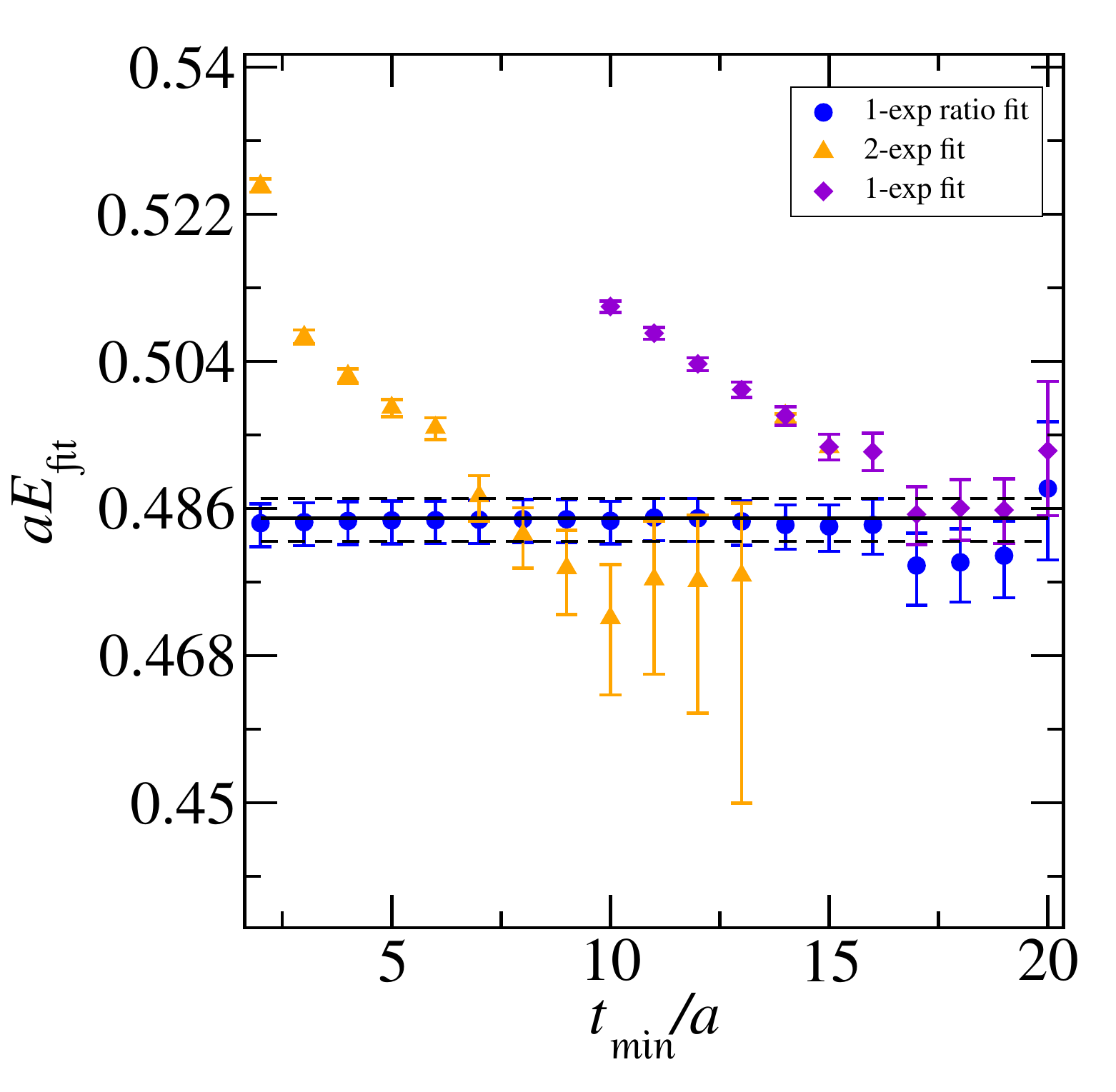}
            \caption{$G(3)$ level 4}
        \end{subfigure}
        \begin{subfigure}{\apwidth}
            \centering
            \includegraphics[width=1.0\linewidth]{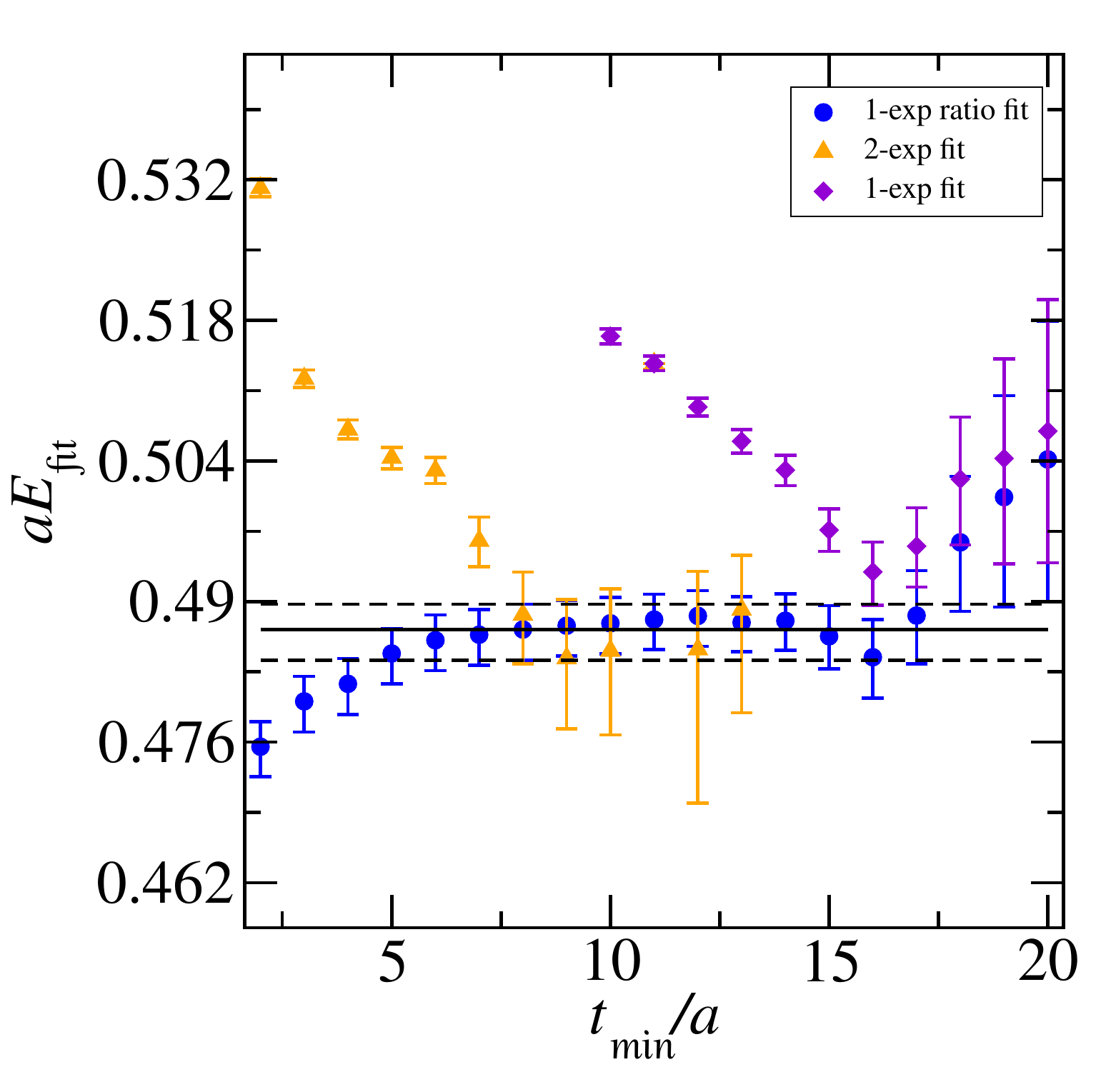}
            \caption{$G(3)$ level 5}
        \end{subfigure}
        \setcounter{subfigure}{3}
        \renewcommand{\thesubfigure}{\Roman{subfigure}}
    \end{subfigure}
\caption{\label{fig:isoquartet_P3_tmin} Same as Fig.~\ref{fig:isoquartet_P0_tmin} for $I=3/2$ except that 
$\boldsymbol{d}^2=3$.} 
\end{figure}

\begin{figure}[p]
    \begin{subfigure}{\textwidth}
        \setcounter{subfigure}{0}
        \renewcommand{\thesubfigure}{\alph{subfigure}}
        \centering
        \begin{subfigure}{\apwidth}
            \centering
            \includegraphics[width=1.0\linewidth]{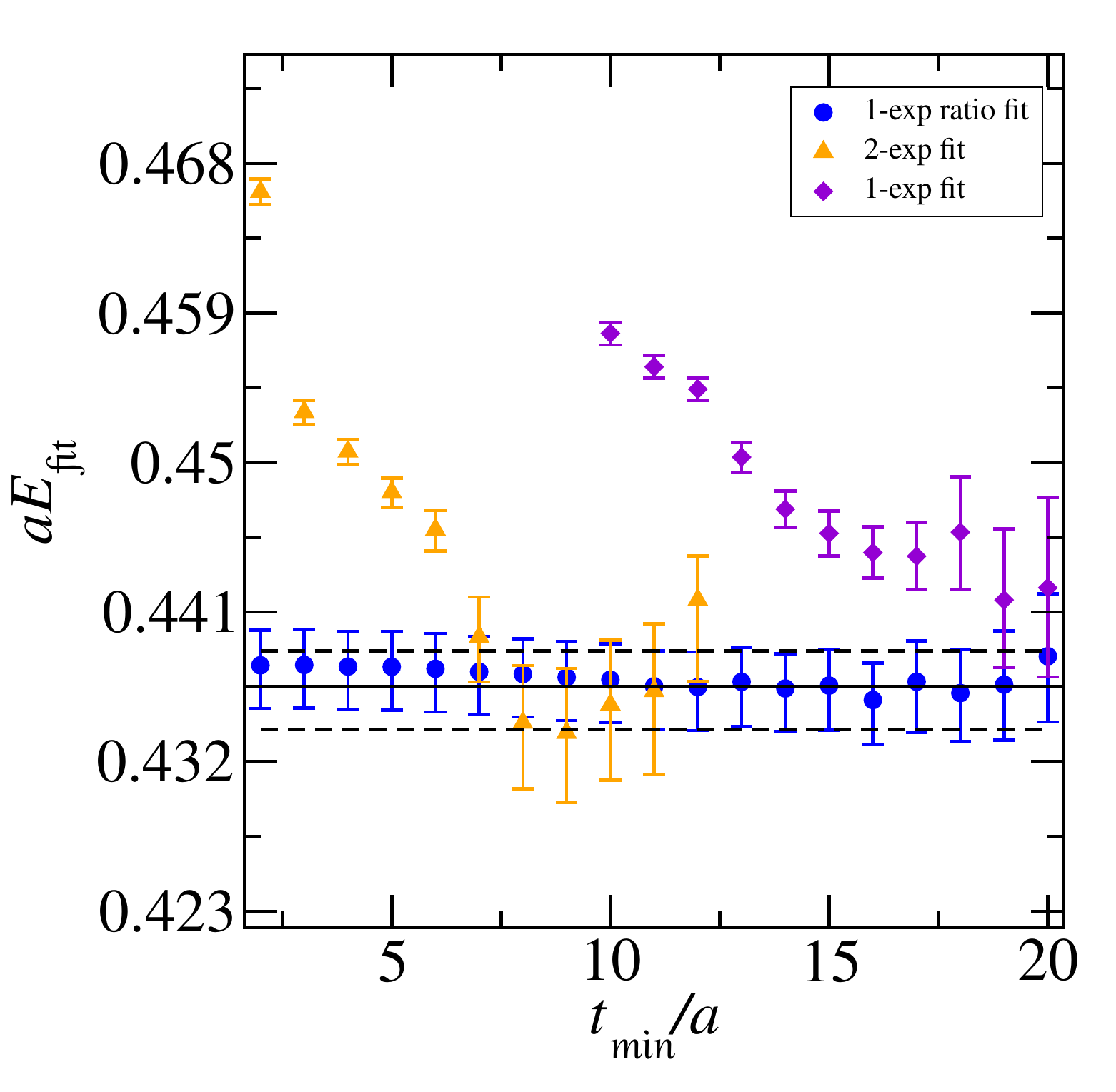}
            \caption{$G_1(4)$ level 0}
        \end{subfigure}
        \begin{subfigure}{\apwidth}
            \centering
            \includegraphics[width=1.0\linewidth]{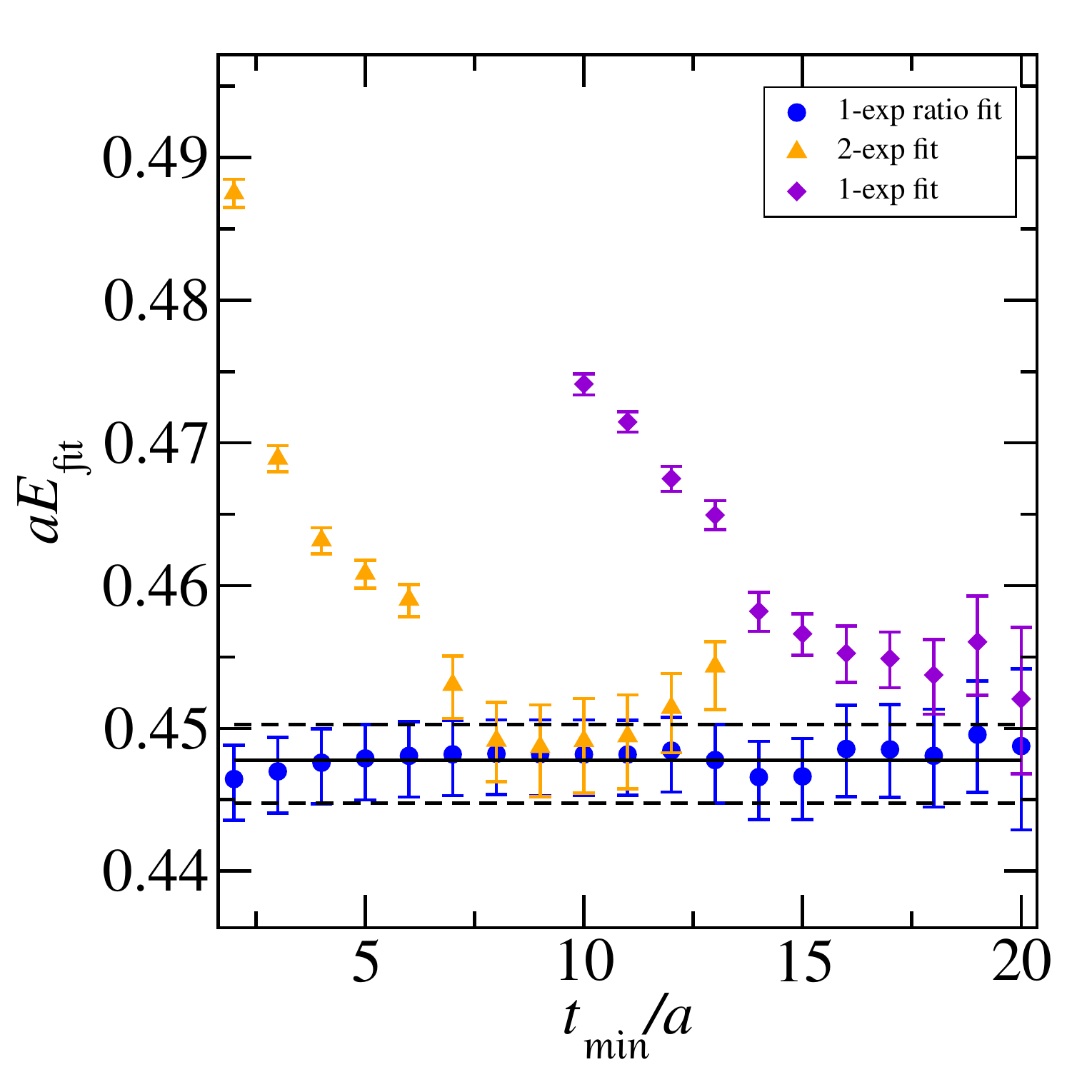}
            \caption{$G_1(4)$ level 1}
        \end{subfigure}
        \begin{subfigure}{\apwidth}
            \centering
            \includegraphics[width=1.0\linewidth]{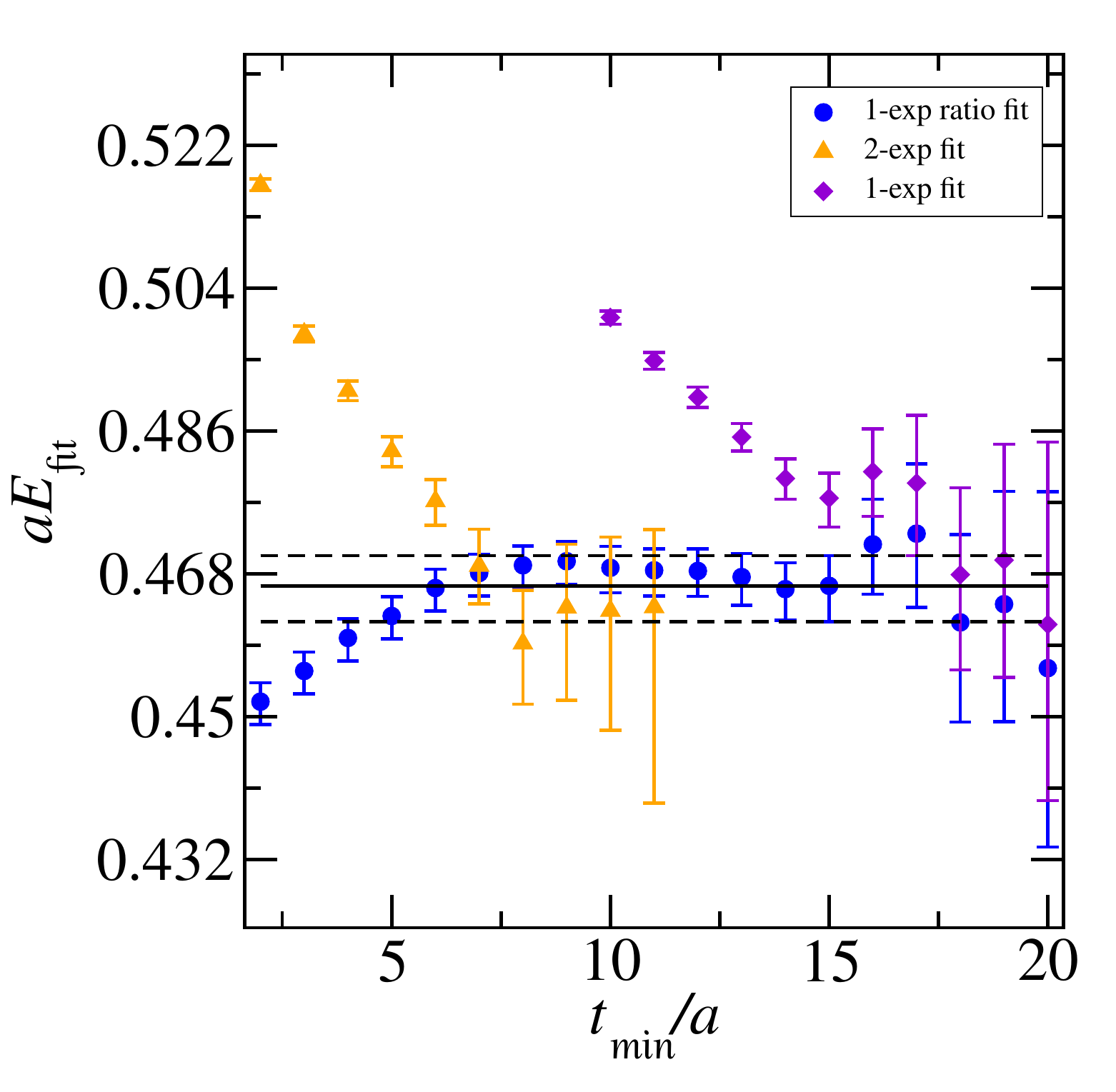}
            \caption{$G_1(4)$ level 2}
        \end{subfigure}
        \begin{subfigure}{\apwidth}
            \centering
            \includegraphics[width=1.0\linewidth]{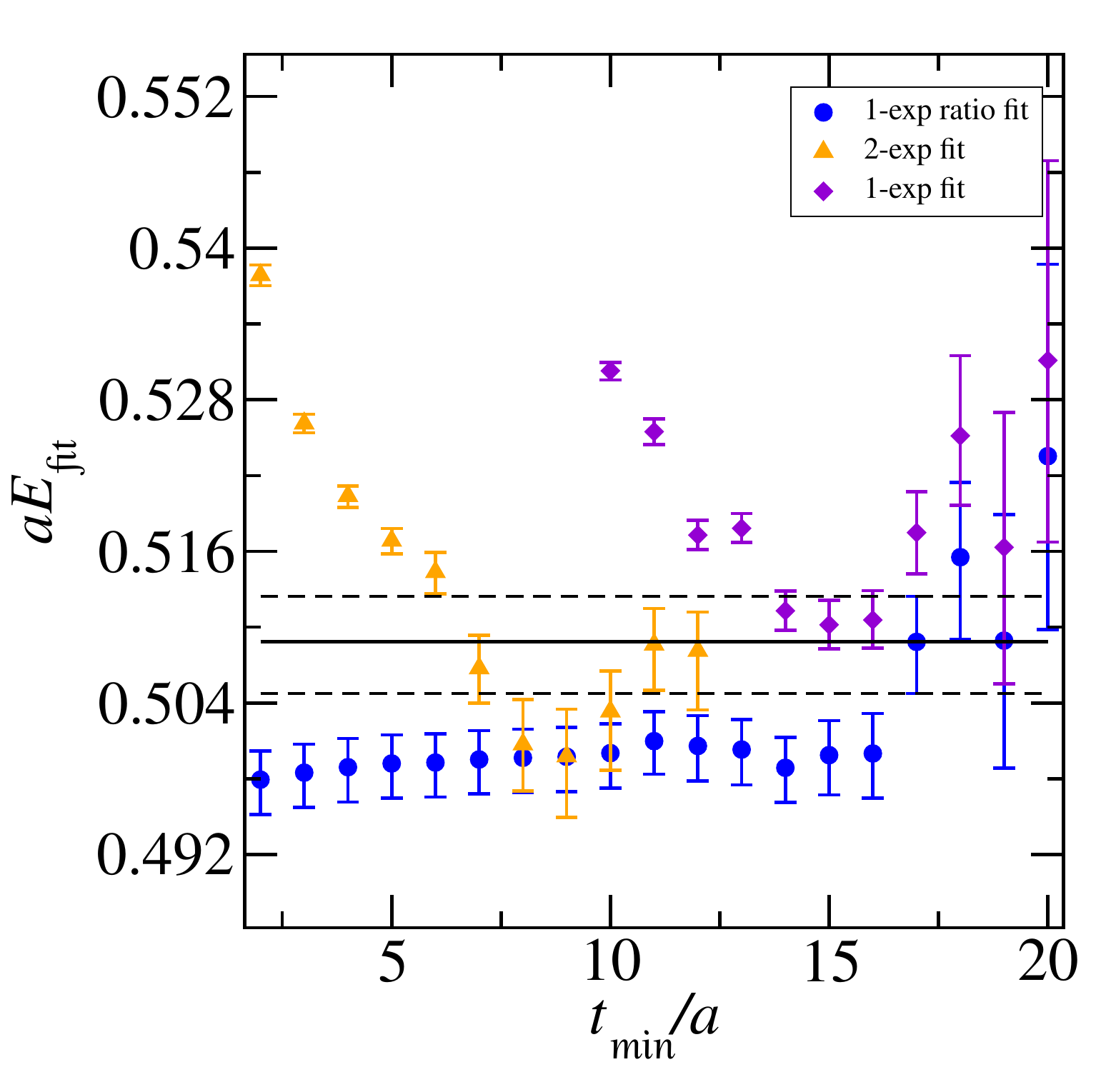}
            \caption{$G_1(4)$ level 3}
        \end{subfigure}
        \begin{subfigure}{\apwidth}
            \centering
            \includegraphics[width=1.0\linewidth]{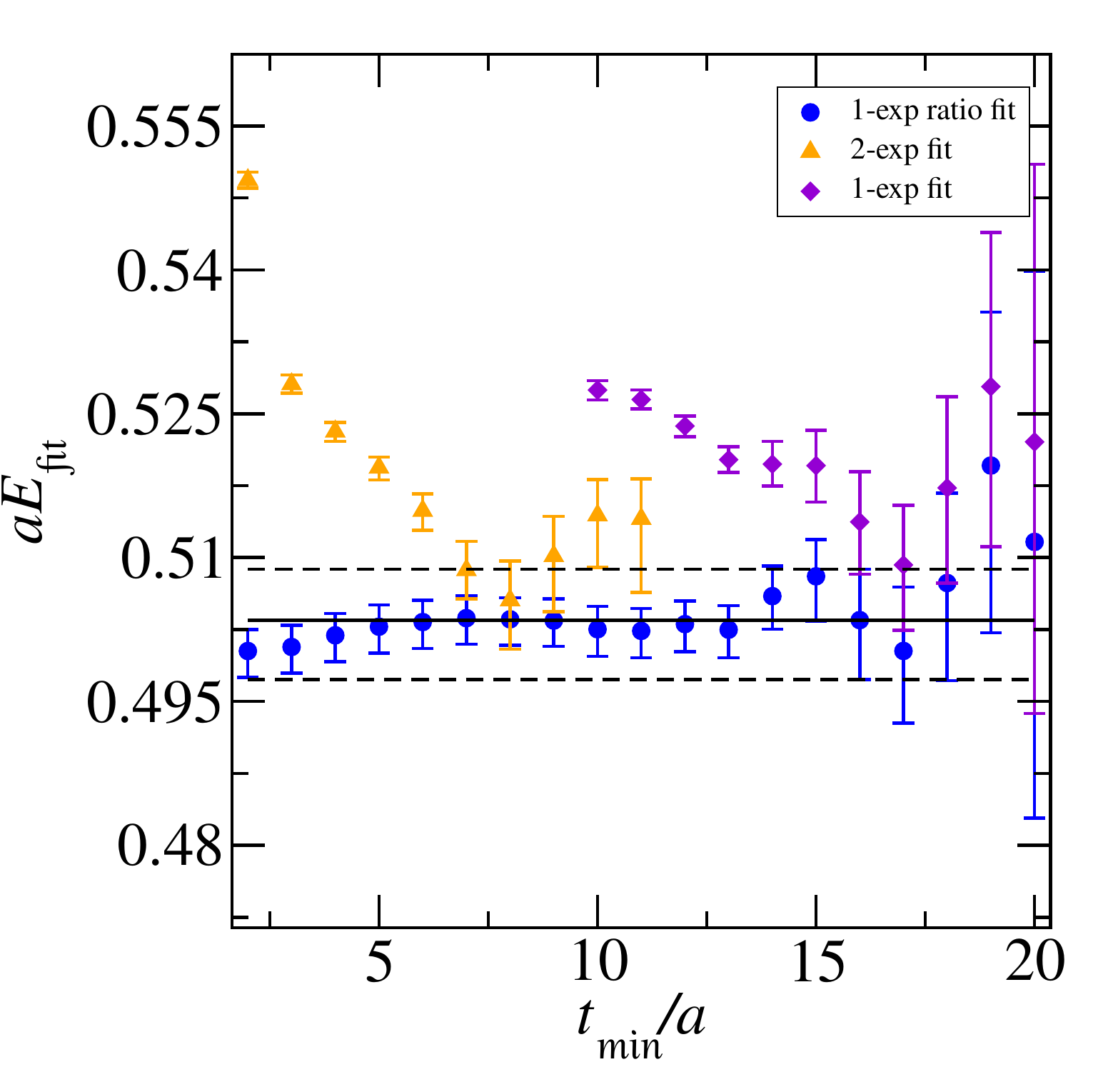}
            \caption{$G_1(4)$ level 4}
        \end{subfigure}
        \begin{subfigure}{\apwidth}
            \centering
            \includegraphics[width=1.0\linewidth]{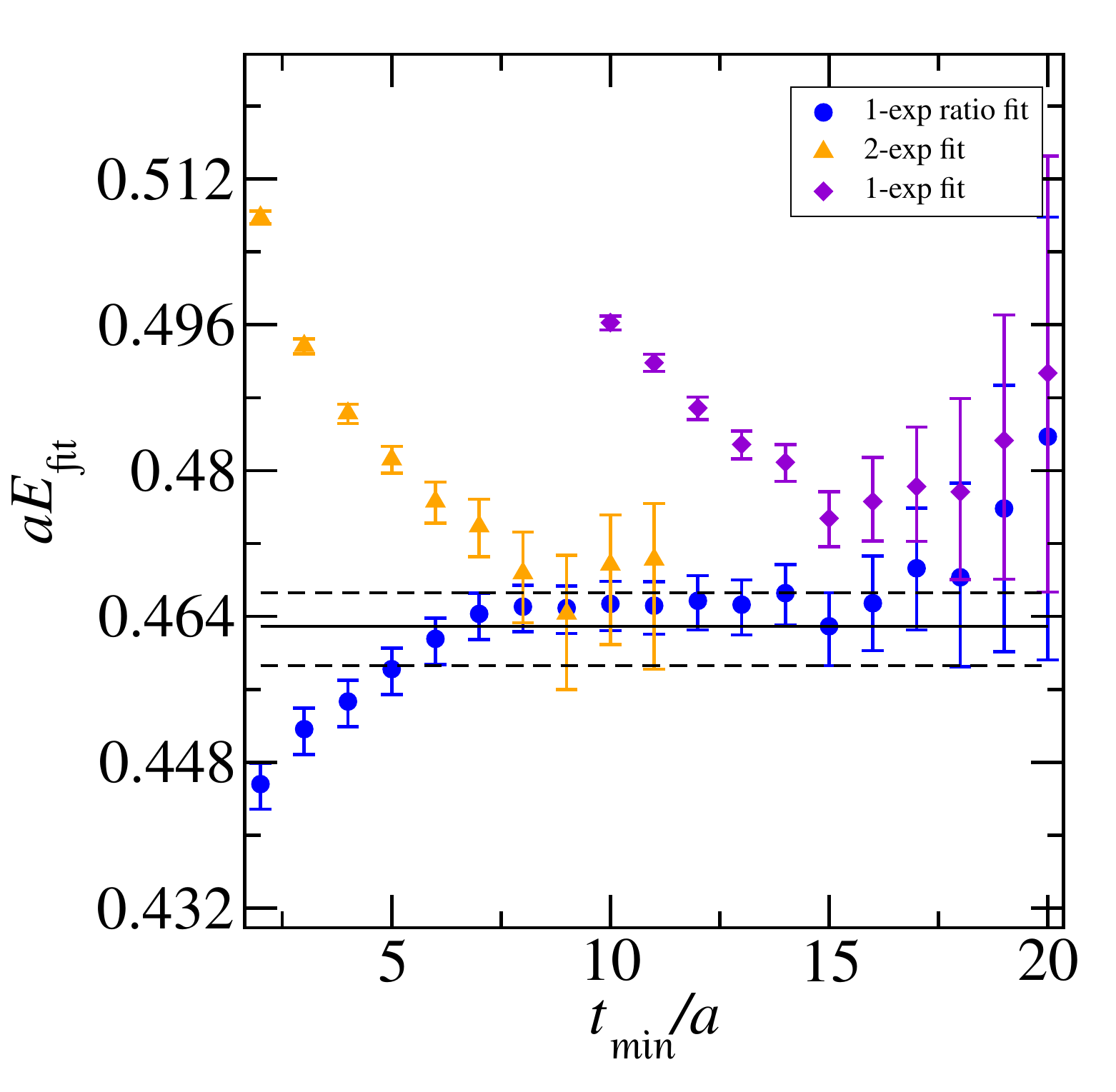}
            \caption{$G_2(4)$ level 0}
        \end{subfigure}
        \begin{subfigure}{\apwidth}
            \centering
            \includegraphics[width=1.0\linewidth]{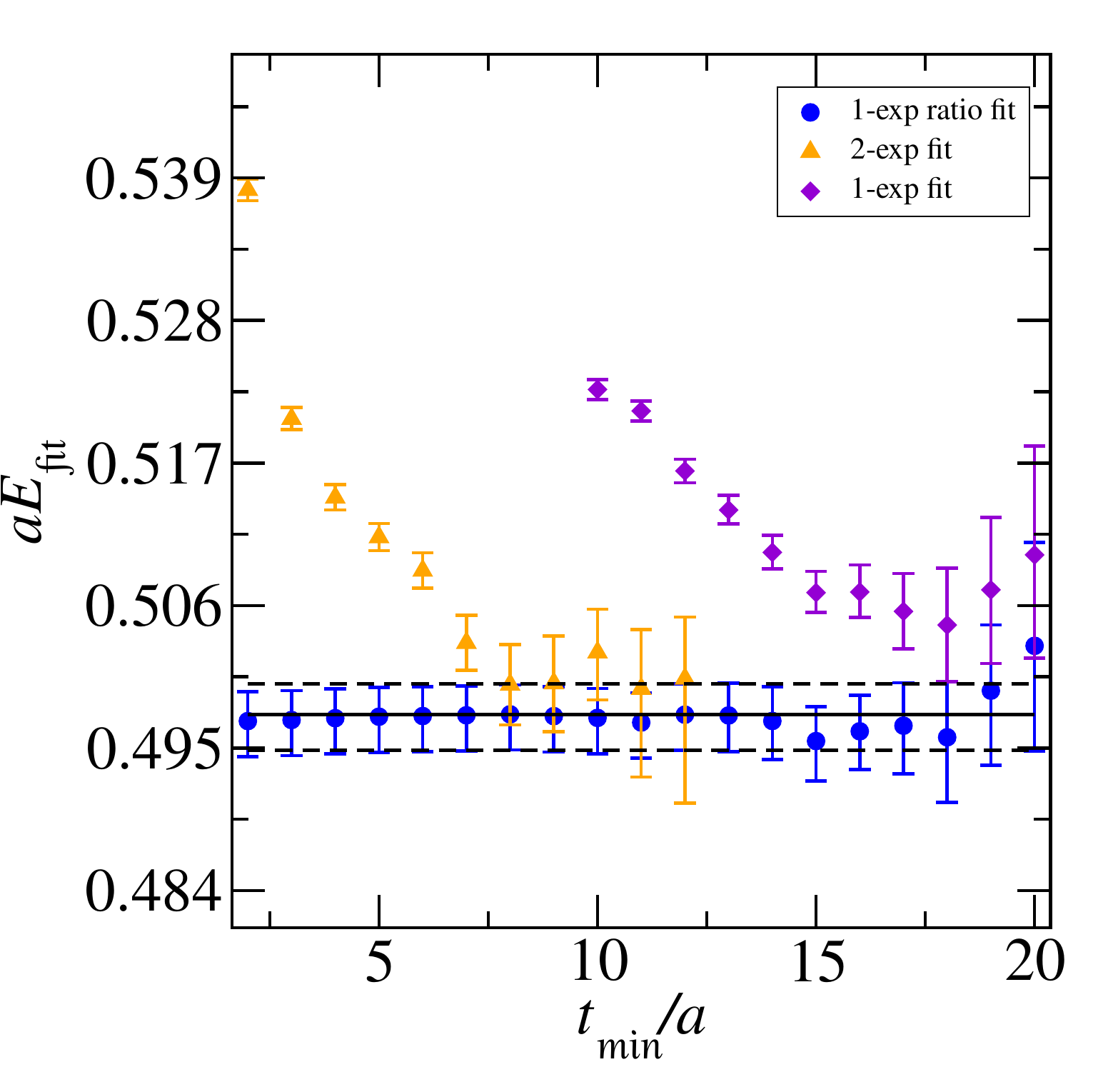}
            \caption{$G_2(4)$ level 1}
        \end{subfigure}
        \begin{subfigure}{\apwidth}
            \centering
            \includegraphics[width=1.0\linewidth]{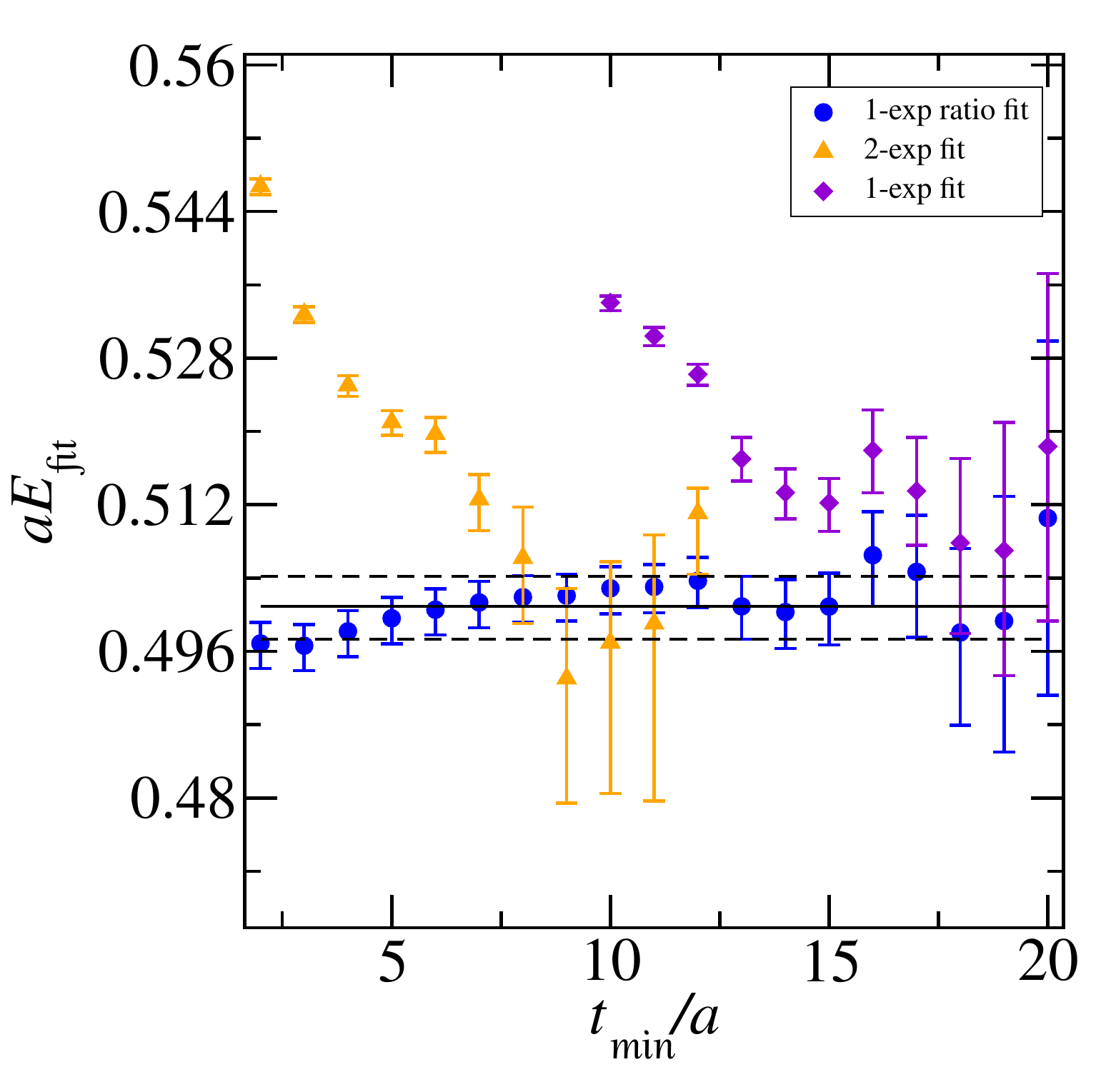}
            \caption{$G_2(4)$ level 2}
        \end{subfigure}\setcounter{subfigure}{4}
        \renewcommand{\thesubfigure}{\Roman{subfigure}}
    \end{subfigure}
    \caption{\label{fig:isoquartet_P4_tmin} Same as Fig.~\ref{fig:isoquartet_P0_tmin} for $I=3/2$ 
    except that $\boldsymbol{d}^2=4$.} 
\end{figure}

As discussed in Sec.~\ref{s:fit}, multiple fit ranges and fit forms are compared for every energy level to ensure 
systematic errors associated with excited state contamination are smaller than the statistical errors. Ultimately, 
single-exponential  fits to the correlator ratios in Eq.~(\ref{e:ratio}) are chosen due to their mild sensitivity to 
$t_\textup{min}$ and good statistical precision. The fit range is chosen to be consistent with the double-exponential 
$t_\textup{min}$ plateau, defined as the range of $t_{\rm min}$ for which the fitted energy exhibits no statistically 
significant variation. Most levels are additionally consistent with the  single-exponential fit plateau, although as 
shown in Fig.~\ref{f:mnpi} for $m_{\rm N}$, these fits may fail to describe correlators with significant excited-state 
contamination.  Plots analogous to the $t_{\rm min}$-plot in Fig.~\ref{fig:spectrum_determination} are shown for 
each of the $I=1/2$ levels in Fig.~\ref{fig:isodoublet_tmin} and the $I=3/2$ levels in 
Figs.~\ref{fig:isoquartet_P0_tmin}-\ref{fig:isoquartet_P4_tmin}.

\end{document}